\documentclass[%
 reprint,
superscriptaddress,
nofootinbib,
 amsmath,amssymb,
 aps,
]{revtex4-1}

\usepackage{graphicx}
\usepackage{dcolumn}
\usepackage{bm}
\usepackage{hyperref}

\usepackage{multirow}
\newcommand\Tstrut{\rule{0pt}{3.3ex}}         
\newcommand\Bstrut{\rule[-2ex]{0pt}{0pt}}   

\hypersetup{
    colorlinks = true
}
\usepackage{aas_macros}

\newcommand{\bk}{\boldsymbol{k}}
\newcommand{\B}[1]{\ensuremath{\boldsymbol{#1}}}
\newcommand{\D}[1]{\ensuremath{\text{d}#1}}

\def\fun#1#2{\lower3.6pt\vbox{\baselineskip0pt\lineskip.9pt
        \ialign{$\mathsurround=0pt#1\hfill##\hfil$\crcr#2\crcr\sim\crcr}}}

\newcommand{\beq}{\begin{equation}}
\newcommand{\eeq}{\end{equation}}
\newcommand{\beqa}{\begin{eqnarray}}
\newcommand{\eeqa}{\end{eqnarray}}

\begin{document}


\title{Bias loop corrections to the galaxy bispectrum}

\author{Alexander Eggemeier}
\email{alexander.eggemeier@durham.ac.uk}
\affiliation{%
 Astronomy Centre, School of Mathematical and Physical Sciences, University of Sussex, Brighton BN1 9QH, United
 Kingdom
}
\affiliation{%
  Institute for Computational Cosmology, Department of Physics, Durham University, South Road, Durham DH1 3LE,
  United Kingdom
}%

\author{Rom\'an Scoccimarro}
\affiliation{
 Center for Cosmology and Particle Physics, Department of Physics, New York University, NY 10003, New York, USA
}%

\author{Robert E. Smith}
\affiliation{%
 Astronomy Centre, School of Mathematical and Physical Sciences, University of Sussex, Brighton BN1 9QH, United
 Kingdom
}%

\date{\today}

\begin{abstract} 

  Combination of the power spectrum and bispectrum is a powerful way of breaking degeneracies between galaxy
  bias and cosmological parameters, enabling us to maximize the constraining power from galaxy surveys. Recent cosmological constraints treat the power spectrum and bispectrum on an uneven footing: they include one-loop bias corrections for the power spectrum but not the bispectrum. To bridge this gap, we develop the galaxy bias description up to fourth order in perturbation theory, conveniently expressed through a basis of Galilean invariants that clearly split contributions that are local and nonlocal in the second derivatives of the linear gravitational potential.  In addition, we consider relevant contributions  from short-range nonlocality (higher-derivative terms), stress-tensor corrections and stochasticity.  To sidestep the usual renormalization of bias parameters that complicates predictions beyond leading order, we
  recast the bias expansion in terms of multipoint propagators, which take a simple form in our split-basis with loop corrections depending only on bias parameters corresponding to nonlocal operators.  We show how to take advantage of Galilean invariance to compute the time evolution of bias and present results for the fourth-order parameters for the first time. We also discuss the possibilities of reducing the bias parameter space by using the evolution of bias and 
  exploiting degeneracies between various bias contributions  in the large-scale limit. Our baseline model allows to verify these simplifications for any application to  large-scale structure data sets.

\end{abstract}

\pacs{Valid PACS appear here}
\maketitle


\section{Introduction}
\label{sec:introduction}

Any mismodeling of galaxy bias --- the relation between galaxies (or any other luminous tracer) and the
underlying matter distribution \citep{Kaiser:1984,PolWis8410,BarBonKai8605,MoWhi9609,SheTor9909,Desjacques:2018}
--- risks an incorrect recovery of cosmological parameters from large-scale structure (LSS)
surveys. Fortunately, although the formation of galaxies involves highly nonlinear, small-scale processes,
recent developments \citep{McDonald:2006,McDonald:2009,Chan:2012, Baldauf:2012, Schmidt:2013, Assassi:2014,
  Mirbabayi:2015,FonRegSee1805} have shown that a perturbative expansion provides a robust treatment of galaxy
bias on sufficiently large scales. This comes at the price of a set of unknown bias parameters, which, once
marginalized over, degrade the statistical power for constraining cosmological models. It is therefore
imperative to combine traditional LSS two-point statistics with higher-order statistics, such as the bispectrum,
which allows us to break the degeneracies that exist between cosmological parameters and galaxy
bias~\cite{FriGaz94,GazFri94,Fry:1994,Matarrese:1997,ScoColFry9803,Sco00,SefCroPue0607}. For this program to
succeed we require consistent models of the galaxy power spectrum and bispectrum at leading (``tree-level'') and
next-to-leading (``one-loop'') order, which should improve the regime of validity and robustness of the
results~\cite{Pollack:2012,Bel:2015,SaiBalVla1405,Chan:2012}. This is particularly important since at present
the analysis of the bispectrum in galaxy surveys~\cite{GilNorVer1407,GilPerVer1606} is done in a manner that is
inconsistent with the treatment of the power spectrum, in that bias is treated at one-loop order for the power
spectrum but tree-level for the bispectrum. The main goal of this study is to present such a unified model that
includes all relevant effects in real space, extending the results of~\citep{Assassi:2014}.

What are the essential elements? First, we need to consider all contributions from the general bias expansion
 up to fourth order in perturbation theory~\citep{Desjacques:2018, Mirbabayi:2015,Assassi:2014}. These are generated by the
gravitational evolution of the dark matter field and include the common linear, and nonlinear bias parameters
\citep{Kaiser:1984, Coles:1993, FryGaz9308}, as well as the second order nonlocal (or tidal field) bias
\citep{Catelan:2000, McDonald:2009, Chan:2012, Baldauf:2012}. Second, it has been argued in
\citep{Fujita:2016,Nadler:2018} that so-called higher-derivative terms, which have been known for a long time~\cite{BarBonKai8605} 
and can be understood as deriving from a dependence of galaxy
formation on the spatial distribution of matter \citep{McDonald:2009}, can be as important or even dominate over
the general bias terms in certain circumstances. Third, on scales of the weakly nonlinear regime stress-tensor
corrections to the evolution of dark matter~\citep{PueSco0908,PieManSav1108,Baumann:2012, Carrasco:2012,BalMerMir1406,AngForSch1406}, might 
potentially become relevant. And finally, the impact of very small scales, in the absence of primordial
non-Gaussianity largely uncorrelated with the previously mentioned large-scale effects, leads to an additional
stochastic bias \citep{Scherrer:1998, Dekel:1999}.

However, a challenge that afflicts all theoretical predictions of galaxy clustering beyond leading order is a
mismatch between the bias parameters from the perturbative expansion, and those an observer would define
through the measurement of correlation functions. This was first  pointed out in
\citep{McDonald:2006}, which showed that appropriate renormalizations of the perturbative bias parameters restore
agreement with the measurements. These renormalizations must be done on a statistic-by-statistic and
parameter-by-parameter basis, and while tractable for the power spectrum \citep{McDonald:2006,McDonald:2009},
this becomes increasingly complicated for higher-order statistics \citep{Assassi:2014}. As a way of
circumventing this procedure altogether, we advocate the use of multipoint propagators.

The multipoint propagator formalism was originally introduced in the context of renormalized perturbation theory
(RPT) and its generalizations \citep{CroSco0603a, Bernardeau:2008, Bernardeau:2010, Bernardeau:2012,
  Crocce:2012, TarNisBer1304} to obtain an accurate description of the dark matter field in the quasi-linear
regime. Multipoint propagators have also been extended to include redshift space distortions and galaxy bias in
\citep{Matsubara:2011, Matsubara:2014, Matsubara:2016}, but without the clear connection to the renormalization
issue that we aim to establish in this paper. In fact, we will demonstrate that the multipoint propagators
correspond to the bias parameters that are commonly identified through cross-correlations of galaxy and matter
fields \citep{Smith:2007,Pollack:2012}. Thus, they are observable and have a well-defined physical meaning,
based on which we argue that they provide the most natural approach towards galaxy bias. As in RPT they further
act as building blocks for the general $N$-point correlation function, and determination of the first three
propagators already fixes the power spectrum and bispectrum at the one-loop level, simplifying the computations
significantly.

Modeling higher-order statistics to the same precision as their lower-order counterparts brings about a quickly
inflating number of terms from the various bias contributions. In order to reduce the growing parameter space, a
further goal of this article is to examine 1) degeneracies between the bias terms, and 2) relations between bias
parameters arising from evolution, which are studied here for the first time at fourth order.

The paper is organized as follows. In Sec.~\ref{sec:galexpansion} we present the central idea of the multipoint
propagator expansion based on the simple and widely known model of local galaxy bias. A complete basis for the
general bias and higher-derivative terms is provided in Sec.~\ref{sec:bare_expansion}. Following
\citep{Chan:2012} this basis makes intuitive use of Galilean invariants of Lagrangian perturbation theory (LPT)
potentials, and is particularly well suited for the computation of the multipoint propagators that we perform in
Sec.~\ref{sec:propformalism}. We also determine their time evolution and present the relations between initial
and evolved bias parameters. Finally, Sec.~\ref{sec:PB} uses these results to compute the power spectrum and
bispectrum, and demonstrates how we incorporate stress-tensor corrections and
stochasticity. Sec.~\ref{sec:conclusions} summarizes the final model components and gives our conclusions. The
appendices~\ref{sec:bias-further-notes} and~\ref{sec:derivation-gamma-recursion} give further details and
include relations between our basis for galaxy bias and others in the
literature. Appendix~\ref{sec:GIbiasevol} shows how to take advantage of Galilean invariance for
  determining the evolution of galaxy bias, while Appendix~\ref{sec:RENbiasevol} demonstrates that the
  corresponding results are unaffected by renormalization.

\section{Basics of the renormalized bias expansion}
\label{sec:galexpansion}

We are interested in the statistical properties of the observed galaxy distribution. These are commonly
quantified by a hierarchy of correlation functions of the density perturbations, which we write as
$\delta_g(\B{x}) = n_g(\B{x})/\bar{n}_g-1$, with $n_g(\B{x})$ denoting the galaxy number density and $\bar{n}_g$
its average.\footnote{Note that while the subscript `g' here stands for galaxies, it could equally well denote
  any other tracer of the mass field, such as quasars, galaxy clusters or the Ly-$\alpha$ forest.}  Model
predictions of the correlation functions require a relation between the galaxy perturbations and matter
fluctuations $\delta(\B{x})$, which is usually written as some functional $\delta_g[\delta(\B{x})]$ that is then
Taylor expanded.

In order to illustrate the main idea pursued in this paper, we start with the simplest and most well-known bias
relation, which considers $\delta_g$ to be a local function of $\delta$ \citep{Coles:1993,FryGaz9308} and can be
written as a Taylor series around $\delta = 0$. Dropping all position arguments we have
\begin{align}
  \label{tracers.LocBias}
  \delta_g = \sum_{n} \frac{1}{n!}\left(\frac{\partial^n \delta_g}{\partial\,
  \delta^n}\right)_{0} \delta^n \equiv \; &\bar{b}_0+ \bar{b}_1\, \delta + {\bar{b}_2\over 2}\, \delta^2 + {\bar{b}_3\over
  3!}\,\delta^3 \nonumber \\ & + \dots\,,
\end{align}
where the bias parameters are identified as $\bar{b}_n = (\partial^n \delta_g / \partial\, \delta^n)_0$ ($n>0$)
with $()_0$ denoting evaluation at $\delta=0$, and $\bar{b}_0$ enforces $\langle \delta_g \rangle=0$. However,
the bias parameters so defined are not observables, as we  measure correlators (expectation values or
ensemble averages of fields), not quantities evaluated at $\delta = 0$. As we will detail below, this is the
reason for difficulties in the computation of galaxy correlation functions that ultimately require us to
redefine (or ``renormalize''~\cite{McDonald:2006}) the bias parameters above. We will further show that a more natural
bias expansion, in the sense that its coefficients are actually measurable, can be defined in terms of
cross-correlations between the galaxy and matter fields. In the language of renormalized perturbation theory
(RPT) and its generalizations~\cite{CroSco0603a,Bernardeau:2008} these cross-correlations correspond to the
so-called \emph{multipoint propagators}.

\subsection{Galaxy clustering statistics in the \\ standard approach}
\label{sec:DMclustering}

We define the two- and three-point correlation functions of the galaxy perturbations in Fourier space --- the
power spectrum and bispectrum --- as follows
\begin{align}
  \langle \delta_g(\B{k}_1)\,\delta_g(\B{k}_2)\rangle &\equiv
  (2\pi)^3\,P_g(k_1)\,\delta_D(\B{k}_{12})\,,   \label{eq:tracers.Pdef} \\
  \langle \delta_g(\B{k}_1)\,\delta_g(\B{k}_2)\,\delta_g(\B{k}_3)\rangle &\equiv
  (2\pi)^3\,B_g(k_1,k_2,k_3)\,\delta_D(\B{k}_{123})\,,   \label{eq:tracers.Bdef}
\end{align}
where $\B{k}_{1\ldots n} \equiv \B{k}_1 + \ldots + \B{k}_n$ and the appearance of the delta distribution is a
manifestation of statistical homogeneity. Statistical isotropy further demands that the power spectrum only
depends on the magnitude of the two wave vectors participating in the correlator, while the bispectrum is a
function of three wave numbers $k_1$, $k_2$ and $k_3$. Finally, to pass from configuration to Fourier space we
have adopted the convention
\begin{align}
  \delta_g(\B{x}) = \int_{\B{k}} \exp{(-i\,\B{k}\cdot\B{x})}\,\delta_g(\B{k})\,,
\end{align}
using a short-hand notation for $k$-space integrals, i.e. $\int_{\B{k}_1,\ldots,\B{k}_n} \equiv \int
\D{^3k_1}/(2\pi)^3 \cdots \D{^3k_n}/(2\pi)^3$.

Analogous definitions hold for the dark matter field, and to begin with let us assume that the bias expansion is
done in Lagrangian space, so that the matter fluctuations can be considered linear and Gaussian (we neglect any
possible primordial non-Gaussianities for the remainder of this paper). In that case all statistical information
is contained in the linear power spectrum, which we denote by $\langle
\delta_L(\B{k}_1)\,\delta_L(\B{k}_2)\rangle = (2\pi)^3\,P_L(k_1)\,\delta_D(\B{k}_{12})$, while the bispectrum and
all other higher-order statistics vanish. Under this assumption, plugging Eq.~(\ref{tracers.LocBias}) into
Eq.~(\ref{eq:tracers.Pdef}) and using Wick's theorem leads to
\begin{equation}
  \label{eq:tracers.Pgbare}
  \begin{split}
    P_g(k) = &\;\bar{b}_1^2\, P_L(k) + \left[ \bar{b}_1 \bar{b}_3 \sigma^2\right] P_L(k) \\ &+
    {\bar{b}_2^2\over 2} \int_{\B{q}} P_L(|\B{k}-\B{q}|)P_L(q) + \dots\,, 
  \end{split}
\end{equation}
where the dots denote contributions of two-loop and higher order, i.e. terms that involve five or more powers of
$\delta$. The variance $\sigma^2 \equiv \langle \delta^2(\B{x})\rangle$, which appears in the first term of the
loop contribution in square brackets, is formally infinite or at least highly sensitive to the nonlinear regime,
depending on the shape of the linear power spectrum\footnote{Imposing a high-$k$ cutoff for the power
    spectrum would only mask the problem, as this would lead to all observables being dependent on an
    arbitrary choice of scale.}. This implies that the large-scale galaxy power spectrum
would be heavily influenced by scales where our perturbative approach is not expected to hold. However, we
notice that this term is also proportional to $P_L(k)$, such that if we redefine the linear bias parameter to be
$b_1\equiv \bar{b}_1 + \bar{b}_3\, \sigma^2/ 2$, we retain the form
\begin{equation}
  P_g(k) = b_1^2\, P_L(k) +  {~\bar{b}_2^2\over 2} \int_{\B{q}}  P_L(|\B{k}-\B{q}|)P_L(q) \,.
  \label{PowerB1Renorm}
\end{equation}
Clearly, as more bias loops are included, the expression for $b_1$ keeps changing, but the principle remains the
same --- the \emph{observed} linear bias is defined as (the square root of) the coefficient in front of $P_L$. 

\begin{figure*}
  \centering
  \vspace*{1em}
  \includegraphics[width=\textwidth]{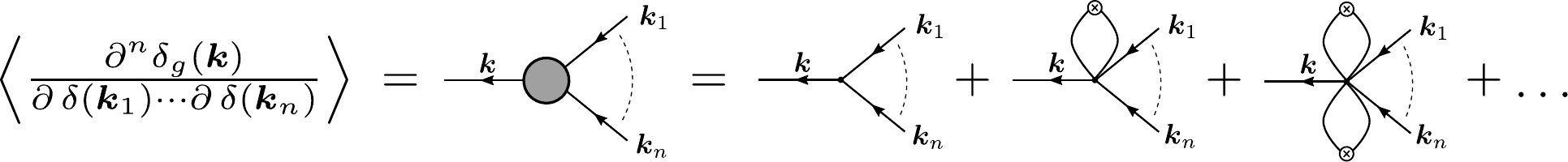}
  \caption{The $n$-th multipoint propagator is given by the sum of all reducible diagrams with $n$ legs of
    incoming momentum $\B{k}_1,\,\ldots,\,\B{k}_n$ (the external lines). The vertices correspond to the kernels
    associated with terms in the bias expansion, e.g. the vertices of the three diagrams on the right-hand side
    are given by $\bar{b}_n$, $\bar{b}_{n+2}$ and $\bar{b}_{n+4}$ for the expansion in Eq.~(\ref{tracers.LocBias}). If terms
    nonlocal in the matter density enter the bias relation, these kernels acquire a scale dependence based on
    all incoming momenta. The crossed circles stand for (linear) power spectra.}
  \label{fig:propagators}
\end{figure*}

A similar situation holds for the quadratic bias parameter $b_2$. To see this let us now consider the bispectrum
up to one-loop order (up to six powers of $\delta$), under the same assumption as above. We then get from
Eq.~(\ref{eq:tracers.Bdef}):
\begin{align}
  \label{eq:tracers.Bgbare}
  B_{g,123}=& \, \bar{b}_2\bar{b}_1^2\, P_1P_2
  +\left[ \bar{b}_1\bar{b}_2\bar{b}_3\,\sigma^2+ {1\over 2}\bar{b}_1^2\bar{b}_4\, {\sigma^2}
  \right] P_1P_2  
  \nonumber \\ & + {1\over 2}\bar{b}_1\bar{b}_2\bar{b}_3\ P_1 \int_{\B{q}}  P_L(|\B{k}_2-\B{q}|)P_L(q) 
  \nonumber \\ & + \bar{b}_2^3 \int_{\B{q}} P_L(|\B{k}_1-\B{q}|)P_L(|\B{k}_2+\B{q}|)P_L(q)
  +{\rm cyc.} \nonumber \\ &+ \dots\,,
\end{align}
where $P_i\equiv P_L(k_i)$ and {\rm cyc.} denotes cyclic permutations of each term over the three wave
vectors. Again, the first term in square brackets corresponds to the renormalization of the linear bias seen in
the power spectrum, but in addition now there is also a second term that can be absorbed by a renormalization of
the {\em quadratic bias} $b_2 \equiv \bar{b}_2 + \bar{b}_4\, \sigma^2/ 2$, so that we can write
\begin{align}
  B_{g,123}=& \, b_2 b_1^2 \, P_1P_2 
 + {1\over 2}\bar{b}_1\bar{b}_2\bar{b}_3 \ P_1    \int_{\B{q}} P_L(|\B{k}_2-\B{q}|)P_L(q)
  \nonumber \\ & 
  + \bar{b}_2^3 \int_{\B{q}} P_L(|\B{k}_1-\B{q}|)P_L(|\B{k}_2+\B{q}|)P_L(q)
  +{\rm cyc.}
  \label{BispB2Renorm}
\end{align}
A number of questions arise from this procedure. 1) Is the $b_2$ renormalization that follows from the one-loop
bispectrum consistent with the one that follows from the two-loop power spectrum? If so, is that true for all
other bias parameters? 2) Is there a way to do all these renormalizations ``automatically'', instead of
calculating statistic by statistic, and order by order? 3) Is there a simple connection between renormalizations
of different $N$-point correlators? As we will see next, the answer to all these questions is
`yes'~\cite{CroSco0603a,CroSco0603b,Bernardeau:2008,Bernardeau:2012}.

\subsection{A bias expansion based on observables}
\label{sec:physical_bias}

First, we notice that the terms that contribute to the renormalization of bias parameters must derive from
factorizable loop corrections, i.e. they can be written as products of lower-order terms and new contributions
(in the example of Sec.~\ref{sec:DMclustering} these new contributions are functions of bias parameters and
$\sigma^2$). In general, any loop correction can be portrayed as a diagram with a fixed number of external lines
and a variable number of internal lines depending on the loop order. Those diagrams that lead to factorizable
contributions are so-called {\em reducible}, meaning they can be decomposed into two or more connected diagrams
by cutting one or more internal lines, while all other ones that do not share this property are classified as
{\em irreducible}.

In order to automatically include renormalizations to arbitrary loop order, it is therefore desirable to
construct the bias expansion in terms of the sum over all reducible diagrams with a given number of external
lines. Such objects are known as multipoint propagators (see~Fig.~\ref{fig:propagators}), and correspond to the
expectation value of functional derivatives of the galaxy field with respect to the matter field: the
derivatives produce the external lines, while the expectation value generates the loops~\cite{CroSco0603a}.

An arbitrarily complicated loop diagram for any statistic can be decomposed easily and uniquely into multipoint
propagators: the reducible subdiagrams are absorbed into the multipoint propagators while the irreducible
subdiagrams are generated by connecting multipoint propagators among themselves. Since a given multipoint
propagator appears in various statistics, this gives us a connection between renormalizations of different
$N$-point correlation functions; e.g. the renormalization of $b_2$ that follows from the one-loop bispectrum is
consistent with the one that follows from the two-loop power spectrum, because both are incorporated into the
two-point propagator (i.e. they correspond to {\em the same} subdiagram in the power spectrum and bispectrum).
As a result of this, $\bar{b}_2$ in Eq.~(\ref{PowerB1Renorm}) can be replaced by $b_2$ and similarly
$\bar{b}_1\bar{b}_2\bar{b}_3 \to b_1b_2b_3$ and $\bar{b}_2^3 \to b_2^3$ in Eq.~(\ref{BispB2Renorm}).

The observed bias parameters of order $n$ that appear in correlators thus correspond to the sum over all
reducible diagrams with $n$ external legs. For example when $n=1$, i.e. linear bias, we simply need to consider
the sum of all reducible diagrams with a single external leg (corresponding to $P_L$ at $k$):
\begin{equation}
  b_1 =  \left<\frac{\partial \delta_g}{\partial\, \delta}\right> = \bar{b}_1 + \bar{b}_3\, {\sigma^2 \over 2} + \ldots
  = \sum_{n=0}^\infty\, \frac{\bar{b}_{2n+1}}{n!}\, \left(\frac{\sigma^{2}}{2 }\right)^n\,,
  \label{b1ren}
\end{equation}
where we use the symbol $\partial$ to denote a functional derivative. Similarly, for quadratic
bias ($n=2$) we have from Eq.~(\ref{tracers.LocBias})
\begin{equation}
  b_2 = \left< \frac{\partial^2 \delta_g}{\partial\, \delta^2}\right> = \bar{b}_2 + \bar{b}_4\,
  {\sigma^2 \over 2} + \ldots = \sum_{n=0}^\infty\, \frac{\bar{b}_{2n+2}}{n!}\, \left(\frac{\sigma^{2}}{2
  }\right)^n\,,
  \label{b2ren}
\end{equation}
that is, the sum over all reducible diagrams with two external legs (corresponding to $P_L$ at $k_1$ and $k_2$
as seen in the leading order bispectrum, Eq.~\ref{BispB2Renorm}). Clearly, the calculations in
Eqs.~(\ref{b1ren}-\ref{b2ren}) are significantly easier than performing the renormalization procedure order by
order and statistic by statistic leading to Eqs.~(\ref{PowerB1Renorm}, \ref{BispB2Renorm}). 

To conclude, we can remove the disconnect between the parameters appearing in the standard bias expansion
(Eq.~\ref{tracers.LocBias}) and those in correlators (Eqs.~\ref{PowerB1Renorm}, \ref{BispB2Renorm}), and
therefore stop thinking about renormalization altogether, if we construct the bias expansion in terms of
reducible diagrams with a given number of external legs. In effect, we trade kernels for multipoint propagators,
\begin{equation}
  \left(\frac{\partial^n \delta_g}{\partial\, \delta^n}\right)_0 \; \longrightarrow \; \left< \frac{\partial^n
    \delta_g}{\partial\, \delta^n}\right>\,,
\end{equation}
which leads to a new expansion of the form (denoted as Gamma expansion in~\cite{Bernardeau:2008,Bernardeau:2012})
\begin{align}
  \delta_g =\;&\left< \frac{\partial \delta_g}{\partial\, \delta}\right>\ \delta + {1\over 2!} \left<
    \frac{\partial^2 \delta_g}{\partial\, \delta^2}\right>\ \Big[ \delta^2-\langle\delta^2\rangle \Big]
  \nonumber \\ & + {1\over 3!} \left<\frac{\partial^3 \delta_g}{\partial\, \delta^3}\right>\ \Big[ \delta^3-3
  \langle\delta^2\rangle \delta-\langle\delta^3\rangle \Big] + \ldots \label{eq:tracers.MPexp_realspace} \\ =\;&
  b_1 \delta + {b_2\over 2!} \Big[ \delta^2-\langle\delta^2\rangle \Big] + {b_3\over 3!}  \Big[ \delta^3-3
  \langle\delta^2\rangle \delta-\langle\delta^3\rangle \Big] \nonumber \\ & + {b_4\over 4!}  \Big[ \delta^4-4
  \langle\delta^3\rangle \delta-6 \langle\delta^2\rangle \delta^2 +3 \langle\delta^2\rangle^2
  -\langle\delta^4\rangle_c \Big] \nonumber \\ & + \ldots
  \label{GammaExp}
\end{align}
The structure of the square brackets is given by $\delta^n$ minus all possible actions of $\langle \rangle$ on
$\delta^n$ with a constant term that respects that the expectation value is zero for non-Gaussian $\delta$, and
we note that Eq.~(\ref{GammaExp}) automatically satisfies $\langle \delta_g \rangle =0$. The second equality
assumes local galaxy bias as we have done so far, but we stress that in this expansion the bias parameters $b_n$
are the renormalized ones, that is, replacing the propagators by numbers gives precisely the renormalized local
bias expansion. 

Furthermore, if the expansion in Eq.~(\ref{GammaExp}) is done with respect to a Gaussian $\delta$, for instance
when using the linear fluctuations, terms such as $\langle \delta^3 \rangle$ and $\langle \delta^4 \rangle_c$
will vanish and the expansion of $\delta_g$ will be given in terms of Hermite polynomials --- a suggestion that
was already put forward in \citep{Szalay:1988}. Another crucial property of the multipoint propagators in this
case is that they can be shown to be the cross-correlation bias between galaxies and matter fluctuations
\citep{CroSco0603b}, e.g. for linear and quadratic bias we have
\begin{align}
  \langle \delta_g\,\delta\rangle &= \left< \frac{\partial \delta_g}{\partial\, \delta}\right>\, \langle
  \delta\,\delta \rangle\,,  \label{PropIsCrossCorr} \\
  \langle \delta_g\,\delta\,\delta \rangle &= \left< \frac{\partial^2 \delta_g}{\partial\,
      \delta^2}\right>\,\langle \delta\,\delta \rangle\,\langle \delta\,\delta
  \rangle\,. \label{PropIsCrossCorr2}
\end{align}
That is, the observables corresponding to the multipoint propagators are no other than the standard
cross-correlation Lagrangian bias coefficients routinely measured in N-body simulations,  e.g.  
~\cite{Jin9904,SheLem99,PorCatLac9903,EliLudPor1204,SheChaSco1304,BiaChuDes1310,ChaSheSco1711,ModCasSel1612,LazSch1712,AbiBal1807,SchElsJas1901,SchSimAss1811}.

Using the expansion in Eq.~(\ref{GammaExp}) allows us to rederive the renormalization procedure by writing it in
the form of Eq.~(\ref{tracers.LocBias}) and matching coefficients of $\delta^n$, e.g. $\bar{b}_1 = b_1 - b_3
\sigma^2/2 $ and $\bar{b}_2 = b_2 - b_4 \sigma^2/2 $, again, without doing any calculations of correlators.
This can become useful if such relations are desired for the more general case of nonlocal bias, as we shall
discuss below. 

Finally, the multipoint propagator expansion has the additional advantage that it simplifies the computation of
correlators considerably. For instance, for the power spectrum there is only one irreducible diagram present at
each loop order, as opposed to an increasing number of diagrams at each loop order for the standard approach; indeed, assuming local
bias for now we simply have: 
\beq 
P_g(k)=(2\pi)^3\,\sum_n {b_n^2 \over n!} \int\displaylimits_{\B{q}_1,\ldots,\,\B{q}_n} [\delta_D]_n\,
P_L(q_1)\ldots P_L(q_n)\,,
\label{loopPower}
\eeq
where $[\delta_D]_n\equiv \delta_D(\B{k}-\sum_{i=1}^n \B{q}_i)$. For the correlation function this leads to,
\beq
\xi_g(r)=\sum_n {b_n^2 \over n!} \ \left[\xi(r)\right]^n\,,
\label{loopXi}
\eeq 
a result first obtained in~\cite{Szalay:1988}. If bias is not local, the functions of wave vectors that
replace $b_n$ are just placed inside the integral in Eq.~(\ref{loopPower}).

\subsection{Renormalized bias expansion with \\ nonlinear evolution}
\label{sec:bias-exp-with-evo}

Let us now consider the renormalized bias expansion in Eulerian space, which requires us to take into account
the nonlinear evolution of the matter field. This computation serves two main objectives. Firstly, it highlights
the complexity in determining contributions to the renormalized bias parameters from evolved fields, and thus
motivates our approach taken in Sec.~\ref{sec:propformalism}, where we instead evolve the multipoint propagators
from their initial conditions. Secondly, it demonstrates that the bias expansion in Eq.~(\ref{tracers.LocBias})
cannot be completely renormalized by the local parameters $b_n$ alone \citep[see also][]{Assassi:2014}, calling
for additional terms in the expansion of $\delta_g$, which we address in Sec.~\ref{sec:bare_expansion}.

On large scales, where we can take the dark matter velocity field\footnote{Note that we work with the
  \emph{scaled} velocity field $\B{v}$, which is related to the \emph{peculiar} velocity via $\B{v} \equiv
  -\B{u}/f{\cal H}$, with $f$ the logarithmic growth rate and ${\cal H}$ the comoving Hubble rate.}
$\B{v}(\B{x})$ to be irrotational, the time evolution is governed by the coupled equations of motion for the
density perturbations and velocity divergence $\theta(\B{x}) \equiv \B{\nabla} \cdot \B{v}(\B{x})$ under
influence of the gravitational potential (see \citep{Bernardeau:2002} for a detailed review). In standard
Eulerian perturbation theory (SPT) these equations are solved as series expansions about the linear density
field $\delta^{(1)}$, which we still assume to be Gaussian. At time $\tau$ we have (neglecting
transients)
\begin{align}
  \delta(\B{x},\tau) = \sum_{n=1}^{\infty} D^n(\tau)\,\delta^{(n)}(\B{x})\,,
\end{align}
where, to very good accuracy, all cosmology dependence is encoded in the linear growth factor $D(\tau)$
\citep{Bouchet:1992,Bouchet:1995,ScoColFry9803}. Note that what we denote as $\delta_L$ is
  actually the linear density field extrapolated to some final time (whose argument we will usually drop),
  i.e. $\delta_L(\B{x}) \equiv D(\tau)\,\delta^{(1)}(\B{x})$. In Fourier space the $n$-th order solution is
constructed out of $n$ powers of the linear density field, which are coupled via the SPT kernels $F_n$:
\begin{align}
  \label{eq:tracers.nSPT}
  \delta^{(n)}(\B{k}) = (2\pi)^3\int\displaylimits_{\B{k}_1,\ldots,\,\B{k}_n}
  &\left[\delta_D\right]_n\,F_n(\B{k}_1,\ldots,\B{k}_n) \nonumber \\
  &\times\,\delta^{(1)}(\B{k}_1)\cdots\delta^{(1)}(\B{k}_n)\,.
\end{align}
The velocity divergence can be expanded in a similar manner and the $n$-th order solutions are obtained by
replacing $F_n$ with $G_n$ in the equation above. Explicit expressions for these kernels can be found in
\citep{Fry84,GorGriRey86,Bernardeau:2002}. We also note that at linear order, we have $\delta = \theta$. 

Due to the mode-coupling in Eq.~(\ref{eq:tracers.nSPT}) the nonlinear (or time-evolved) density field becomes
non-Gaussian. In order to retain the usual linear (as opposed to nonlinear) spectra in the loop integrals when
computing clustering statistics, we however still define the multipoint propagators with respect to the linear
Gaussian matter fluctuations, i.e. derivatives are taken with respect to $\delta_L$. However,
  this implies that the kernels $(\partial^n \delta_g / \partial\, \delta_L^n)_0$ are not just numbers but
  acquire a scale dependence, in which case it becomes more convenient to express the multipoint propagators in
  Fourier space,
\begin{align}
  \left<\frac{\partial^n \delta_g}{\partial\, \delta_L^n}\right> \; \longrightarrow \; \left<
  \frac{\partial^n \delta_g(\bk)}{\partial\, \delta_L({\bk}_1) \cdots \partial\, \delta_L({\bk}_n)}\right> \,,
\end{align}
as was already indicated in Fig.~\ref{fig:propagators}. While we present the resulting Fourier space
equivalent of Eq.~(\ref{eq:tracers.MPexp_realspace}) in Sec.~\ref{sec:hermite-expansion}, in what follows it
suffices to notice that the Fourier space multipoint propagators are functions of the $n$ momenta
$\B{k}_1,\ldots,\B{k}_n$ and include the factor $(2\pi)^3\,\delta_D(\B{k}-\B{k}_{1\ldots n})$. Primed
angle brackets indicate that we have dropped this common factor.
 
Returning now to the calculation of the renormalized linear bias parameter, we see that because of second-order
SPT (2SPT) we have an effective cubic term inside the quadratic bias contribution, $\bar{b}_2\, \delta^2/2$,
from Eq.~(\ref{tracers.LocBias}). Accordingly, the one-point propagator has the following extra term,
\begin{align}
  {1\over 2}\left< {\partial \delta^2(\B{k}) \over \partial \delta_L(\B{k}_1)} \right>' &= \left<\left[\delta_L \ast
    {\partial \delta^{(2)} \over \partial \delta_L}\right](\B{k}_1)\right>' + \ldots \nonumber \\ &\hspace{-0.7em}= 2
\underbrace{\int_{\B{q}} F_2(\B{k}_1,\B{q})\,P_L(q)}_{\displaystyle \bar{F}_2\,\sigma^2} + \ldots = {34 \over 21} \sigma^2 +\ldots\,,
\label{b1renormBYb2}
\end{align}
where $\ast$ denotes a convolution and $\bar{F_2}$ is the angular average of the 2SPT kernel
(i.e. the contribution from spherical collapse to $F_2$). This leads to
$b_1= \bar{b}_1+(34\, \bar{b}_2/21+\bar{b}_3/2)\, \sigma^2 +\ldots$ \cite{McDonald:2006}. More broadly, since each
term in the bias expansion of Eq.~(\ref{tracers.LocBias}) appears at all orders in perturbation theory when
$\delta$ corresponds to the Eulerian matter field, the renormalization of each bias parameter depends on all
$\bar{b}_n$ ($n\geq 2$). Thus, the simplicity of Eqs.~(\ref{b1ren}-\ref{b2ren}) is lost.
 
Let us consider what that implies for the renormalization of the quadratic bias, where we now have an effective
quartic term (again due to 2SPT) inside the $\bar{b}_3\, \delta^3/3!$ contribution in
Eq.~(\ref{tracers.LocBias}), extending the two-point propagator by
\begin{align} 
  {1\over 3!}\left< {\partial^2 \delta^3 \over \partial \delta^L_1 \partial\delta^L_2} \right>' &= \left< \delta
    \hspace{-0.07em}\ast\hspace{-0.07em} {\partial \delta \over \partial \delta^L_1}
    \hspace{-0.07em}\ast\hspace{-0.07em} {\partial \delta \over \partial \delta^L_2} \right>' + {1\over 2}
  \left< \delta^2 \hspace{-0.07em}\ast\hspace{-0.07em} {\partial^2 \delta \over \partial
      \delta^L_1\partial\delta_2^L} \right>' \nonumber \\ &= \Big[4 \bar{ F_2} + F_2(\B{k}_1,\B{k}_2)\Big]\,
  \sigma^2 + \ldots
  \label{b2renorm}
\end{align}
with $\delta^L_i \equiv \delta_L(\B{k}_i)$. The second term depending on the full $F_2$ kernel simply gives the
desired contribution to the renormalization of the linear bias parameter by $\bar{b}_3$ as found for the
one-point propagator before. That is because \beq \bar{b}_1\,\left< {\partial^2 \delta \over \partial
    \delta^L_1 \partial\delta^L_2} \right>' = \bar{b}_1\, 2F_2(\B{k}_1,\B{k}_2) + \ldots\,,
 \label{Gamma2b1}
\eeq 
such that when calculating $\langle \partial^2 \delta_g/\partial \delta^L_1\partial^L_2 \rangle$ the second term
in Eq.~(\ref{b2renorm}) corresponds to changing $\bar{b}_1 \to \bar{b}_1+ \bar{b}_3\,\sigma^2/2$ in
Eq.~(\ref{Gamma2b1}). The first term in Eq.~(\ref{b2renorm}) on the other hand is the contribution to the
renormalization of the quadratic bias parameter from nonlinear evolution, which now reads $b_2= \bar{b}_2+(68
\,\bar{b}_3/21+\bar{b}_4/2)\, \sigma^2 +\ldots$. In addition, quadratic bias contributes to the renormalization
of $b_2$ itself, since
\begin{alignat}{2} 
 \label{b2renormsItself}
{1\over 2}\left< {\partial^2 \delta^2 \over \partial \delta^L_1 \partial\delta^L_2} \right>' &= \;
&&\left< {\partial \delta \over \partial \delta^L_1} \ast {\partial \delta \over \partial \delta^L_2} \right>' + \left<
  \delta \ast {\partial^2 \delta \over \partial \delta^L_1\partial\delta_2^L} \right>' \nonumber \\ &\hspace{-0.7em}= &&\hspace{-0.7em}1 + \left<
  {\partial \delta^{(2)} \over \partial \delta^L_1} \ast {\partial \delta^{(2)} \over \partial \delta^L_2} \right>' + \left<
  \delta_L \ast {\partial^2 \delta^{(3)} \over \partial \delta^L_1\partial\delta_2^L} \right>' \nonumber \\ & &&\hspace{-0.7em} + \left<
  {\partial \delta^{(3)} \over \partial \delta^L_1}+{\partial \delta^{(3)} \over \partial \delta^L_2} \right>' + \ldots 
\end{alignat}
The last term in this expression dresses the external lines to include the one-loop propagator due to nonlinear
evolution (i.e. the standard $P_{13}$ contribution from the one-loop matter power spectrum \citep{Bernardeau:2002}),
while the remaining two terms give rise to the renormalization of the quadratic bias we are interested in, and
they read
\begin{widetext} 
  \begin{equation}
    \begin{split}
      \left< {\partial \delta^{(2)} \over \partial \delta^L_1} \ast {\partial \delta^{(2)} \over \partial
          \delta^L_2} \right>' + \left< \delta_L \ast {\partial^2 \delta^{(3)} \over \partial
          \delta^L_1\partial\delta_2^L} \right>'
      &= \int_{\B{q}} \Big[4 F_2(-\B{q},\B{k}_1) F_2(\B{q},\B{k}_2) + 6 F_3(\B{q},\B{k}_1,\B{k}_2)\Big]P_L(q) \\
      &= {68\over 21}F_2(\B{k}_1,\B{k}_2) \sigma^2 + {8126\over 2205} \sigma^2 + {254 \over 2205}
      K(\B{k}_1,\B{k}_2) \sigma^2 + \int_{\B{q}} {\cal F}(\B{k}_1,\B{k}_2,q) P_L(q)
    \end{split}
  \label{b2renormsb1b2g2}
  \end{equation}
  with ($\mu \equiv \hat{\B{k}}_1\cdot\hat{\B{k}}_2$, $K(\B{k}_1,\B{k}_2) \equiv \mu^2-1$, and ${\cal L}_\ell$ denoting Legendre polynomials)
  \beqa {\cal F}(\B{k}_1,\B{k}_2,q) &\equiv & {(k_1^2-q^2)^3 \over 168\, k_1^5\,
    q^3}\Big[(q^2-k_1^2){\cal L}_2(\mu)-9k_1k_2\mu \Big] \ln \left| \frac{k_1+q}{k_1-q}\right| + k_1 \leftrightarrow
  k_2 + \left({{k_1^2+k_2^2} \over {7 k_1 k_2}} -{19 k_1 k_2 \over 84\, q^2} - {3(k_1^4+k_2^4)q^2\over 56\, k_1^3\,
    k_2^3} \right) \mu \nonumber \\ && + \left[ {73\over 630} + {{k_1^2+k_2^2}\over 84\, q^2} -{11q^2\over
    252}\left(\frac{1}{k_1^2}+\frac{1}{k_2^2}\Big)+ {q^4\over 84}\Big(\frac{1}{k_1^4}+\frac{1}{k_2^4}\right)\right]\,
  {\cal L}_2(\mu)\,.
  \label{MonsterInt} 
  \eeqa
\end{widetext}
While the integral over the function ${\cal F}$ in Eq.~(\ref{MonsterInt}) represents the finite part (going as
$1/q^{2}$ as $q \to \infty$) of the one-loop two-point propagator due to quadratic bias, the three terms
proportional to $\sigma^2$ in the second line of Eq.~(\ref{b2renormsb1b2g2}) must be absorbed by
renormalizations. We recognize that the first of these corresponds to a linear bias renormalization by
$\bar{b}_2$, leading to $\bar{b}_1 \to b_1 = \bar{b}_1 + (34 \bar{b}_2/21 + \bar{b}_3/2)\,\sigma^2$ in
Eq.~(\ref{Gamma2b1}), which is consistent with the result found for the one-point propagator in
Eq.~(\ref{b1renormBYb2}). The second term renormalizes the quadratic bias, giving
$b_2= \bar{b}_2+(8126\, \bar{b}_2/2205+68 \,\bar{b}_3/21+\bar{b}_4/2)\, \sigma^2$~\cite{Assassi:2014}. Finally,
the third term has a different scale dependence compared to the previous ones, encoded by $K(\B{k}_1,\B{k}_2)$,
such that it cannot be absorbed by a redefinition of any local bias parameter $b_n$. This illustrates that
nonlinear evolution gives rise to additional effects that enter the relation between the galaxy and matter
density and thus renders the local-in-matter expansion incomplete. In particular, we will see in
Appendix~\ref{sec:RENbiasevol} that the term proportional to $K(\B{k}_1,\B{k}_2)\,\sigma^2$ corresponds to a
renormalization of the nonlocal quadratic bias
$\gamma_2=\bar{\gamma}_2 + 127\, \bar{b}_2 \sigma^2/2205$~\cite{Assassi:2014}, which is associated to the
gravitational tidal field.

As we alluded to at the beginning of this section, compared to the calculation when the bias relation is written
in Lagrangian space, the steps taken here are a lot more complicated (and we just scratched the surface, since
we only assumed local galaxy bias so far). On the other hand, if we would like to maintain the simplicity of the
Lagrangian expansion we would instead have to evolve the Lagrangian propagators by nonlinear evolution, which is where the
complications may surface again. However, this is not the case for the main reason that the evolution from
Lagrangian to Eulerian bias by conserving the number of objects cannot generate unphysical terms going like
$\sigma^2$. Therefore, the time-evolved propagators do not require  extra renormalizations from nonlinear
dynamics, and we obtain the `finite parts' such as Eq.~(\ref{MonsterInt}) automatically.
 
It is further worth noting that the evolved propagators connect the Lagrangian bias parameters with the (late
time) Eulerian ones, i.e. they give us the time evolution of the \emph{observable} bias parameters, as opposed
to the evolution of the \emph{bare} bias parameters.  We discuss this in some detail in
Appendix~\ref{sec:RENbiasevol}, as to whether renormalization affects the time evolution of bias parameters.

\section{A complete Galilean invariant basis for galaxy bias}
\label{sec:bare_expansion}

We have already seen how the renormalization of local quadratic bias, when applied in Eulerian space, requires
the existence of at least one additional term in the bias expansion, which was not initially included in
Eq.~(\ref{tracers.LocBias}). That is because gravitational evolution automatically generates a variety of terms
nonlocal in $\delta$, such as the tidal field, which in principle can affect the formation of galaxies and
should consequently be taken into account in the perturbative description of galaxy
bias~\cite{Fry9604,Catelan:2000,McDonald:2009,Chan:2012,Assassi:2014}.

We will refer to any such terms (both local and nonlocal in $\delta$) as \emph{operators}, and it is our aim in
this section to provide a complete basis of operators required by the bispectrum at one-loop order. By basis we
mean a set of linearly independent operators at each order of perturbation theory. We will largely follow up on
the earlier work of \citep{Chan:2012,Assassi:2014, Mirbabayi:2015, Desjacques:2018}, but distinguish between
operators local in second derivatives of the linear potential and those which are not. The former will be
generated even if nonlinear evolution is entirely local (Zel'dovich approximation), whereas the latter derive
from corrections at second-order Lagrangian perturbation theory (LPT) and beyond. Accordingly, we denote these
as ``local evolution'' (LE) and ``nonlocal evolution'' \mbox{(NLE) operators.}
  
Our choice of basis will make this distinction explicit and therefore differs in the type of operators from that
given in \citep{McDonald:2009,Sen1406,Mirbabayi:2015, Desjacques:2018}. The rationale for this choice is:
\mbox{1) to} give the multipoint propagators a particularly simple form, and 2) to provide a hierarchy of
approximations based on which one can reduce the total number of bias parameters. While the usual local
Lagrangian bias approximation (where only LE operators which are also local in $\delta$ are present at the
initial conditions) may well be too restrictive for cosmological parameter estimation, it might prove useful in
practice to relax this at least to all LE operators. This approximation is more accurate but still reduces the
number of free bias parameters compared to the case when we also allow for NLE operators at the initial
conditions. Other bases in the literature mix our two sets of operators, so one might not be able to see these
subtleties when measuring bias parameters from simulations or data. We discuss their relation to ours in
Appendix~\ref{sec:basis-relation}.

\subsection{Galileons as general basis operators}
\label{sec:galileons}

Let us begin with two physical scales important for the process of galaxy formation: 1) the spatial extent
$R_{*}$ on which this process depends on the precise distribution of matter, and 2) the typical time $T_{*}$ it
takes for this matter distribution to collapse into a bound object. While the latter is a significant fraction
of the Hubble time $H^{-1}$, the scale $R_{*}$ usually corresponds to the Lagrangian radius of the galaxy's host
halo, which is of the order $\sim 1~\text{Mpc}$. If we are interested in the clustering of galaxies on scales
\mbox{$r \gg R_{*}$}, then we can consider galaxy formation as essentially local \emph{in space}. For now we
will take this to be the case, before relaxing this assumption in Sec.~\ref{sec:higherD-galaxy-bias}.

As stated at the beginning of this section, other properties than density of the matter field, such as the tidal
field, must enter the bias relation~\cite{Fry9604,Catelan:2000,McDonald:2009,Chan:2012}. More generally, we can
argue that these should depend on the (scaled) gravitational and velocity potentials, defined by
\begin{align}
  \nabla^2\,\Phi(\B{x},\tau) &\equiv \delta(\B{x},\tau)\,, \label{eq:tracers.Poisson} \\
  \nabla^2\,\Phi_v(\B{x},\tau) &\equiv \theta(\B{x},\tau)\,,
\end{align}
as these drive the time evolution in the regime where the dark matter flow is irrotational. According to the
equivalence principle all leading local gravitational effects must stem from second derivatives, which we write
as $\nabla_{ij}\Phi(\B{x},\tau) \equiv \partial_i\partial_j\Phi(\B{x},\tau)$. Similarly, if we assume that dark
matter and galaxies are comoving (i.e. no velocity bias), Galilean invariance of the equations of motion
\citep{Scoccimarro:1996} implies that only second derivatives of the velocity potential are allowed to appear.

Furthermore, as $\delta_g$ is a scalar and therefore invariant under spatial coordinate transformations, we can
limit ourselves to all scalar invariants of the tensors $\nabla_{ij}\Phi$ and $\nabla_{ij}\Phi_v$. In three
dimensions the Cayley-Hamilton theorem~\cite{HofKun71} guarantees that there can only be three such invariants, which can be
expressed by the so-called \mbox{\emph{Galileons}~\citep{Chan:2012}} (repeated indices are summed over):
\begin{align}
  {\cal G}_1(\Phi) \equiv \,&\nabla^2\Phi\,, \\
  {\cal G}_2(\Phi) \equiv \,&\left(\nabla_{ij}\Phi\right)^2 -
  \left(\nabla^2\Phi\right)^2\,, \label{eq:tracers.G2} \\ 
  {\cal G}_3(\Phi) \equiv \,&\left(\nabla^2\Phi\right)^3 +
  2\nabla_{ij}\Phi\,\nabla_{jk}\Phi\,\nabla_{ki}\Phi \nonumber \\ &-
  3\left(\nabla_{ij}\Phi\right)^2\nabla^2\Phi\,, \label{eq:tracers.G3}
\end{align}
and similarly for $\Phi_v$. We note that their leading SPT expressions are of first, second, and third order,
respectively. By inverting the Poisson equation (Eq.~\ref{eq:tracers.Poisson}) we can derive their Fourier space
analogs and for the latter two we obtain:
\begin{alignat}{2}
  {\cal G}_2(\B{k}|\Phi) &= &&\int\displaylimits_{\B{k}_1,\,\B{k}_2} \left[\delta_D\right]_2\,
  K(\B{k}_1,\B{k}_2)\,\delta(\B{k}_1)\,\delta(\B{k}_2)\,, \label{eq:tracers.G2k} \\
  {\cal G}_3(\B{k}|\Phi) &= &&\int\displaylimits_{\B{k}_1,\,\B{k}_2,\,\B{k}_3}\left[\delta_D\right]_3\,L(\B{k}_1,\B{k}_2,\B{k}_3)\,
  \delta(\B{k}_1)\,\delta(\B{k}_2)\,\delta(\B{k}_3)\,, \label{eq:tracers.G3k}
\end{alignat}
where we have defined the following two kernel functions,
\begin{align}
  K(\B{k}_1,\B{k}_2) &\equiv \mu_{12}^2 - 1\,, \label{eq:tracers.K}\\
  L(\B{k}_1,\B{k}_2,\B{k}_3) &\equiv 2\,\mu_{12}\,\mu_{23}\,\mu_{31} - \mu_{12}^2 - \mu_{23}^2 - \mu_{31}^2 +
  1\,, \label{eq:tracers.L}
\end{align}
with $\mu_{ij} \equiv \B{k}_i \cdot \B{k}_j / k_i\,k_j$. Similar expressions hold for $\Phi_v$ by replacing
$\delta$ with $\theta$ in Eqs.~(\ref{eq:tracers.G2k}) and (\ref{eq:tracers.G3k}). Note that these kernels vanish
as $k^2$ when $k=|\sum_i \B{k}_i|\to 0$, and therefore $\langle {\cal G}_2 \rangle = \langle {\cal G}_3 \rangle
=0$ \citep[see e.g.][]{Chan:2012,Assassi:2014}. In addition $\langle \partial {\cal G}_3/\partial \delta \rangle
=0$ since $L(\B{k},\B{q},-\B{q})=0$, which means that ${\cal G}_3$ cannot contribute to the one-loop galaxy
propagator, as we shall discuss in section~\ref{sec:one-point-propagator} below. All these properties make using
Galileons as basis functions better suited for calculations compared to other choices that do not obey these
relations, e.g.~\cite{McDonald:2009}.

\subsection{Local evolution (LE) operators}
\label{sec:lagrangian-basis}

Let us consider the bias relation on some initial time slice in the far past. In that case we are only dealing
with linear fluctuations and the single degree of freedom is $\Phi_L$, as at linear order we have $\Phi =
\Phi_{v}$. If we assume that objects can be identified by a local procedure on $\nabla_{ij}\Phi_L$, the only
terms that can appear in the bias relation for objects at the initial conditions are ${\cal G}_m(\Phi_L)$
($m\leq 3$), such that there will be $n$ basis operators at $n$-th order in the expansion ($n\leq 5$), i.e.
\begin{alignat}{2}
  \label{eq:tracers.ICbasis}
  \begin{aligned}
    &1\text{st:} \, && {\cal G}_1\,, \\
    &2\text{nd:} \, && {\cal G}_1^2\,,\;{\cal G}_2\,, \\
    &3\text{rd:} \, && {\cal G}_1^3\,,\;{\cal G}_1\,{\cal G}_2\,,\;{\cal G}_3\,, \\
    &4\text{th:} \, && {\cal G}_1^4\,,\;{\cal G}_1^2\,{\cal G}_2\,, 
    {\cal  G}_1\,{\cal G}_3\,,\;{\cal G}_2^2\,,
  \end{aligned}
\end{alignat}
and we assign a free bias parameter to each of these operators. We stress that the corresponding tracer density
at initial time will thus be a local function of $\nabla_{ij}\Phi_L$, which is similar in spirit with more
phenomenological approaches, such as the peak and excursion set bias models
~\cite{BarBonKai8605,MoWhi9609,MoJinWhi97,ParShe1211}. In reality though, tracers are not identified at the
initial time, but in the late-time, nonlinear field, which means that the basis in
Eq.~(\ref{eq:tracers.ICbasis}) will be insufficient if gravitational instability produces terms nonlocal in
$\nabla_{ij}\Phi_L$. This is the case starting with second-order corrections to the Zel'dovich approximation
\citep{KofPog9503}, as we will see in Sec.~\ref{sec:eulerian-basis}. Therefore, even when traced back to the
initial conditions such terms can in principle not be ignored. If we do so, the nonlocal terms are still
produced during the evolution process, but with their bias coefficients fixed in terms of the parameters
associated to the operators in Eq.~(\ref{eq:tracers.ICbasis}) \citep{Chan:2012}. For that reason, this provides
us with a very useful approximation (with fewer free parameters) that can be tested against numerical
simulations~\cite{BiasLoops2}.

\subsection{Nonlocal evolution (NLE) operators}
\label{sec:eulerian-basis}

Due to nonlinear evolution, the gravitational and velocity potentials will no longer be identical. Following our
previous arguments, it thus seems obvious to simply double the number of operators at each order by including a
set of Galileons for both, the evolved $\Phi$ and $\Phi_v$, and also allow for their combinations, in order to
complete the basis in Eq.~(\ref{eq:tracers.ICbasis}). Unfortunately, this produces a lot of redundancy as many
of these operators are degenerate, so our task will be to identify those, which are linearly independent. We
follow the strategy first developed in \citep{Chan:2012}.

At first order in SPT we have already established that ${\cal G}_1(\Phi) = {\cal G}_1(\Phi_v)$, and we choose
the former, i.e. the matter fluctuation itself, as our first basis operator. Likewise, the two second-order
Galileons are degenerate at second order in SPT and furthermore, we have
\begin{align}
  \label{eq:tracers.G1diff}
  {\cal G}_1^{(2)}(\Phi) - {\cal G}_1^{(2)}(\Phi_v) = \delta^{(2)}(\B{x}) - \theta^{(2)}(\B{x}) =
  -\frac{2}{7}\,{\cal G}_2(\Phi_L)\,,
\end{align}
proving that the basis in Eq.~(\ref{eq:tracers.ICbasis}) is complete up to that order. The need for an
additional operator occurs for the first time at third order. Using the notation $\Delta_n{\cal G}_m \equiv
{\cal G}_m^{(n)}(\Phi)-{\cal G}_m^{(n)}(\Phi_v)$ for the difference between the $m$-th Galileons evaluated at
$n$-th order in SPT, we see that in addition to the ones already written in Eq.~(\ref{eq:tracers.ICbasis}) there
are the following four combinations:
\begin{align}
  \Delta_3{\cal G}_1\,,\; \Delta_3{\cal G}_2\,,\; \delta\,\Delta_2{\cal G}_1\,,\; \Delta_3\left({\cal G}_1^2\right)\,.
\end{align}
From Eq.~(\ref{eq:tracers.G1diff}) follows that the latter two are degenerate with $\delta\,{\cal G}_2(\Phi_v)$,
while $\Delta_3{\cal G}_1$ contains a contribution that cannot be written in terms of second derivatives of
$\Phi_v$ and is thus not Galilean invariant. This only leaves the second combination, which gives
\begin{align}
  \label{eq:tracers.G2diff}
  \Delta_3{\cal G}_2 = -\frac{4}{7} \Big[\nabla_{ij}\Phi_L\,\nabla_{ij} \nabla^{-2} {\cal G}_2(\Phi_L) -
  \delta\,{\cal G}_2(\Phi_L)\Big]\,,
\end{align}
demonstrating, as claimed above, that the additional basis operators induced by gravity can no longer be
expressed as local functions of the linear gravitational potential, i.e. $\nabla_{ij}\Phi_L$. As these effects
are precisely captured by second-order LPT and beyond, instead of explicitly calculating the differences between
Galileons of $\Phi$ and $\Phi_v$, we can follow an alternative strategy (which builds on~\cite{Chan:2012}) that
will prove particularly useful for extending the basis beyond third order. Let us consider LPT, which summarizes
all of the dynamics in its Lagrangian displacement field
\begin{align}
  \B{\Psi}(\B{q},\tau) = \; &D_1(\tau)\B{\Psi}^{(1)}(\B{q}) + D_2(\tau)\B{\Psi}^{(2)}(\B{q}) + \ldots\,,
\end{align}
that moves particles from their initial positions $\B{q}$ to their final destinations $\B{x} = \B{q} +
\B{\Psi}(\B{q},\tau)$ at conformal time $\tau$. The functions $\B{\Psi}^{(n)}(\B{q})$ are the $n$-th order contributions and $D_n(\tau)$
are the corresponding growth factors ($D_1 \equiv D$ is the linear growth factor). At any order both the
gravitational and velocity potentials can be expressed in terms of these $\B{\Psi}^{(n)}$, which are in turn
given by the LPT potentials $\varphi_n$~\cite{Buc9301,Bouchet:1995}, e.g.
\begin{align}
  \B{\nabla} \cdot \B{\Psi}^{(1)} &= \nabla^2\,\varphi_1 = -\delta\,, \label{eq:tracers.phi1} \\
  \B{\nabla} \cdot \B{\Psi}^{(2)} &= \nabla^2\,\varphi_2 = - {\cal G}_2(\varphi_1)\,. \label{eq:tracers.phi2}
\end{align}
Any set of linearly independent operators induced by gravity must therefore be connected to combinations of the
LPT potentials. In order to guarantee that these are still Galilean invariant, we can generalize the definition
of the Galileons to~\citep{Chan:2012}
\begin{align}
  {\cal G}_2(A,B) \equiv \nabla_{ij}A\,\nabla_{ij}B - \nabla^2A\,\nabla^2B\,,
\end{align}
and similarly for ${\cal G}_3(A,B,C)$. From Eq.~(\ref{eq:tracers.phi1}) we have $\varphi_1 = -\Phi_L$, so that
the first new combination appears at third order of perturbation theory: ${\cal
  G}_2(\varphi_2,\varphi_1)$. Evaluating this Galileon using Eq.~(\ref{eq:tracers.phi2}) shows that it is
precisely related to the only gravity induced operator that we previously identified at third-order,
i.e.~\cite{Chan:2012} 
\beq
\Delta_3{\cal G}_2 = -{4\over 7}\,{\cal G}_2(\varphi_2,\varphi_1).
\eeq

\begin{table*}
  \centering
  \setlength{\tabcolsep}{8pt}
  \caption{Overview of basis operators for galaxy bias, along with their associated bias
    parameters. Each single column presents a different order in SPT and we have categorized operators into groups, which are local
    and nonlocal in second derivatives of the  linear gravitational potential (left and middle columns, respectively), and which
    contain higher than second derivatives of the potentials. Note that to simplify notation, we have
    relabeled some bias parameters compared to \citep{Chan:2012} --- what was formerly $\gamma_3^{\times}$ is
    now $\gamma_2^{\times}$ and $\gamma_3^-$ has become $-7/4\,\gamma_{21}$.  A superscript $\times$ denotes that a Galileon field has been multiplied by $\delta$.     Operators related to noise terms are not listed here, see section~\ref{sec:shot-noise}.}
  \begin{ruledtabular}
    \begin{tabular}{cccc|cc|cc}
      \multicolumn{4}{c|}{Local Evolution Operators} & \multicolumn{2}{c|}{Nonlocal Evolution Operators} & \multicolumn{2}{c}{Higher-derivative}
      \Tstrut\Bstrut\\
      \multicolumn{1}{c}{$1$st} & \multicolumn{1}{c}{$2$nd} & \multicolumn{1}{c}{$3$rd} &
      \multicolumn{1}{c|}{$4$th} & \multicolumn{1}{c}{$3$rd} & \multicolumn{1}{c|}{$4$th} &
      \multicolumn{1}{c}{$3$rd} & \multicolumn{1}{c}{$4$th} \Bstrut\\ 
      \hline
      $b_1\,\delta$ & $b_2\,\delta^2/2$ & $b_3\,\delta^3/3!$ & $b_4\,\delta^4/4!$ & $\gamma_{21}\,{\cal
        G}_2(\varphi_2,\varphi_1)$ & $\gamma_{21}^{\times}\,\delta\, {\cal G}_2(\varphi_2,\varphi_1)$ &
      $\beta_1\,\nabla^2\delta$ & $\beta_{2,1}\,\nabla^2\delta^2$ \Tstrut\Bstrut \\
      & $\gamma_2\,{\cal G}_2(\Phi_v)$ & $\gamma_2^{\times}\,\delta\,{\cal G}_2(\Phi_v)$ &
      $\gamma_2^{\times\times}\,\delta^2\,{\cal G}_2(\Phi_v)$ & & $\gamma_{211}\,{\cal
        G}_3(\varphi_2,\varphi_1,\varphi_1)$ & & $\beta_{2,2}\,\big(\B{\nabla}\delta\big)^2$
      \Bstrut \\
      & & $\gamma_3\,{\cal G}_3(\Phi_v)$ & $\gamma_3^{\times}\,\delta\,{\cal G}_3(\Phi_v)$ & &
      $\gamma_{22}\,{\cal G}_2(\varphi_2,\varphi_2)$ & & $\beta_{2,3}\,\nabla^2{\cal
        G}_2(\Phi_v)$ \Bstrut \\ 
      & & & $\gamma_2^{\mathrm{sq}}\,{\cal G}_2(\Phi_v)^2$ & & $\gamma_{31}\,{\cal G}_2(\varphi_3,\varphi_1)$ &
      & $\beta_{2,4}\,{\cal G}_2(\B{\nabla}\Phi_v)$ \Bstrut \\
    \end{tabular}
  \end{ruledtabular}
  \label{tab:basis}
\end{table*}

Following this line of argument we can now easily determine the additional operators at fourth order: apart from
the combination $\delta\, {\cal G}_2(\varphi_2,\varphi_1)$, we can construct the following three invariants out of the LPT
potentials
\begin{align}
  {\cal G}_2(\varphi_2,\varphi_2)\,,\; {\cal G}_2(\varphi_3,\varphi_1)\,,\; {\cal
    G}_3(\varphi_2,\varphi_1,\varphi_1)\,.
\end{align}
However, beyond second order the LPT solutions are no longer purely potential anymore and at third order in
particular it consists of two scalar and a transverse vector potential, all with different time dependencies:
\begin{align}
  \left.\B{\Psi}(\B{q},\tau)\right|_{3\mathrm{rd}} = \; &D_3^{(a)}(\tau)\,\B{\Psi}^{(3,a)}(\B{q}) +
                                                          D_3^{(b)}(\tau)\,\B{\Psi}^{(3,b)}(\B{q}) \nonumber \\
                                                        & + D_3^{(c)}(\tau)\,\B{\Psi}^{(3,c)}(\B{q})\,,
\end{align}
where in the Einstein-de Sitter (EdS) approximation the growth factors are given by
$D_3^{(a)}(\tau) = 1/18\,D(\tau)^3$, $D_3^{(b)}(\tau) = 5/42\,D(\tau)^3$ and
$D_3^{(c)}(\tau) = 1/14\,D(\tau)^3$. The functions $\Psi^{(3,a/b/c)}$ and their corresponding potentials satisfy
the following Poisson equations~\cite{Buc9404}
\begin{alignat}{2}
  \B{\nabla} \cdot \B{\Psi}^{(3,a)} &= \nabla^2\,\varphi_3^{(a)} &&= -{\cal G}_3(\varphi_1)\,, \label{eq:tracers.p3a_def} \\
  \B{\nabla} \cdot \B{\Psi}^{(3,b)} &= \nabla^2\,\varphi_3^{(b)} &&= -{\cal G}_2(\varphi_2,\varphi_1)\,, \label{eq:tracers.p3b_def} \\
  -\B{\nabla} \times \B{\Psi}^{(3,c)} &= \nabla^2  \B{A}_3 &&= - \hat{e}_i\, \epsilon_{ijk}\,\left(\nabla_{jl}\,\varphi_1\right)\,\left(\nabla_{kl}\,\varphi_2\right)\,,  \label{eq:tracers.A3_def}
\end{alignat}
where $\epsilon_{ijk}$ denotes the fully anti-symmetric Levi-Civita symbol and $\hat{e}_i$ the unit vector in direction $i$. The combination of the third and
first order LPT potentials is thus made up of three pieces and factoring out $D(\tau)^3$ from the EdS
solutions, we define
\begin{align}
  \label{eq:tracers.G2p3p1_def}
  {\cal G}_2(\varphi_3,\varphi_1) \equiv \;&\frac{1}{18}{\cal G}_2(\varphi_3^{(a)},\varphi_1) +
  \frac{5}{42}{\cal G}_2(\varphi_3^{(b)},\varphi_1) \nonumber \\ &+ \frac{1}{14} \nabla_i \left(\B{\nabla}
    \times \B{A}_3\right)_j\,\nabla_{ij}\,\varphi_1\,.
\end{align}
Due to the different time dependencies we should in principle allow these three pieces to enter the bias basis
individually, but in practice the departures from the EdS approximation to growth factors are below the percent-level, so for all
purposes of this paper we are safe to ignore this complication.

To conclude, our choice of a complete Galilean invariant basis for the evolved galaxy perturbations is given by
a set of 15 operators up to fourth order, which are summarized in the first two columns of
Table~\ref{tab:basis}. We have separated what we denote as local evolution operators (first column), which are
local in $\nabla_{ij}\,\Phi_L$, from those that correspond to nonlinear corrections to the gravitational and
velocity potentials that are nonlocal in $\nabla_{ij}\,\Phi_L$ (middle column).  With respect to the number of
basis operators we are thus in agreement with~\citep{Desjacques:2018}, the relation between our and their basis
up to fourth order is given in Appendix~\ref{sec:basis-relation}.

\subsection{Higher-derivative galaxy bias}
\label{sec:higherD-galaxy-bias}

Although some of the basis operators we derived in Sec.~\ref{sec:lagrangian-basis} and \ref{sec:eulerian-basis}
are nonlocal in the matter fluctuations and gravitational potential, we made the central assumption that the
formation of galaxies is \emph{spatially-local}, meaning it is determined by the value of these operators at a
single point in space. On small scales this approximation must break down because galaxies form due to matter
collapsing from a finite region of size $R_*$, which is of the order of the Lagrangian radius of the host dark
matter halo. As shown in \citep{McDonald:2009,Sen1406,Desjacques:2018} this gives rise to additional operators in the
bias expansion that contain higher than second derivatives of the gravitational and velocity potentials\footnote{These have been known for a very long time, e.g.~\cite{BarBonKai8605} since they naturally appear in peak models of biased tracers.}. 

As these operators must still be scalars, the simplest one involves four derivatives of $\Phi$ and is given by
$\nabla^2\,\delta$. Its effect can be interpreted as an emerging scale dependence of the linear bias parameter,
because upon Fourier transformation and grouping all terms linear in $\delta$, we get
\begin{equation}
  b_1(k) = b_1 - \beta_1\,k^2 + {\cal O}(k^4)\,,
\end{equation}
where $\beta_1$ is the bias parameter associated with $\nabla^2\,\delta$ and ${\cal O}(k^4)$ stands for
contributions involving even higher derivatives. Note that $\beta_1$ has units of length squared, and as we can
expect it to scale with $R_{*}$, we see that on scales much larger than the Lagrangian radius, i.e.
$k\,R_* \ll 1$, the higher-derivative contributions are suppressed by powers of $(k\,R_*)^2$ compared to the
spatially-local linear term. This is similar to the nonlinear bias terms, where an increase in nonlinear order
is suppressed by powers of $\delta\simeq(k/k_{\rm nl})^{(n_{\rm eff}+3)/2}$ at large scales, with
$k_{nl}^3 P(k_{nl})/(2\pi^2)\equiv1$ and $n_{\rm eff}$ the effective spectral index at the nonlinear
scale. However, comparing the nonlinear and higher-derivative contributions against each other \emph{a priori}
is difficult, since this depends on the relative size of $R_*$ (which in turn depends on the biased tracer in
question) and $k_{nl}$ and $n_{\rm eff}$ (which depend on the linear spectrum and redshift), in addition to the
size of the bias parameters. We therefore take the following approach: we consider \emph{a priori} each
higher-derivative factor in correlators as a nonlinear bias loop, counting each additional derivative acting on
the gravitational or velocity potentials as an increase of the SPT order by one. The operator $\nabla^2\,\delta$
would thus be considered as third order. After performing a measurement of bias parameters from clustering data,
one can reassess \emph{a posteriori} whether bias loops of derivatives are more important for a particular
biased tracer. In~\cite{Nadler:2018}, the authors conclude that higher-derivative biases are more important than
loops for tracers of a very wide range of masses. We revisit this \mbox{issue in~\cite{BiasLoops2}}, but
Figs.~\ref{fig:Ptreevsloop} and~\ref{fig:Btreevsloop} below already suggest that loop corrections are as
important for the bispectrum as for the power spectrum and that they matter on scales commonly used in the
analysis of galaxy surveys.

Based on our counting, the fourth-order higher-derivative terms will simply be given by acting with two
derivatives on the spatially-local quadratic bias operators. This allows only for the following four independent
combinations
\begin{equation}
  \label{eq:tracers.hd4th}
  \nabla^2\delta^2\,,\; \big(\B{\nabla} \delta\big)^2\,,\; \nabla^2{\cal G}_2(\Phi_v)\,,\;
    {\cal G}_2(\nabla_i\Phi_v,\nabla_i\Phi_v)\,,
\end{equation}
where $\delta\,\nabla^2\,\delta$ as well as ${\cal G}_2(\Phi_v,\nabla^2\Phi_v)$ do not enter individually, as
they can be expressed through combinations of the operators above. In total we thus obtain a set of five
higher-derivative operators, which are summarized in Table~\ref{tab:basis} according to their classification in
terms of SPT order as discussed above. Note that Ref. \cite{Desjacques:2018} gives a total of ten operators up
to the same order. We find that the set presented here is equivalent with the first five operators in their
Eq.~(2.74), while the last five can be expressed in terms of the other ones, making them superfluous.

\section{{Multipoint propagators in Lagrangian and Eulerian space}}
\label{sec:propformalism}

We now return to the main idea presented in Sec.~\ref{sec:physical_bias}: in order to guarantee that the bias
parameters corresponding to all of the Galilean basis operators are observable quantities, we should construct
the galaxy density field out of multipoint propagators,
\begin{align}
  \left<\frac{\partial^n \delta_g(\B{k})}{\partial\,\delta_L(\B{k}_1) \cdots \partial\,\delta_L(\B{k}_n)}\right>\,,
\end{align}
as opposed to the usual kernel functions that we obtain from
$\left[\partial^n \delta_g(\B{k}) /\partial\,\delta_L(\B{k}_1) \cdots \partial\,\delta_L(\B{k}_n)\right]_0$. Our
goal is to compute the complete multipoint propagators at initial and final time, while paying particular
attention to the evolution of the bias parameters. Before that we illustrate their relation to the
Wiener-Hermite functionals, which formalizes the multipoint propagator expansion in Fourier space and will prove
useful for determining their time evolution.

\subsection{Relating multipoint propagators and the Wiener-Hermite expansion}
\label{sec:hermite-expansion}

For a linear Gaussian dark matter density field, the probability density function (PDF) for a mode in Fourier
space is given by
\begin{align}
  \label{eq:MP.PDF}
  {\cal P}[\delta_L] = N\,\exp{\left[-\frac{1}{2}\int_{\B{q}}\frac{|\delta_L(\B{q})|^2}{P_L(q)}\right]}\,,
\end{align}
with normalization factor $N$. The $n$-th generalized Wiener-Hermite functional ${\cal H}_n$ is then defined by
taking $n$ functional derivatives of the PDF \citep{Matsubara:1995}:
\begin{align}
  \label{eq:MP.defhermite}
  \frac{{\cal H}_n(\B{k}_1,\ldots,\B{k}_n)}{\prod_{i=1}^n P_L(\B{k}_i)} \equiv\;&\frac{(-1)^n}{{\cal P}[\delta_L]}
  \frac{\partial^n\,{\cal P}[\delta_L]}{\partial\delta_L(\B{k}_1)\,\cdots\,\partial\delta_L(\B{k}_n)}\,.
\end{align}
Using this definition, we obtain the following first three functionals (suppressing the momentum arguments):
\begin{alignat}{2}
  \label{eq:MP.hermite1-3}
  \begin{aligned}
    {\cal H}_1 &= &&\hspace*{-0.5em}\delta_L^*(\B{k})\,, \\[0.25em]
    {\cal H}_2 &= &&\hspace*{-0.5em}\delta_L^*(\B{k}_1)\,\delta_L^*(\B{k}_2) -
    \langle\delta_L(\B{k}_1)\,\delta_L(\B{k}_2)\rangle\,, \\[0.25em]
    {\cal H}_3 &=
    &&\hspace*{-0.5em}\delta_L^*(\B{k}_1)\,\delta_L^*(\B{k}_2)\,\delta_L^*(\B{k}_3) \\ & &&\hspace*{-0.5em}- \Big[
    \langle\delta_L(\B{k}_1)\,\delta_L(\B{k}_2)\rangle\,\delta_L^*(\B{k}_3) + \text{cyc.}\Big]\,,
  \end{aligned}
\end{alignat}
where a superscript $*$ denotes complex conjugation. Like the standard Hermite polynomials they satisfy an
orthogonality relation, which can be shown to be (see \citep{Matsubara:1995} and
App.~\ref{sec:hermite_products}):
\begin{align}
  \label{eq:MP.hermite_orthonality}
  \langle &{\cal H}_n(\B{k}_1,\ldots,\B{k}_n)\,{\cal
    H}_m^*(\B{q}_1,\ldots,\B{q}_n)\rangle = (2\pi)^{3n}\,\delta_{nm}^{\rm K} \nonumber \\
  &\times\,\Big[\delta_D(\B{k}_1-\B{q}_1)\cdots\delta_D(\B{k}_n-\B{q}_n) + \text{sym.}\Big] \prod_{i=1}^n
  P_L(\B{k}_i)
\end{align}
where $\delta_{nm}^{\rm K}$ is the Kronecker delta, and ``sym.'' stands for the remaining $(n!-1)$ combinations
of the arguments. Expanding the Fourier space galaxy overdensity in terms of Hermite functionals gives a
convolution over ${\cal H}_n$ at each order, such that
\begin{align}
  \label{eq:MP.deltag_hermite}
  \delta_g(\B{k}) = \sum_n \frac{(2\pi)^3}{n!} \int\displaylimits_{\B{k}_1,\ldots,\,\B{k}_n}
  &\left[\delta_D\right]_n\,\Gamma_g^{(n)}(\B{k}_1,\ldots,\B{k}_n)\, \nonumber \\ &\times\,{\cal
    H}_n^*(\B{k}_1,\ldots,\B{k}_n)\,,
\end{align}
where the coefficients $\Gamma_g^{(n)}(\B{k}_1,\ldots,\B{k}_n)$ are scale-dependent functions (kernels), that
can be interpreted as the corresponding bias parameters. Multiplying both sides of
Eq.~(\ref{eq:MP.deltag_hermite}) with ${\cal H}_m$ and using the orthogonality relation
(Eq.~\ref{eq:MP.hermite_orthonality}), we see that
\begin{align}
  \label{eq:MP.hermite_product}
  \frac{\langle{\cal H}_n(\B{k}_1,\ldots,\B{k}_n)\,\delta_g(\B{k})\rangle}{\prod_{i=1}^nP_L(\B{k}_i)} =
  \;&(2\pi)^3\,\Gamma_g^{(n)}(\B{k}_1,\ldots,\B{k}_n) \nonumber \\
  &\times\,\delta_D(\B{k}-\B{k}_{1\cdots n})\,,
\end{align}
which generalizes Eqs.~(\ref{PropIsCrossCorr}-\ref{PropIsCrossCorr2}). Cross-correlating the galaxy density
field with Hermite polynomials to measure local bias parameters has been proposed in a similar way
in~\cite{ParSefChu1305}. Moreover, this is also how propagators for the mapping from linear to nonlinear
fluctuations have been measured in~\cite{CroSco0603b,Bernardeau:2008,Bernardeau:2012}. Note that in general for
nonlocal bias this cross-correlation has to be further decomposed in terms of the structures inside
$\Gamma_g^{(n)}$ to end up with parameter estimates (i.e. one has to separate $b_2$ from $\gamma_2$
contributions etc.), e.g.~\cite{SchBalSel1411,Bel:2015,HofBelGaz1702,LazSch1712,AbiBal1807}, but that procedure is basis-dependent.

We can derive a different relation by plugging in Eq.~(\ref{eq:MP.defhermite}) into
Eq.~(\ref{eq:MP.hermite_product}) and replacing the ensemble average by its definition --- the functional
integral of all modes $\delta_L$ over their joint PDF:
\begin{align}
  \label{eq:MP.hermite_derivative}
  &\frac{\langle{\cal
      H}_n(\B{k}_1,\ldots,\B{k}_n)\,\delta_g(\B{k})\rangle}{\prod_{i=1}^nP_L(\B{k}_i)} \nonumber \\
  &\hspace{1.5em}= (-1)^n \int {\cal D}[\delta_L]\,\left[\frac{\partial^n\,{\cal
        P}[\delta_L]}{\partial\delta_L(\B{k}_1)\,\cdots\,\partial\delta_L(\B{k}_n)}\right] \delta_g(\B{k})
  \nonumber \\
  &\hspace{1.5em}= \int {\cal D}[\delta_L]\,{\cal
    P}[\delta_L]\,\frac{\partial^n\,\delta_g(\B{k})}{\partial\delta_L(\B{k}_1)\,\cdots\,\partial\delta_L(\B{k}_n)}
  \nonumber \\
  &\hspace{1.5em}=
  \left<\frac{\partial^n\,\delta_g(\B{k})}{\partial\delta_L(\B{k}_1)\,\cdots\,\partial\delta_L(\B{k}_n)}\right>\,,
\end{align}  
where we have integrated by parts $n$ times in going from the second to the third line and ignored any surface
terms as ${\cal P}[\delta_L] \to 0$ (same as its derivatives) for $\delta_L \to \pm\infty$. Thus, it follows
that the kernels of the Wiener-Hermite expansion are indeed the multipoint propagators:
\begin{align}
  \label{eq:MP.defGamma}
  \left<\frac{\partial^n\,\delta_g(\B{k})}{\partial\delta_L(\B{k}_1)\,\cdots\,\partial\delta_L(\B{k}_n)}\right>
  =\;&(2\pi)^3\,\Gamma_g^{(n)}(\B{k}_1,\ldots,\B{k}_n) \nonumber \\ &\times\,\delta_D(\B{k}-\B{k}_{1\cdots n})\,.
\end{align}
This equation is equivalent to the definition of multipoint propagators in the previous work
of~\citep{Bernardeau:2008,Bernardeau:2012}, which considered the nonlinear matter fluctuations in
place of the galaxy fluctuations. As they showed that the multipoint propagators function as basic building
blocks for constructing arbitrary $N$-point spectra, the same holds for the $\Gamma_g^{(n)}$ and we will use
this fact to compute the galaxy power spectrum and bispectrum in Sec.~\ref{sec:PB}.

\subsection{Initial conditions}
\label{sec:galaxy-propagators}

We now determine the first three multipoint propagators at an initial time where nonlinearities in the dark
matter field can be ignored, which implies we can set all SPT kernels $F_n$ and $G_n$ for $n \geq 2$ to zero. We
include all of the basis operators summarized in Table~\ref{tab:basis} and work up to fourth order in SPT, as
required by the one-loop bispectrum. For reasons explained below, the general structure of the multipoint
propagators in our basis is very simple: they are given by a tree-level contribution consisting of all basis
operators that correspond to the order of the propagator itself, in addition to loop corrections involving only
NLE operators.

\subsubsection{The one-point propagator}
\label{sec:one-point-propagator}

To compute the one-point propagator, we need to take a single derivative of $\delta_g$ and take the expectation
value, which implies due to Gaussianity of $\delta_L$ that only odd orders in the bias expansion enter. More
generally, as each derivative cancels exactly one factor of $\delta_L$, we see that odd (even) numbered
propagators can only contain terms stemming from odd (even) orders of the bias expansion.

The first term in the bias expansion gives just linear bias, thus to compute $\Gamma_g^{(1)}$ to one loop we
need the derivative of a generic third order term ${\cal O}_B^{(3)}$, which can be written as the convolution
\begin{align}
  \label{eq:MP.delta3rd_generic}
  {\cal O}_B^{(3)}(\B{k}) = (2\pi)^3\int\displaylimits_{\B{k}_1,\B{k}_2,\B{k}_3} &\left[\delta_D\right]_3\,{\cal
    K}_B^{(3)}(\B{k}_1,\B{k}_2,\B{k}_3)\, \nonumber \\ &\times\,\delta_L(\B{k}_1)\,\delta_L(\B{k}_2)\,\delta_L(\B{k}_3)\,,
\end{align}
with $B \in \{\delta^3,\,\delta\,{\cal G}_2,\,{\cal G}_3,\,{\cal G}_2(\varphi_2,\varphi_1)\}$ and the kernels
${\cal K}_B^{(3)}$ are given in Eqs.~(\ref{eq:appbasis.K3d3}) to (\ref{eq:appbasis.K3Gp2p1}). Making use of the
symmetry of ${\cal K}_B^{(3)}$, we obtain:
\begin{align}
  \label{eq:MP.Gamma1_3rdorder}
  \left<\frac{\partial\,{\cal O}_B^{(3)}(\B{k})}{\partial\,\delta_L(\B{k}')}\right>' = 3
  \int_{\B{q}}{\cal K}_B^{(3)}(\B{k},\B{q},-\B{q})\,P_L(q) \,,
\end{align}
using the same notation for ensemble averages as in Sec.~\ref{sec:bias-exp-with-evo}. We note that
Eq.~(\ref{eq:MP.Gamma1_3rdorder}) is analogous to the result obtained for the one-loop contribution to the
matter propagator, which gives rise to $P_{13}/2P_L$ \citep{CroSco0603a}.

As pointed out already, most of the possibilities for third-order kernel contributions are trivial or zero in
our choice of basis functions for bias. In fact, no basis function local in $\nabla_{ij}\Phi$ (denoted as LE
operators in Table~\ref{tab:basis}) can give a non-trivial contribution to the loop corrections of
$\Gamma_g^{(1)}$ (with trivial contributions meaning renormalizations of the linear bias parameter, as discussed
in section~\ref{sec:physical_bias}). The reason is that in propagator loop integrals no wavevector angles can
appear inside power spectra (due to being reducible diagrams), thus one can always perform angular integrations
over momenta of the resulting kernels, which for operators local in $\nabla_{ij}\Phi$ simply give rise to
numerical prefactors (many of them zero in our choice of basis). Therefore such loops are only functions of
$\sigma^2$, e.g. for $B = \delta\,{\cal G}_2$,
\begin{align}
  \left<\frac{\partial\,{\cal O}_{\delta{\cal G}_2}^{(3)}(\B{k})}{\partial\,\delta_L(\B{k}')}\right>' &= 2 \int_{\B{q}}
  K(\B{k},\B{q})\,P_L(q)  =
  -\frac{4}{3}\,\sigma^2\, ,
\end{align}
similarly to the case $B=\delta^3$, while $\langle \partial {\cal G}_3/\partial \delta_L \rangle =0$ since $L(\B{k},\B{q},-\B{q}) = 0$.
On the other hand, for the only NLE operator (see Table~\ref{tab:basis})  at third order $B={\cal G}_2(\varphi_2,\varphi_1)$, we get
\begin{align}
    \left<\frac{\partial\,{\cal O}_{{\cal G}_2(\varphi_2,\varphi_1)}^{(3)}(\B{k})}{\partial\,\delta_L(\B{k}')}\right>' &=
    2\int_{\B{q}} K(\B{k}-\B{q},\B{q})\,K(\B{k},\B{q})\,P_L(q)\,.
\end{align}
Collecting all these results, the first galaxy propagator at initial time is given by
\begin{align}
  \label{eq:MP.Gamma1}
  \Gamma_g^{(1)}(\B{k}) = b_1 - \beta_1\, k^2 +
   2\gamma_{21} \int_{\B{q}} K(\B{k}-\B{q},\B{q})\,K(\B{k},\B{q})\,P_L(q)\,,
\end{align}
where
\beq
  b_1 \equiv \bar{b}_1 + \left[\frac{1}{2}\bar{b}_3-\frac{4}{3}\bar{\gamma}_2^{\times}\right]\sigma^2\,, \ \ \ \ \ \ 
  \gamma_{21} \equiv \bar{\gamma}_{21}\,,
  \label{b1renFull}
\eeq 
and we have added the higher-derivative term. As noted in Sec.~\ref{sec:physical_bias}, Eq.~(\ref{b1renFull}) can
also be derived from the renormalized expansion in Eq.~(\ref{GammaExp}) without explicitly computing the loops
for the terms local in $\nabla_{ij}\Phi$.  

In the limit $k \to 0$, the $\gamma_{21}$ integral scales as
\begin{equation}
  \int_{\B{q}} K(\B{k}-\B{q},\B{q})\,K(\B{k},\B{q})\,P_L(q) = \frac{8}{15}\,k^2\,\int_{\B{q}}
  \frac{P_L(q)}{q^2} + {\cal O}(k^4)\,,
\end{equation}
which ensures that $\Gamma_g^{(1)}$ indeed corresponds to the linear bias parameter on large scales. Moreover,
we note that if terms of order $k^4$ and beyond are negligible, the $\gamma_{21}$ contribution is entirely
degenerate with the higher-derivative term \citep[see also][]{McDonald:2009,BiaDesKeh1408,SaiBalVla1405}. As long
as we restrict ourselves to sufficiently large scales, this implies it is not necessary to include the
higher-derivative parameter in excess of $\gamma_{21}$. However, \emph{a priori} it is difficult to tell what
``sufficient'' means for a given data set, and therefore requires careful testing using mock data.

\subsubsection{The two-point propagator}
\label{sec:two-point-propagator}

In a similar manner we can now derive all remaining multipoint propagators. The two-point propagator gets
contributions from second and fourth order bias operators, which, analogous to
Eq.~(\ref{eq:MP.delta3rd_generic}), we write as the symmetric kernels ${\cal K}_B^{(2)}$ and ${\cal
  K}_B^{(4)}$. The corresponding expressions are given in Eqs.~(\ref{eq:appbasis.K2d2}) and
(\ref{eq:appbasis.K2G2}), and Eqs.~(\ref{eq:appbasis.K4d4}) to (\ref{eq:appbasis.K4Gp3p1}), respectively.
Differentiating a generic second or fourth order contribution twice results in
\begin{align}
  \left<\frac{\partial^2 {\cal O}_B^{(2)}(\B{k})}{\partial\,\delta_L(\B{k}_1)\,\partial\,\delta_L(\B{k}_2)}\right>' = 2\,
  {\cal K}_B^{(2)}(\B{k}_1,\B{k}_2)\,,
\end{align}
and
\begin{align}
  \left<\frac{\partial^2 {\cal O}_B^{(4)}(\B{k})}{\partial\,\delta_L(\B{k}_1)\,\partial\,
      \delta_L(\B{k}_2)}\right>' = 12 \int_{\B{q}}{\cal K}_B^{(4)}(\B{k}_1,\B{k}_2,\B{q},-\B{q})\,P_L(q)\,.
\end{align}
As explained above, the non-trivial loop corrections to the two-point galaxy propagator can only consist of the
NLE operators and so we get:
\begin{align}
  \label{eq:MP.Gamma2}
  \Gamma_g^{(2)}(\B{k}_1,\B{k}_2) &=  b_2 + 2 \gamma_2\,K(\B{k}_1,\B{k}_2) \nonumber \\ &+ 12
  \int_{\B{q}}\Big[\gamma_{21}^{\times}\,{\cal K}_{\delta{\cal G}_2(\varphi_2,\varphi_1)}^{(4,\text{F})} +
  \gamma_{211}\,{\cal K}_{{\cal G}_3(\varphi_2,\varphi_1,\varphi_1)}^{(4)}\Big. \nonumber \\ &\Big.+
  \gamma_{22}\,{\cal K}_{{\cal G}_2(\varphi_2,\varphi_2)}^{(4)} + \gamma_{31}\,{\cal K}_{{\cal
      G}_2(\varphi_3,\varphi_1)}^{(4)}\Big]\,P_L(q) \nonumber \\ &
      - \beta_{2,1}\, k_{12}^2  - \beta_{2,2}\, (\B{k}_1\cdot \B{k}_2)   
\nonumber \\ & - \big[\beta_{2,3}\, k_{12}^2+\beta_{2,4}\,(\B{k}_1\cdot \B{k}_2)\big]\,K(\B{k}_1,\B{k}_2)
\end{align}
where $k_{12}^2\equiv |\B{k}_1+\B{k}_2|^2$ and the bracket inside the loop integral is evaluated at
$(\B{k}_1,\B{k}_2,\B{q},-\B{q})$. Furthermore, the kernel
${\cal K}_{\delta{\cal G}_2(\varphi_2,\varphi_1)}^{(4,\text{F})}$ denotes the finite part of
${\cal K}_{\delta{\cal G}_2(\varphi_2,\varphi_1)}^{(4)}$ (i.e. removing contributions proportional to
$\sigma^2$) and we have added the higher-derivative terms in the last two lines.  For the same reasons mentioned
above, all of the initial operators again only contribute to renormalizations of
\begin{align}
\label{b2renFull}
  b_2 &\equiv \bar{b}_2 + \left[\frac{1}{2}\bar{b}_4 -
    \frac{16}{3}\bar{\gamma}_2^{\times\times} + \frac{32}{15}\bar{\gamma}_{21}^{\times} +
    \frac{64}{15}\bar{\gamma}_2^{\mathrm{sq}}\right]\sigma^2\,, \\
  \gamma_2 &\equiv \bar{\gamma}_2 +
      \left[\bar{\gamma}_2^{\times\times} + \frac{2}{5}\bar{\gamma}_{21}^{\times} -
        \frac{1}{2}\bar{\gamma}_{3}^{\times} + \frac{8}{15}\bar{\gamma}_2^{\mathrm{sq}}\right]\sigma^2\,.
        \label{g2ren}
\end{align}
These are the observable quadratic and tidal tensor bias parameters, which match those in Eqs.~(3.16-3.17)
in~\citep{Assassi:2014} up to the signs of the $\gamma_{21}^{\times}$ and $\gamma_3^{\times}$ terms in the
expression for $\gamma_2$ (note that in their notation $b_{\Gamma_3\delta} = -4/7\,\gamma_{21}^{\times}$ and
$b_{{\cal G}_3\delta} = \gamma_{3}^{\times}$). A more direct comparison with~\citep{Assassi:2014} including
renormalization due to nonlinear time evolution is discussed in Appendix~\ref{sec:RENbiasevol}. Again, the
relations in Eqs.~(\ref{b2renFull}-\ref{g2ren}) can be obtained directly from Eq.~(\ref{GammaExp}) without
computing the loops explicitly.

The large-scale behaviour of the nonlocal evolution terms that we found for the one-point propagator in the
previous section also applies to the two-point propagator. More precisely, it can be shown that the four loop
integrals in Eq.~(\ref{eq:MP.Gamma2}) are fully expressable in terms of combinations of the four
higher-derivative operators (and vice versa), and we present the corresponding relations in
App.~\ref{sec:lowk_NLE}. As noted above, this means that the higher-derivative operators become degenerate with
4th-order contributions from the general bias expansion, and their impact is automatically accounted for by the
latter, as long as terms of order $k_1^4$, $k_2^4$, etc. are negligible.

\subsubsection{The three-point propagator}
\label{sec:three-point-propagator}

Finally, for the three-point propagator, we compute three derivatives of ${\cal O}_B^{(3)}$, giving
\begin{align}
  \left<\frac{\partial^3 {\cal
        O}_B^{(3)}(\B{k})}{\partial\,\delta_L(\B{k}_1)\,\partial\,\delta_L(\B{k}_2)\,\partial\,\delta_L(\B{k}_3)
    }\right>' = 6\,{\cal K}_B^{(3)}(\B{k}_1,\B{k}_2,\B{k}_3)\,.
\end{align}
At the order of SPT we are working in, we only need the tree-level expression for $\Gamma_g^{(3)}$ and we simply get:
\begin{align}
  \label{eq:MP.Gamma3}
  \Gamma_g^{(3)}(\B{k}_1,\B{k}_2,\B{k}_3) =\;&b_3 + 2\gamma_2^{\times}\left[K(\B{k}_1,\B{k}_2) +
    \text{cyc.}\right] \nonumber \\
  &+2\gamma_{21}\left[K(\B{k}_1,\B{k}_2)\,K(\B{k}_{12},\B{k}_3) + \text{cyc.}\right] \nonumber \\
  &+6\gamma_3\,L(\B{k}_1,\B{k}_2,\B{k}_3)\,.
\end{align}

\subsection{Time evolution}
\label{sec:time-evolution}

As we already discussed in Sec.~\ref{sec:bias-exp-with-evo} for the case of local galaxy bias, until the time
when galaxies are observed, all of the basis operators will have evolved and thus developed nonlinear
corrections that are of higher SPT orders. Consequently, if we were to compute the galaxy multipoint propagators
at late times, we would have to account for these corrections, which lead to numerous new renormalizations (see
discussion related to Eqs.~\ref{b2renormsb1b2g2}-\ref{MonsterInt}). Especially when pushed to fourth order in
SPT this approach becomes very cumbersome.

For that reason this section presents an alternative --- we start from the multipoint propagator expansion at
initial time using the propagators derived in Sec.~\ref{sec:galaxy-propagators} and then evolve this expansion
instead of resorting to the usual SPT solutions. Because the mapping from the initial time to the final time
conserves the number of tracers, no new divergences going like $\sigma^2$ can arise, and this bypasses the need
to deal with these complexities. Therefore, the only bias renormalizations done in our approach are trivial
ones, done at initial time, and in fact dealt simply by using the multipoint propagators: in this sense, in our
approach one does not have to think about renormalization at all.

Since time evolution from the initial conditions also gives rise to an evolution of the bias parameters, we
first consider them separately from the propagators. This allows us to illustrate precisely how various
assumptions made about the initial bias relation affect the late-time galaxy overdensity, and thus generalizes
previous results in the literature up to fourth order. In order to distinguish initial from evolved quantities,
we will make the following changes to notation:
\begin{equation}
  b,\, \gamma \to b_{{\cal L}},\,\gamma_{{\cal L}} \quad\text{and}\quad \Gamma^{(n)} \to \Gamma^{(n)}_{{\cal L}}\,,
\end{equation}
for the initial bias parameters and propagators, respectively.

\subsubsection{Bias parameters}
\label{sec:bias-evo}

One of the key principles underpinning the bias relation is that it must be Galilean invariant
  if there is no velocity bias. We can exploit this symmetry to determine the evolution of the bias parameters
  in a significantly simpler way by restricting all quantities that appear in intermediate steps of the
  calculation to a Galilean invariant basis of operators, i.e. the basis we presented in
  Sec.~\ref{sec:bare_expansion}. To illustrate this novel technique we start from the evolution equations for
  conserved tracers, which in the absence of velocity bias can be directly integrated to give (see Eq.~48 in
  \citep{Chan:2012})
  \begin{equation}
    \label{eq:MP.integrated_evo}
    1 + \delta_g(\B{x}) = \frac{1 + \delta_g^{\mathrm{IC}}(\B{q})}{1 + \delta^{\mathrm{IC}}(\B{q})}\,\Big[1 +
    \delta(\B{x})\Big]\,,
  \end{equation}
  where Lagrangian fields (with argument $\B{q}$) are evaluated at the initial time when $\delta =
  \delta_L$. Since we are not interested in decaying modes, we can neglect $\delta^{\rm IC}(\B{q})$ in the
  denominator because it is suppressed by one growth factor compared to the Eulerian fields, leading to the
  well-known expression for the evolution of bias~\cite{CatLucMat9807}. Similarly, we can ignore decaying modes
  and write the initial bias relation in terms of the extrapolated (to final time) linear density fluctuations
  and Lagrangian bias parameters (identical to the parameters in Sec.~\ref{sec:galaxy-propagators}),
  \vspace{0.3em}
  \begin{align}
      \delta_g^\text{IC} = \; & b_{1,{\cal L}}\, \delta_L + \frac{b_{2,{\cal L}}}{2}\, \delta_L^2 + \gamma_{2,{\cal L}}\,
      {\cal G}_2 +{b_{3,{\cal L}}\over 6}\, \delta_L^3 + \gamma_{3,{\cal L}}\, {\cal G}_3 \nonumber \\
      &+ \gamma_{2,{\cal L}}^\times \, \delta_L\, {\cal G}_2 + \gamma_{{21},{\cal L}} \, {\cal G}_2(\varphi_2,\varphi_1) 
      + \ldots \,,  \label{deltaGinitial} \\[-0.2em] \nonumber
  \end{align}
  where all RHS fields are linear Gaussian evaluated at the final time. To proceed we would then have to use
  $\B{q} = \B{x}-\B{\Psi}$ to relate initial to final positions and expand all quantities around
  $\B{x}$. However, this can be done trivially by realizing that in order for the LHS of
  Eq.~(\ref{eq:MP.integrated_evo}) to be Galilean invariant, so must be the RHS, which implies that the dipole
  terms that arise from the $\B{q}$ to $\B{x}$ mapping must cancel against those in the nonlinear matter
  density. All we need to do is thus to consider the Galilean invariant contributions to the nonlinear density
  and velocity expressed in terms of our bias basis, which is done in Appendix~\ref{sec:SPTgal}. Indicating this
  by the superscript ``GI'', it follows that \beq \delta_g(\B{x})= \delta_g^\text{IC}(\B{x})
  +\delta^\text{GI}(\B{x}) + \delta_g^\text{IC}(\B{x})\, \delta^\text{GI}(\B{x})
  \label{dgEvol2quoted}
  \eeq
  and by using our bias basis at each order we can then find the coefficients (bias parameters) for the
  Eulerian expansion in terms of those in the initial expansion. For instance, at linear order we have
  $\delta_g^\text{IC}(\B{x}) = b_{1,{\cal L}}\,\delta_L(\B{x})$ and plugging this into Eq.~(\ref{dgEvol2quoted})
  gives
  \begin{equation}
    b_1 = 1 + b_{1,{\cal L}}\,,
  \end{equation}
  which agrees with the result from full evolution \citep{Fry9604} when neglecting transients. For higher-order
  bias evolution we proceed as in \citep{Chan:2012}, i.e. to find bias parameters at a given order we subtract
  the contributions expected from the lower order bias parameters. The full details of this computation are
  given in Appendix~\ref{sec:BiasEvol} and at second order we get the following well known relations
  \begin{equation}
    \label{eq:B2evo}
    \ b_2=b_{2,{\cal L}},\ \ \ \ \ \gamma_2=-{2\over 7}\, b_{1,{\cal L}} + \gamma_{2,{\cal L}}\,,
  \end{equation}
  while we obtain
\begin{widetext}
\beq
\label{B3evo}
b_3=b_{3,{\cal L}}-3 b_2,\ \ \ \ \ \gamma_3 = -{1\over 9} b_{1,{\cal L}} - \gamma_2 + \gamma_{3,{\cal L}},\ \ \
\ \ \gamma_2^\times=-{2\over 7} b_2+\gamma_{2,{\cal L}}^\times,\ \ \ \ \ \gamma_{21}={2\over 21}b_{1,{\cal L}}
+{6\over 7}\gamma_2 + \gamma_{21,{\cal L}}
\eeq
for cubic bias (note that there was a typo in Eq.~(116) of \citep{Chan:2012}, which should have had a minus sign
for its second term), and finally
\beqa b_4&=&b_{4,{\cal
    L}} -12\, b_2 -8\, b_3,\ \ \ \ \ \gamma_2^{\times\times}=-{3\over 7}b_2-{1\over 7}b_3 -\gamma_2^\times +
\gamma_{2,{\cal L}}^{\times\times},\ \ \ \ \ \gamma_3^\times=-{b_2\over 9}-\gamma_2^\times+\gamma_{3,{\cal
    L}}^\times ,\ \ \ \ \
\label{B4resA2} \\
\gamma_2^\text{sq}&=&-{2\over 49}b_2 -{2\over 7} \gamma_2^\times + \gamma_{2,{\cal L}}^\text{sq},\ \ \ \ \ 
\gamma_{21}^\times={2\over 21}b_2+{6\over 7}\gamma_2^\times + \gamma_{21,{\cal L}}^\times,\ \ \ \ \ 
\gamma_{31}=-{4\over 11} b_{1,{\cal L}} - 6 \gamma_2 +\gamma_{31,{\cal L}},\ \ \ \ \  
\label{B4resB2}\\
\gamma_{22}&=&-{6\over 539}b_{1,{\cal L}} - {9\over 49} \gamma_2 + \gamma_{22,{\cal L}} ,\ \ \ \ \  
\gamma_{211}={5\over 77}b_{1,{\cal L}}+{15\over 14}\gamma_2+\gamma_{21}-{9\over 7}\gamma_3 + \gamma_{211,{\cal L}},
\label{B4resC2}
\eeqa
\end{widetext}
for quartic bias. Appendix~\ref{SphAvgSubtle} shows how to recover the familiar spherically symmetric results
from these expressions, which serves as a robust consistency check. Although the relations above have been
derived for the \emph{bare} bias parameters, they are equally valid for the renormalized ones. This is
demonstrated explicitly in Appendix~\ref{sec:RENbiasevol}, which considers what happens if one applies
Eq.~(\ref{dgEvol2quoted}) to the renormalized bias expansion. However, as we mentioned above, in this paper we
advocate a different, simpler route which evolves the initial multipoint propagators, as we discuss next.

\subsubsection{Propagators}
\label{sec:propagators-evo}

As a first step we write a combined evolution equation for matter, its velocity divergence and galaxies, and for
that purpose we define the three-component vector
\begin{align}
  \B{\Psi}(\B{k},\tau) \equiv \left(\,\delta(\B{k},\tau),\, \theta(\B{k},\tau)/f\,{\cal H},\, \delta_g(\B{k},\tau)\,\right)\,.
\end{align}
In terms of $\B{\Psi}$ and by using the logarithm of the linear growth rate as our new time variable, i.e. $\eta
\equiv \ln{D(\tau)}$, the evolution equations can be recast as (see~\citep{Chan:2012})
\begin{align}
  \label{eq:MP.PsiEvolution}
  \hspace{-0.3em}\frac{\partial \Psi_a(\B{k},\eta)}{\partial \eta} + \Omega_{ab}\,\Psi_b(\B{k},\eta) &=
  (2\pi)^3\int\displaylimits_{\B{k}_1,\B{k}_2}\left[\delta_D\right]_2\gamma_{abc}(\B{k}_1,\B{k}_2)\nonumber \\
                                                                                      &\times\,\Psi_b(\B{k}_1,\eta)\,\Psi_c(\B{k}_2,\eta)\,, 
\end{align}
where we follow the convention that repeated indices are summed over, and we assume galaxies move with the dark matter, i.e. no velocity bias. The matrix
\begin{align}
  \Omega_{ab} \equiv \frac{1}{2}
  \left[
    \begin{array}{ccc}
      0 & -2 & 0 \\
      -3 & 1 & 0 \\
      0 & -2 & 0
    \end{array}
  \right]
\end{align}
describes the coupling between densities and velocities, while $\gamma_{abc}$ encodes the nonlinear interactions
between different Fourier modes. Its only non-zero components are given by
\begin{align}
  \gamma_{121} = \gamma_{323} &= \alpha(\B{k}_1,\B{k}_2) \equiv \frac{\B{k}_{12} \cdot \B{k}_1}{k_1^2}\,, \\
  \gamma_{222} &= \beta(\B{k}_1,\B{k}_2) \equiv \frac{k_{12}^2\,(\B{k}_1 \cdot \B{k}_2)}{2\,k_1^2\,k_2^2} \,,
\end{align}
and $\gamma_{112}(\B{k}_1,\B{k}_2) = \gamma_{121}(\B{k}_2,\B{k}_1)$. Given some arbitrary initial conditions
$\Psi_a(\B{k},\eta=0)\equiv \phi_a(\B{k})$, Eq.~(\ref{eq:MP.PsiEvolution}) has the integral
solution~\citep{Scoccimarro:1998b}
\begin{align}
  \label{eq:MP.Psi_intsolution}
  \Psi_a(\B{k},\eta) &= g_{ab}(\eta)\,\phi_b(\B{k}) + (2\pi)^3\int_0^{\eta}\D{\eta'}\,g_{ab}(\eta-\eta') \nonumber \\
  &\times\,\int\displaylimits_{\B{k}_1,\B{k}_2}\left[\delta_D\right]_2\gamma_{bcd}(\B{k}_1,\B{k}_2)\,\Psi_c(\B{k}_1,\eta')\,\Psi_d(\B{k}_2,\eta')\,.
\end{align}
This expression depends on the linear propagator $g_{ab}(\eta)$, which solves the linearized equations of motion
(i.e. setting the righthand side of Eq.~(\ref{eq:MP.PsiEvolution}) to zero) and presents a mixture of growing
and decaying, as well as time independent modes \citep{Scoccimarro:1998b, Chan:2012}:
\begin{align}
  g_{ab}(\eta) = \;
  &\frac{\mathrm{e}^{\eta}}{5} \left[
    \begin{array}{ccc}
      3 & 2 & 0 \\
      3 & 2 & 0 \\
      3 & 2 & 0
    \end{array}
  \right] - \frac{\mathrm{e}^{-3\eta/2}}{5}\left[
    \begin{array}{ccc}
      -2 & 2 & 0 \\
      3 & -3 & 0 \\
      -2 & 2 & 0 \\
    \end{array}
  \right] \nonumber \\
  &+\left[
    \begin{array}{ccc}
      0 & 0 & 0 \\
      0 & 0 & 0 \\
      -1 & 0 & 1
    \end{array}
  \right]\,.
\end{align}
As we did for the galaxy overdensity in Sec.~\ref{sec:hermite-expansion}, we now expand $\Psi(\B{k},\eta)$ in
terms of generalized Wiener-Hermite functionals, 
\begin{align}
  \label{eq:MP.Psi_hermite}
  \Psi_a(\B{k},\eta) = \sum_n \frac{(2\pi)^3}{n!} \int\displaylimits_{\B{k}_1,\ldots,\,\B{k}_n}
  &\left[\delta_D\right]_n\,\Gamma_a^{(n)}(\B{k}_1,\ldots,\B{k}_n\,;\,\eta) \nonumber \\ &\times\,{\cal
    H}_n^*(\B{k}_1,\ldots,\B{k}_n)\,,  
\end{align}
where $\Gamma_3^{(n)} \equiv \Gamma_g^{(n)}$. An equivalent expansion holds at initial time $\eta=0$ and we
denote the corresponding propagators by the symbol $\Gamma_{{\cal L}}$. Since the dark matter and velocity fields
are linear at that time, we have $\Gamma_{1,{\cal L}}^{(1)} = 1 = \Gamma_{2,{\cal L}}^{(1)}$, while all
higher-order propagators must vanish, i.e.
\begin{equation}
  \Gamma_{a,{\cal L}}^{(n)} =
  \delta^K_{n,1}\,\left[\begin{array}{c}
          1 \\
          1 \\
          0
        \end{array}\right] + \Gamma_{g,{\cal L}}^{(n)}\,\left[
        \begin{array}{c}
          0 \\
          0 \\
          1
        \end{array}\right]\,.
\end{equation}
The initial galaxy propagators, on the other hand, are given by the
expressions from Sec.~\ref{sec:galaxy-propagators}. As mentioned in the introduction to Sec.~\ref{sec:bare_expansion}, it is here where one might take advantage of some simplifying assumptions, for instance, removing fourth-order NLE operators from the initial conditions. This sets
\beq
\gamma_{21,{\cal L}}^\times=\gamma_{31,{\cal L}}=\gamma_{22,{\cal L}}=\gamma_{211,{\cal L}}=0
\eeq
in Eqs.~(\ref{B4resB2}-\ref{B4resC2}), which  does not imply that we ignore such NLE operators completely because, as demonstrated by these equations, time evolution  will generate amplitudes $\gamma_{21}^\times,\gamma_{31},\gamma_{22},\gamma_{211}$  fixed through lower order bias parameters (note this approximation is more general than the local Lagrangian approximation, which sets {\em all} nonlocal in $\delta$ Lagrangian bias parameters to zero).

By multiplying both sides of Eq.~(\ref{eq:MP.Psi_intsolution}) with ${\cal H}_n$ and taking the ensemble
average, we can derive a recursion relation for the time evolved multipoint propagators. Following the steps
detailed in App.~\ref{sec:derivation-gamma-recursion}, we arrive at the expression:
\begin{widetext}
  \begin{align}
    \label{eq:MP.Gamma_evolution}
    \Gamma_a^{(n)}(\B{k}_1,\ldots,\B{k}_n\,;\,\eta) =
    g_{ab}(\eta)\,\Gamma_{b,{\cal L}}^{(n)}(\B{k}_1,\ldots,\B{k}_n) + \sum_{r=0}\;\sum_{m = \max{\{r,1}\}}^{n+r-\delta^K_{r,0}} \frac{1}{r!}\,\int_0^\eta
    \D{\eta'}\,g_{ab}(\eta-\eta')\,\Gamma^{(n,m,r)}_b(\B{k}_1,\ldots,\B{k}_n\,;\,\eta')\,,
  \end{align}
  where the quantity $\Gamma^{(n,m,r)}_a$ represents $r$ loop integrals over propagators of orders $m$ and $n-m+2r$:
  \begin{align}
    \label{eq:MP.Gamma_recursion}
    \Gamma^{(n,m,r)}_a(\B{k}_1,\ldots,\B{k}_n\,;\,\eta') \equiv \int\displaylimits_{\B{q}_1,\ldots,\B{q}_r}
    &\Bigg[\gamma_{abc}(\B{k}_{1\ldots 
      m-r}+\B{q}_{1\ldots r},\B{k}_{m-r+1\ldots n}-\B{q}_{1\ldots r})\,
    \Gamma_b^{(m)}(\B{k}_1,\ldots,\B{k}_{m-r},\B{q}_1,\ldots,\B{q}_r\,;\,\eta') \Bigg.\nonumber \\
    &\Bigg.\times\,\Gamma_c^{(n-m+2r)}(\B{k}_{m-r+1},\ldots,\B{k}_{n},-\B{q}_1,\ldots,-\B{q}_r\,;\,\eta')\prod_{i=1}^r    
    P_L(\B{q}_i) + \text{sym}.(\B{k}_i) \Bigg]\,,
  \end{align}
\end{widetext}
with $\text{sym.}(\B{k}_i)$ standing for the symmetrization over all possibilities of building a subset of $m-r$
$\B{k}$-modes from a total group of $n$, i.e. $\binom{n}{m-r}$ terms.

Furthermore, Eq.~(\ref{eq:MP.Gamma_recursion}) illustrates the point we made at the beginning of this section
--- the time evolved multipoint propagators are free of potentially divergent contributions proportional to $\sigma^2$, 
which means we do not have to perform any additional renormalization steps. This is a
consequence of the scale dependence of the mode coupling kernels $\gamma_{abc}$, which contribute to
$\Gamma_g^{(n)}$ only via the symmetrized combination $\alpha^s(\B{k}_1,\B{k}_2) =
1/2[\alpha(\B{k}_1,\B{k}_2)+\alpha(\B{k}_2,\B{k}_1)]$ coming from the continuity equation (conservation of tracers). 

Plugging in the wavevectors from
Eq.~(\ref{eq:MP.Gamma_recursion}) and expanding in inverse powers of $q_{1\cdots r}$, we find that to leading
order
\begin{align}
  \alpha^s(\B{K}-\B{q}_{1\ldots r},\B{k}_{1\ldots n}-\B{K}-\B{q}_{1\ldots r}) \sim \left(\frac{k_{1\ldots
        n}}{q_{1\ldots r}}\right)^2 + \ldots \,,
\end{align}
where $\B{K} = \B{k}_{1\ldots m-r}$. In this paper we are interested in corrections only up to the one-loop
level, so we are led to consider expressions of the form
\begin{align}
  \sim k_{1\ldots n}^2 \int_{\B{q}} &\frac{P_L(q)}{q^2}\,\Gamma^{(m)}(\B{k}_1,\ldots,\B{k}_{m-1},\B{q}) \nonumber
  \\ &\times\,\Gamma^{(n-m+2)}(\B{k}_{m},\ldots,\B{k}_n,-\B{q})\,,
\end{align}
where both propagators are evaluated at tree-level. That means the propagators remain finite when the loop
momentum $q$ becomes large, which in turn implies that the overall integrand scales, at most, as $1/q^3$ in this
limit (using that $P_L(q) \sim 1/q^3$ for $q \rightarrow \infty$). This guarantees that any integral of the
above type is quickly convergent. 

For higher than one-loop corrections this argument no longer holds, as each loop adds an additional power
spectrum to the integrand, while the scaling with the loop momenta remains the same. However, the scale
dependence on the external momenta, i.e. $\sim k_{1\cdots n}^2$, suggests that these terms can be absorbed by
redefinitions of the higher-derivative bias parameters, which display the exact same scale dependence, as shown
in Sec.~\ref{sec:higherD-galaxy-bias}. Using higher-derivative terms to absorb sensitivities to the highly
nonlinear regime was already presented in a slightly different manner in \cite{SchJeoDes1307,Avi1810}. We note
that this behavior of the loop integrals is well known in the context of matter
perturbations~\citep{Bernardeau:2012,PajZal1308,BlaGarKon1309,BerTarNis1401,BalMerZal1507}, where the small
scale sensitivity can be understood in terms of non-zero stress tensor corrections. Indeed, we will show in
Sec.~\ref{sec:nonlocal-bias-vlasov} that these are completely degenerate with the contributions from
higher-derivative bias.

Using Eq.~(\ref{eq:MP.Gamma_evolution}) it is straightforward to compute the evolved multipoint propagators by
starting from lowest order at tree-level and constructing all higher-order solutions recursively. One part of
the of the solution is due to the evolution of the bias parameters and can be simply obtained by replacing the
Lagrangian parameters in the initial multipoint propagators by their Eulerian analogs (according to the
relations given in Sec.~\ref{sec:bias-evo}). Here we focus on the remaining corrections from nonlinear
evolution, which we write as $\Delta\Gamma_g^{(n)}$, and neglecting all but the fastest growing mode and
expressing the result back in terms of the SPT kernels, we find that these corrections for the one-point
propagator at one-loop ($r=1$) are given by
\begin{align}
  \label{eq:MP.Gamma1loop_evo}
  \Delta\Gamma_g^{(1)}\Big|_{\text{1-loop}} &= 3b_1 \int_{\B{q}} F_3(\B{k},\B{q},-\B{q})\,P_L(q) \nonumber \\
                                            &+ 4\gamma_2 \int_{\B{q}} K(\B{k}-\B{q},\B{q})\,G_2(\B{k},-\B{q})\,P_L(q)\,.
\end{align}
The two- and three-point propagators are already affected at tree-level by nonlinear evolution and the
resulting corrections are
\begin{align}
  \Delta\Gamma_g^{(2)}\Big|_{\text{tree}} &= 2b_1\,F_2(\B{k}_1,\B{k}_2)\,. \label{eq:MP.Gamma2tree_evo} \\
  \Delta\Gamma_g^{(3)}\Big|_{\text{tree}} &= 6b_1\,F_3(\B{k}_1,\B{k}_2,\B{k}_3) + 2\Big[
  b_2\,F_2(\B{k}_1,\B{k}_2) \nonumber \\ &+ 2\gamma_2\,K(\B{k}_1+\B{k}_2,\B{k}_3)\,G_2(\B{k}_1,\B{k}_2) +
  \text{cyc.}\Big]\,. \label{eq:MP.Gamma3tree_evo}
\end{align}
For the one-loop bispectrum we also require nonlinear corrections of $\Gamma_g^{(2)}$ at one-loop order, which
are, however, not easily expressed in terms of SPT kernels, so we give the full result in terms of the initial
multipoint propagators in Appendix~\ref{sec:time-evolved-prop}.

\section{Power Spectrum and Bispectrum}
\label{sec:PB}

This section details the final step of this paper: the computation of the galaxy power spectrum and bispectrum
in terms of the multipoint propagators. In addition, we focus on residual sensitivities of the loop corrections
on the highly nonlinear regime, where our perturbative approach breaks down. These affect the statistics not
only on small scales, but also on asymptotically large scales, and we discuss how they can be regularized by
the addition of physically motivated terms.

\subsection{Reconstructing correlators from \\ multipoint propagators}
\label{sec:reconstruction}

\begin{figure}
  \centering
  \includegraphics[width=0.87\columnwidth]{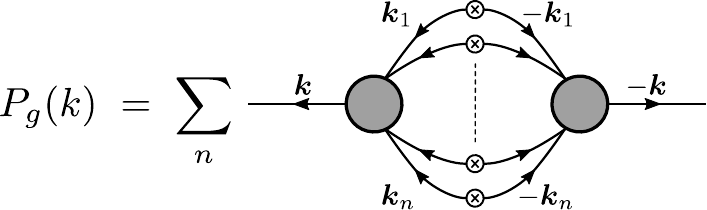}
  \caption{Galaxy power spectrum, reconstructed from multipoint propagators, which are represented by the shaded
    circles with incoming and outgoing momentum $\B{k}$. The sum runs over the number of connected internal
    lines, each of which produces a linear power spectrum depicted by a crossed circle.}
  \label{fig:powerspec}
\end{figure}
\begin{figure}
  \centering
  \includegraphics[width=\columnwidth]{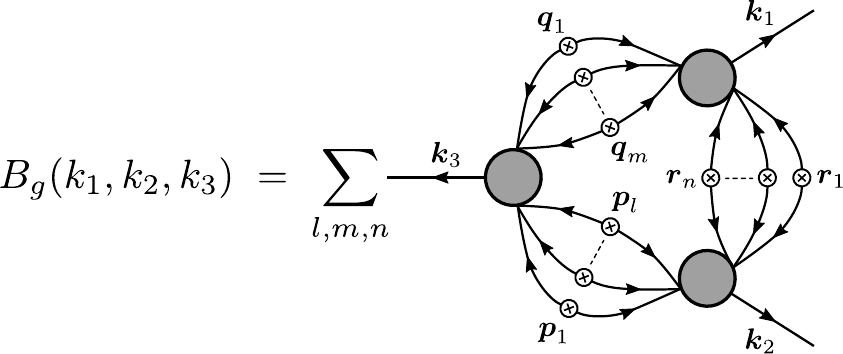}
  \caption{Galaxy bispectrum, expressed through multipoint propagators. The sum runs over the number of
    connected internal lines of each pair of propagators (shaded circles). At most one of the three indices can
    be zero, so that the overall diagram remains a connected graph.}
  \label{fig:bispec}
\end{figure}

The multipoint propagators serve as the basic building blocks for computing $N$-point spectra. This follows
easily from the orthogonality relations of the generalized Wiener-Hermite functionals (see
App.~\ref{sec:hermite_products}), and was already shown for the dark matter density in \citep{Bernardeau:2008},
whose results we can directly apply to the present case of galaxy clustering.

In particular, by evaluating $\langle\Psi_a(\B{k})\Psi_b(\B{k}')\rangle$ and using
Eq.~(\ref{eq:MP.hermite_orthonality}) one finds that the power spectrum is given by a series of two contracted
multipoint propagators of the same order. Diagramatically this can be represented by glueing together two of the
objects shown in Fig.~\ref{fig:propagators}, where each combination of the incoming lines gives rise to a
(linear) power spectrum. This is demonstrated in Fig.~\ref{fig:powerspec} and as the shaded circles include all
vertex loop corrections (i.e. vertex renormalizations) we only need to consider one distinct diagram for the
power spectrum at one-loop level, compared to the usual two in the standard treatment. The galaxy power spectrum
is thus given by \citep{Bernardeau:2008}
\begin{align}
  \label{eq:PB.Pg1loop}
  P_g(k) &= \left[\Gamma_g^{(1)}(k)\right]^2 P_L(k) + \frac{1}{2}\int_{\B{q}}\,
  \left[\Gamma_g^{(2)}(\B{k}-\B{q},\B{q})\right]^2 \nonumber \\ &\times\,P_L(|\B{k}-\B{q}|)\,P_L(q)\,,
\end{align}
where $\Gamma_g^{(1)}$ is evaluated up to one-loop order, while tree-level terms are sufficient for
$\Gamma_g^{(2)}$.

We proceed in a similar manner for the bispectrum, which is obtained from
$\langle\Psi_a(\B{k}_1)\Psi_b(\B{k}_2)\Psi_c(\B{k}_3)\rangle$ and application of
Eq.~(\ref{eq:Grec.3Hproduct}). The complete solution can be found in \citep{Bernardeau:2008}, but its
diagrammatic depiction in Fig.~\ref{fig:bispec} is straightforward --- a combination of three multipoint
propagators with a varying number of connecting lines between each pair (maximally one such pair is allowed to
be disconnected to avoid an overall disconnected graph). Taking care of the appropriate symmetry factors that
arise in these various combinations, it follows that \citep{Bernardeau:2008}:
\begin{widetext}
  \begin{align}
    \label{eq:PB.Bg1loop}
    B_g(k_1,k_2,k_3) &= \Gamma_g^{(2)}(\B{k}_1,\B{k}_2)\, \Gamma_g^{(1)}(k_1)\, \Gamma_g^{(1)}(k_2)\, P_L(k_1)P_L(k_2)+ {\rm cyc.} \nonumber \\
    &+ \Bigg[\int_{\B{q}} \Gamma_g^{(2)}(\B{k}_1-\B{q},\B{q})\, \Gamma_g^{(2)}(\B{k}_2+\B{q},-\B{q}) \,
    \Gamma_g^{(2)}(\B{k}_1-\B{q},\B{k}_2+\B{q})\, P_L(|\B{k}_1-\B{q}|)\, P_L(|\B{k}_2+\B{q}|)\, P_L(q)  \nonumber \\
    &+ \frac{1}{2} \int_{\B{q}} \Gamma_g^{(3)}(\B{k}_3,\B{k}_2-\B{q},\B{q})\, \Gamma_g^{(2)}(\B{k}_2-\B{q},\B{q})\,
    \Gamma_g^{(1)}(k_3)\, P_L(|\B{k}_2-\B{q}|)\,P_L(q)\,P_L(k_3)+ {\rm cyc.}  \Bigg]\,.
  \end{align}
\end{widetext}
Therefore, for the one-loop galaxy bispectrum we require both, $\Gamma_g^{(1)}$ and $\Gamma_g^{(2)}$, up to
one-loop order in the first term of Eq.~(\ref{eq:PB.Bg1loop}), but tree-level expressions for them and
$\Gamma_g^{(3)}$ are enough in the loop integrals, i.e. in the second and third line. Note that there are only
two one-loop diagrams instead of the four in SPT; this is because there are two diagrams in SPT that are
reducible, and thus incorporated into the one-loop $\Gamma_g^{(1)}$ (giving the 321-II  diagram in the notation
of~\cite{Sco97}) and the one-loop $\Gamma_g^{(2)}$ diagram (giving the 411 diagram).

\begin{figure}
  \centering
  \includegraphics[width=\columnwidth]{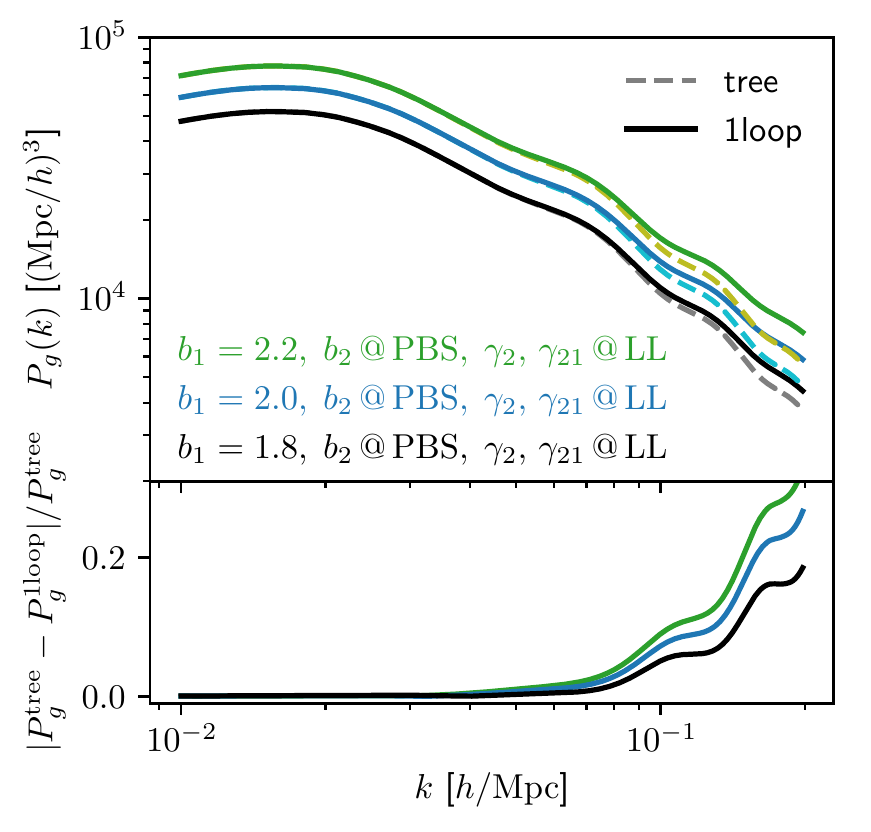}
  \caption{Top panel: comparison of the tree-level and one-loop galaxy power spectrum for three different values
    of the linear bias parameter. Second and third order bias parameters are fixed in terms of $b_1$ by means of
    the peak-background split (PBS) and local Lagrangian (LL) predictions. Bottom panel: relative difference
    between the tree-level and one-loop models for the same three cases.}
  \label{fig:Ptreevsloop}
\end{figure}

From comparing Eqs.~(\ref{eq:PB.Pg1loop}) and (\ref{eq:PB.Bg1loop}) we note that the two-point propagator
contributes to the leading order bispectrum, while showing up as a loop correction for the power spectrum. This
structure extends to consecutively higher orders, for instance, the three-point propagator which appears as a
one-loop expression in the bispectrum, will enter at tree-level for the trispectrum. That suggests that
constraints on bias parameters required to fit the small-scale behaviour of a given correlator (and thus
cosmological parameters, too) will already highly benefit from the large-scale information of the next order
correlator.

\begin{figure*}
  \centering
  \includegraphics[width=\textwidth]{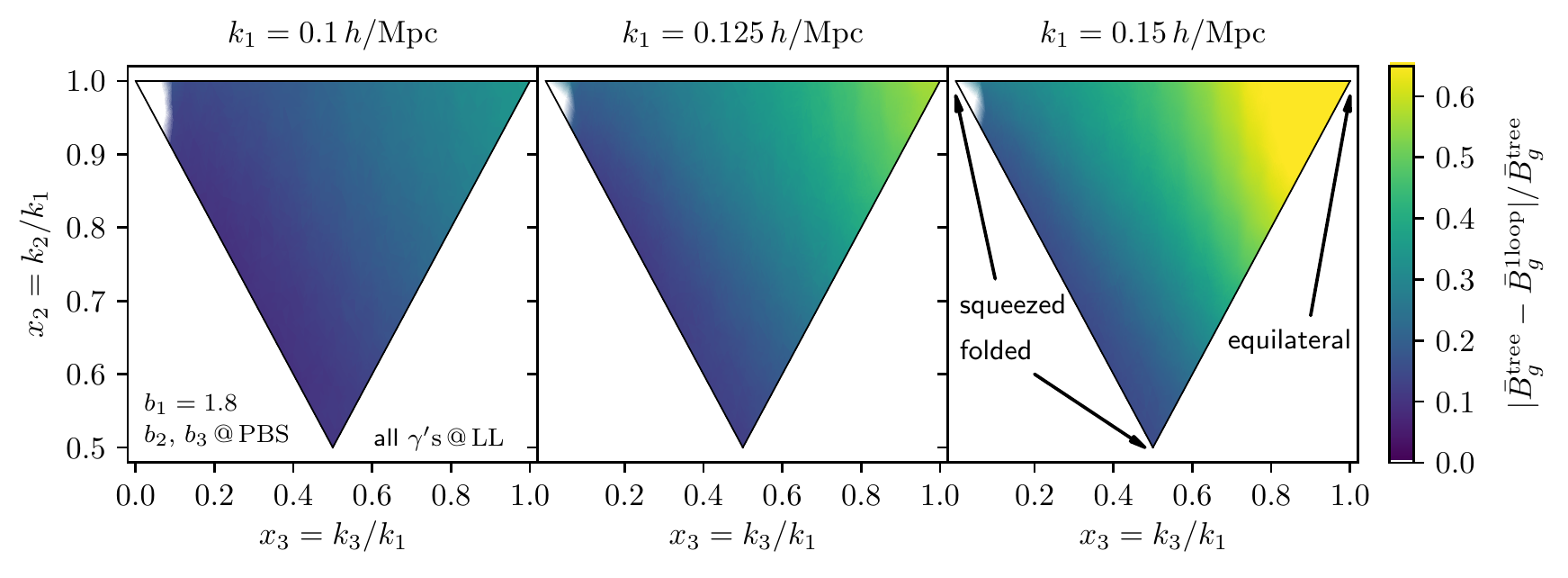}
  \caption{Relative difference between the tree-level and one-loop galaxy bispectrum as a function of $x_2 =
    k_2/k_1$ and $x_3 = k_3/k_1$, integrated over a thin shell centered on three different values of $k_1$. The
    linear bias parameter is given by $b_1 = 1.8$, while $b_2$ and $b_3$ are determined from the peak-background
    split (PBS) relations, and all other parameters are fixed using the local Lagrangian (LL)
    approximation. Note that Eq.~(\ref{eq:PB.Bx2x3}) was evaluated using a grid of bispectra with fixed binning
    for $k_1$, $k_2$ and $k_3$, leading to an absence of very squeezed configurations.}
  \label{fig:Btreevsloop}
\end{figure*}

When splitting each expression in Eq.~(\ref{eq:PB.Bg1loop}) into its individual contributions, one finds that in
total there are 40 additional terms in the one-loop galaxy bispectrum compared to tree-level. Each of these
terms is multiplied by a combination of various bias parameters, making a comparison of the individual terms not
particularly meaningful. However, given a set of bias parameters, it is interesting to consider whether the
bispectrum loop contributions become relevant on similar scales as for the power spectrum.  To that end, in
Fig.~\ref{fig:Ptreevsloop} we show a comparison of the one-loop galaxy power spectrum evaluated from
Eq.~(\ref{eq:PB.Pg1loop}) and its tree-level prediction $P_g^{\mathrm{tree}} = b_1^2\,P_L$, where different
colors correspond to a different choice of the linear bias parameter. In order to adopt representative values
for the higher-order bias parameters, we fix them in terms of $b_1$ by making use of the local Lagrangian
approximation (obtained by setting all parameters with subscript '${\cal L}$' in
Eqs.~(\ref{eq:B2evo}-\ref{B4resC2}) to zero), as well as the peak-background split relations for $b_2$ and
$b_3$.\footnote{Note that the power spectrum and bispectrum models in Fig.~\ref{fig:Ptreevsloop} and
  \ref{fig:Btreevsloop} include the stochastic corrections in the low-$k$ limit to be discussed in
  Sec.~\ref{sec:shot-noise}, but with all noise parameters set to zero.} These were calibrated using separate
universe simulations in \cite{LazWagBal1602}, yielding
\begin{align}
  b_2(b_1) &= 0.412 - 2.143 b_1 + 0.929 b_1^2 + 0.008 b_1^3\,, \\
  b_3(b_1) &= -1.028 + 7.646 b_1 - 6.227 b_1^2 + 0.912 b_1^3\,.
\end{align}
As is demonstrated by the lower panel of Fig.~\ref{fig:Ptreevsloop}, the one-loop corrections to the galaxy
power spectrum become relevant on scales $k \gtrsim 0.1\,h/\text{Mpc}$. A similar trend can be observed for the
galaxy bispectrum, whose relative difference between the tree-level and one-loop models for $b_1 = 1.8$ is
plotted in Fig.~\ref{fig:Btreevsloop}. To show its configuration dependence we evaluate the bispectrum as a
function of $x_2 = k_2/k_1$ and $x_3 = k_3/k_1 $, averaged over a thin shell centered at a given value of $k_1$,
i.e.
\begin{equation}
  \label{eq:PB.Bx2x3}
  \bar{B}_g(k_1,x_2,x_3) = \frac{1}{\Delta k}\int\displaylimits_{k_1 -\Delta k/2}^{k_1 + \Delta k/2} \D{q}\,B_g(q,x_2\,q,x_3\,q)\,,
\end{equation}
where we have chosen $\Delta k = 0.02\,h/\text{Mpc}$ and three different values for $k_1$ (see
Fig.~\ref{fig:Btreevsloop}). We notice that for all triangle configurations the relative difference surpasses
$10\,\%$ for $k_1 \gtrsim 0.1\,h/\text{Mpc}$, with the strongest deviations to be found for equilateral
triangles. Of course, the precise numerical values obtained in this demonstration depend sensitively on the set
of bias parameters, but we do not expect any big impact on the overall conclusion: bias loop corrections affect
the power spectrum and bispectrum starting from comparable scales.

\subsection{Stress-tensor corrections and their degeneracy with higher-derivative bias}
\label{sec:nonlocal-bias-vlasov}

While the introduction of multipoint propagators has automatically removed potentially divergent contributions proportional
to $\sigma^2$ (i.e. $(k/q)^0$ with $k$ the external momentum and $q$ the loop momentum), the loop integrals remain sensitive to the nonlinear regime through terms
scaling as powers of $(k/q)^2$ (and $(k/q)^4$, etc). This is true for loop corrections of individual multipoint propagators, as
discussed in Sec.~\ref{sec:galaxy-propagators} and at the end of Sec.~\ref{sec:time-evolution}, but also for the
loop integrals over tree-level propagators appearing in Eqs.~(\ref{eq:PB.Pg1loop}) and
(\ref{eq:PB.Bg1loop}). Sensitivity of loop integrals to the nonlinear regime is obviously a problem since
perturbation theory does not hold at small scales. 

At such small scales the dark matter field can no longer be treated as a pressureless perfect fluid, as
is assumed in SPT. Initially, or on large scales, this is a good approximation as dark matter particles tend to
move within single coherent flows, which implies a vanishing stress tensor $\sigma_{ij}$. At later times,
multi-streaming induces non-zero stresses, which can  have an impact on quasi-linear scales, with the same $(k/q)^2$ scaling 
as one-loop SPT  and one-loop bias terms~\cite{PueSco0908}. 

These stress tensor corrections have been computed in the framework of the effective field theory (EFT)
\citep{Baumann:2012, Carrasco:2012,BalMerMir1406,AngForSch1406} and directly from the Vlasov equation
\citep{StressTensor} following~\cite{PueSco0908,Pue0810}. This leads to additional terms in the power spectrum
and bispectrum that at lowest order scale with powers of $k^2$, identically to one-loop corrections sensitive to
the nonlinear regime. In the particularly economic parametrization of \citep{StressTensor} (which is otherwise
equivalent to the EFT calculations mentioned above) they read,
\begin{align}
  P_{\sigma}(k) &= -2 \beta_P\,k^2\,P_L(k) \label{eq:PB.Pvlasov} \\
  B_{\sigma,123} &= - \Big[\Big(\beta_{B,a}\,\left(k_1^2 + k_2^2\right) +
  \beta_{B,b}\,k_3^2\Big)\,F_2(\B{k}_1,\B{k}_2) \nonumber \\ &+ \Big(\beta_{B,c}\,\left(k_1^2 + k_2^2\right) +
  \beta_{B,d}\,k_3^2\Big)\,K(\B{k}_1,\B{k}_2)\Big] \nonumber \\ &\times\,P_L(k_1)\,P_L(k_2) +
  \text{cyc.}\,, \label{eq:PB.Bvlasov}
\end{align}
where $\beta_P$ and $\beta_{B,i}$ are numbers that result from integrating over stress tensor components
weighted by functions of time (growth factors). Since the time-dependence of the stress tensor is not calculable
in a generic $\Lambda$CDM universe, these numbers are free parameters. Crucially though, they can absorb and
regularize all contributions from SPT one-loop integrals that are sensitive to a range of scales where our
perturbative approach breaks down and have the same $k^2$ scaling.

\begin{figure*}
  \centering
  \includegraphics[width=0.95\textwidth]{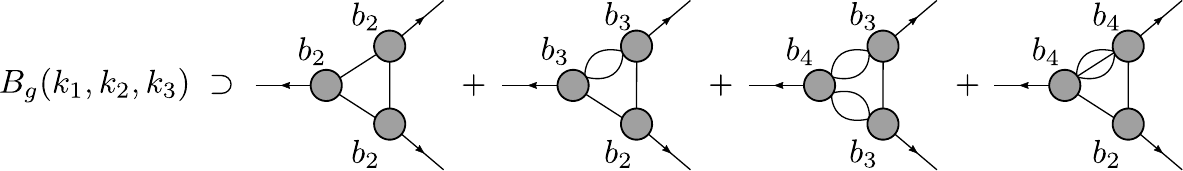}
  \caption[Subset of diagrams contributing the a non-zero large-scale limit of the galaxy bispectrum.]{Subset of
    diagrams that contribute to a non-zero large-scale limit ($k_1,k_2,k_3 \rightarrow 0$) of the galaxy
    bispectrum. The first diagram appears at one-loop level, the second at two-loop and the last two at
    three-loop. The bias constants $b_i$ indicate the value of the multipoint propagators (shaded circles),
    i.e. constants that do not vanish in the large-scale limit.}
  \label{fig:shotnoise}
\end{figure*}

As we have seen in Sec.~\ref{sec:higherD-galaxy-bias}, the higher-derivative bias terms display the same
momentum scaling and thus give rise to very similar terms. Ignoring all precoefficients, evaluation of
Eqs.~(\ref{eq:PB.Pg1loop}) and (\ref{eq:PB.Bg1loop}) shows that the higher-derivative contribution to the power
spectrum is exactly degenerate with Eq.~(\ref{eq:PB.Pvlasov}), while for the bispectrum we get five different terms:
\begin{align}
  \left.\begin{array}{lc}
      \rm{i}) & k_3^2 \\[0.2em]
      \rm{ii}) & \B{k}_1 \cdot \B{k}_2 \\[0.2em]
      \rm{iii}) & k_3^2\,F_2(\B{k}_1,\B{k}_2) \\[0.2em]
      \rm{iv}) & k_3^2\, K(\B{k}_1,\B{k}_2) \\[0.2em]
      \rm{v}) & (\B{k}_1 \cdot \B{k}_2)\, K(\B{k}_1,\B{k}_2)
  \end{array}\right\}\,\times\,P_L(k_1)\,P_L(k_2) + \text{cyc.}\,.
\end{align}
The contributions iii) and iv) are clearly degenerate with terms in Eq.~(\ref{eq:PB.Bvlasov}) and using that
$\B{k}_1 \cdot \B{k}_2 = 1/2\left(k_3^2-k_1^2-k_2^2\right)$ we see that v) can be written as a combination where
$\beta_{B,d} = 1/2 = -\beta_{B,c}$. Furthermore, one can show that
\begin{align}
  k_3^2 &= -\left[k_1^2 + k_2^2 - k_3^2\right]\,F_2(\B{k}_1,\B{k}_2) \nonumber \\ &- \left[\frac{5}{7}\left(k_1^2+k_2^2\right)
    + \frac{2}{7}\,k_3^2\right]\,K(\B{k}_1,\B{k}_2)\,,
\end{align}
such that a combination of all four terms in Eq.~(\ref{eq:PB.Bvlasov}) with $\beta_{B,a} = 1$, $\beta_{B,b} =
-1$, $\beta_{B,c} = 5/7$ and $\beta_{B,d} = 2/7$ can also accommodate for i). Only ii) cannot be expressed
through the stress tensor terms and must consequently enter the galaxy bispectrum as an independent contribution. In
total, this demonstrates that the stress tensor corrections to the bispectrum are also completely degenerate with those
from higher-derivative bias. For that reason we consider them collectively, using the following basis:
\begin{equation}
  \begin{split}
    \label{eq:cha7_PB.Bsigmanabla}
    B_{\sigma+\nabla,123} = - \Big\{&\Big[\beta_{B,a}\,\left(k_1^2 + k_2^2\right) +
    \beta_{B,b}\,k_3^2\Big]\,F_2(\B{k}_1,\B{k}_2) \\ + &\Big[\beta_{B,c}\,\left(k_1^2 + k_2^2\right) +
    \beta_{B,d}\,k_3^2\Big]\,K(\B{k}_1,\B{k}_2) \\ + \hspace{0.005em} &\hspace{0.4em}\beta_{B,e}\,\B{k}_1 \cdot
    \B{k}_2\Big\}\,P_L(k_1)\,P_L(k_2) + \text{cyc.}\,,
  \end{split}
\end{equation}
which reduces the number of free parameters to five, in addition to another one for the power spectrum. Note
that if we had considered only higher-derivative bias, then we would have only five parameters in total for
the power spectrum and bispectrum, since in that case $2\beta_P=\beta_{B,a}+\beta_{B,b}$. However that is not the
case for stress tensor contributions since $\beta_P$ and $\beta_B$'s result from integrating the stress tensor
weighted with different powers of the growth factor~\cite{StressTensor}. In addition, it's worth noting that some of the contributions of velocity bias are also degenerate with higher-derivative bias (see~\cite{DesCroSco1011}). Finally, note that the contributions in
Eq.~(\ref{eq:cha7_PB.Bsigmanabla}) are degenerate with one-loop galaxy bias contributions in the low-$k$ limit,
see Appendix~\ref{sec:lowk_NLE} for an explicit discussion of this.

\subsection{Stochastic contributions}
\label{sec:shot-noise}

The effects discussed in the last section become relevant towards smaller scales, but the bias loop corrections
can also have an impact on the large-scale power spectrum and bispectrum. As already pointed out in
\citep{Heavens:1998,McDonald:2006}, that is because a subset of the bias loop integrals do not vanish in the
large-scale limit, and thus come to dominate for values of $k^{-1}$ above a certain scale. Taking the limit $k
\rightarrow 0$ of Eq.~(\ref{eq:PB.Pg1loop}) shows that the galaxy power spectrum approaches a constant:
\begin{equation}
  \begin{split}
    \label{eq:PB.Pshot}
    \lim_{k \to 0}\,P_g(k) &= \frac{1}{2} \int_{\B{q}} \left[\Gamma_g^{(2)}(\B{-q},\B{q})\right]^2 P_L(q)^2 \\
    &= \frac{b_2^2}{2} \int_{\B{q}} P_L(q)^2\,,
  \end{split}
\end{equation}
where we have assumed that the linear power spectrum falls off to zero on large scales and used that
$K(-\B{q},\B{q}) = 0$. Crucially, corrections from successively higher orders of SPT add in comparable measures
to this large-scale limit, meaning that this low-$k$ limit is not controlled by any particular order in
perturbation theory and thus one must introduce a new parameter that describes its size.

The same extends to the loop corrections of the galaxy bispectrum, and in the limit of all three triangle sides
approaching zero, Eq.~(\ref{eq:PB.Bg1loop}) equally becomes constant:
\begin{equation}
  \begin{split}
    \label{eq:PB.Bshot1}
    \lim_{k_1, k_2 \to 0}\,B_g(k_1,k_2,k_3) &=
    \int_{\B{q}} \left[\Gamma_g^{(2)}(-\B{q},\B{q})\right]^3 P_L(q)^3 \\ &= b_2^3\,\int_{\B{q}} P_L(q)^3\,.
  \end{split}
\end{equation}
As for the power spectrum the value of this constant receives corrections from higher-order SPT terms, which are
shown schematically up to the three-loop level by the diagrams in Fig.~\ref{fig:shotnoise}. Although diagrams
which contain a one-point propagator at one of their external legs do not contribute to this limit, they are
affected by large-scale loop corrections in a similar manner. The corresponding terms can be identified as
follows:
\begin{equation}
  \begin{split}
    &\lim_{k_1, k_2 \to 0} \frac{B_g(k_1,k_2,k_3)\big|_{\propto P_L(k_1)}}{P_L(k_1)} \\ &\hspace{2.5em}=
    b_1\,b_2\,\left[\frac{115}{42}b_2 + b_3 - \frac{8}{3}\gamma_2^{\times}\right]\,\int_{\B{q}}
    P_L(q)^2\,,
  \end{split}
\end{equation}
where $B_g(k_1,k_2,k_3)\big|_{\propto P_L(k_1)}$ denotes all terms in the galaxy bispectrum that are
proportional to $P_L(k_1)$. After cyclic permutations their total contribution to the one-loop bispectrum is
thus given by
\begin{equation}
  \label{eq:PB.Bshot2}
  \begin{split}
    b_1\,b_2\,&\left(\frac{115}{42}b_2 + b_3 - \frac{8}{3}\gamma_2^{\times}\right) \\ &\times\,\Big[P_L(k_1) + P_L(k_2) +
    P_L(k_3)\Big]\,\int_{\B{q}} P_L(q)^2\,.
  \end{split}
\end{equation}
These effects can be interpreted as the impact of small-scale perturbations on the formation of galaxies that
cannot be captured by any perturbative bias model, and were originally introduced as a ``stochastic'' bias in
the literature \citep{Scherrer:1998,Dekel:1999}. Its defining property is that it must be mostly uncorrelated on
large scales (assuming Gaussian initial conditions as we do), meaning it will manifest in the same way as shot
noise. Inspecting Eq.~(\ref{eq:PB.Pshot}) and Eqs.~(\ref{eq:PB.Bshot1}, \ref{eq:PB.Bshot2}), we notice that this
is indeed the case for the large-scale limit of the one-loop power spectrum and bispectrum. We could have chosen
to incorporate these effects by including stochastic fields in the expansion for $\delta_g$ (see for instance
\citep{Schmidt:2016,Ginzburg:2017,Desjacques:2018}), instead we now introduce \emph{a posteriori} three
effective shot noise parameters \citep{McDonald:2006,McDonald:2009} for the power spectrum and bispectrum, such
that
\begin{align}
  \label{Pshot}
    P_{\text{shot}}(k) &= N_0\,, \\
    B_{\text{shot}}(k_1,k_2,k_3) &= \epsilon_0 + \eta_0\,\big[P_L(k_1) + P_L(k_2) + P_L(k_3)\big]\,.
    \label{Bshot}
\end{align}
Similar to the stress tensor parameters, these parameters are able to absorb any residual large-scale
contributions stemming from the bias loops.  We stress that the values of $N_0$, $\epsilon_0$ and $\eta_0$ are
typically not given by their Poissonian shot noise predictions, i.e. $N_0 = 1/\overline{n}$,
$\epsilon_0 = 1/\overline{n}^2$ and $\eta_0 = 1/\overline{n}$~\cite{Pee80} for an average number density of
galaxies $\overline{n}$, but must be determined from the data itself.

Finally, in order to take into account a slight correlation of the stochastic bias on large scales, we can think
of Taylor expanding its contributions in powers of $k^2$, where in general the expansion coefficients must also
be considered as free parameters. This is motivated by requiring that not only the leading order terms in the
large-scale limits above can be absorbed, but also their next-to-leading order (NLO) contributions. These terms
exclusively scale as powers of $k^2$ (or $k_1^2$, $k_2^2$ and $k_3^2$ in case of the bispectrum), and it is
straightforward to show that the allowed terms can be summarized as follows:
\begin{align}
    \label{PshotNLO}
    P_{\text{shot,NLO}}(k) &= N_2\,k^2 \\
    B_{\text{shot,NLO}}(k_1,k_2,k_3) &= \epsilon_2\big(k_1^2 + k_2^2 + k_3^2\big) \nonumber \\ &\hspace{-3em}+ \Big[\eta_{2,1}\,k_1^2 +
    \eta_{2,2}\big(k_2^2 + k_3^2\big)\Big]\,P(k_1) + \text{cyc.}\,,
    \label{BshotNLO}
\end{align}
which leads to an additional four parameters.

\section{Discussion and conclusions}
\label{sec:conclusions}

In this paper, we presented the complete set of one-loop (next-to-leading order) galaxy bias corrections for the
real-space galaxy bispectrum. These corrections serve to increase the range of scales where the bispectrum can
be robustly used to extract information from the clustering of galaxies (or any other tracer), and brings the
bispectrum to the same state-of-the-art as the galaxy power spectrum, meaning joint analyses can be performed
consistently. We carry this out in detail and compare against numerical simulations of biased tracers in a
follow-up paper~\cite{BiasLoops2}

Our perturbative bias model systematically combines a variety of effects that become relevant on scales of the
weakly non-linear regime. These effects include contributions from: 1) the general bias expansion, generated by
gravitational evolution of the matter field; 2) higher-derivative terms, due to the spatial nonlocality of
galaxy formation, and a non-zero stress tensor; and 3) stochasticity, resulting from the impact of non-perturbative physics on the large-scale
galaxy distribution.

\begin{table}
  \centering
  \setlength{\tabcolsep}{5pt}
  \begin{ruledtabular}
    \begin{tabular}{ccccc}
      & \multirow{2}{*}{\begin{minipage}{2cm}General bias expansion\end{minipage}} &
      \multicolumn{2}{c}{Stochastic} & \multirow{2}{*}{\begin{minipage}{3cm}Stress-tensor \& Higher-derivative\end{minipage}} \Tstrut\Bstrut \\ 
      &  & ${\cal O}(k^0)$ & ${\cal O}(k^2)$ &  \Bstrut \\ \hline
      $P_g$ & \multirow{2}{*}{11} & 1 & 1 & 1\Tstrut \Bstrut \\
      $B_g$ & & 2 & 3 & 5
    \end{tabular}
  \end{ruledtabular}
  \caption{Number of bias parameters required for the galaxy power spectrum and bispectrum up to one-loop order
    in perturbation theory, organized according to the various contributions to the bias expansion. Of the 11
    general bias parameters for $P_g+B_g$ description, 4 are enough to describe $P_g$ alone
    ($b_1,b_2,\gamma_2,\gamma_{21}$), while adding $B_g$ requires 7 more parameters
    ($\gamma_2^\times,\gamma_{21}^\times,b_3,\gamma_3,\gamma_{22},\gamma_{31},\gamma_{211}$).\vspace*{-1em}}
  \label{tab:num_parameters}
\end{table}

In all generality, we found that a joint one-loop power spectrum and bispectrum analysis in real-space requires
24 free parameters. In agreement with \citep{Assassi:2014}, we showed that 11 parameters are due to the general
bias expansion, which are shared among the power spectrum and bispectrum. They include the usual linear bias
parameter at first order ($b_1$), the local and nonlocal quadratic parameters at second ($b_2,\gamma_2$) etc. To account for
higher-derivative effects, which were not considered by \citep{Assassi:2014}, we demonstrated that we need to
introduce an additional five parameters for the bispectrum and one for the power spectrum, five less than
given in~\citep{Desjacques:2018}. We also found that the stress-tensor corrections are
entirely degenerate with those from higher derivatives and therefore do not consider them separately. Finally,
at leading order stochasticity contributes with two terms (one term) to the bispectrum (power spectrum), whereas
at next-to-leading order we identified a further three (one). An overview of these numbers is given in
Table~\ref{tab:num_parameters}.

Obviously, it would be bad news if a joint analysis of power spectrum and bispectrum demands that each of the 24
parameters enters the model with a freely adjustable amplitude. In that case the gain from including smaller
scales in the analysis could be easily cancelled out by the loss in constraining power from having to
marginalize over so many nuisance parameters. From that point of view the work presented in this paper is best
regarded as the theory on which we can analyze various ways of reducing the parameter space, as will be
discussed in detail in a follow-up paper~\cite{BiasLoops2}. For example, it is known in the case of the one-loop
galaxy power spectrum that higher-derivatives are degenerate with bias
loops~\cite{McDonald:2009,SaiBalVla1405,BiaDesKeh1405,SanScoCro1701}, and as now shown here (see
Appendix~\ref{sec:lowk_NLE}) the structure of one-loop bias at low-$k$ is also degenerate with higher-derivative
bias for the bispectrum. However, it remains to be seen how the higher-derivative terms compare to those from
the general bias expansion. While \citep{Fujita:2016,Nadler:2018} suggest that the former dominate for
tracers residing in very massive halos, this is not necessarily correct for certain types of tracers that are
targeted by upcoming galaxy surveys such as DESI. This includes, for instance, emission line galaxies, which are
usually found in less massive halos.

Apart from testing the significance of the
higher-derivative contributions, another clear possibility of reducing the parameter space is to check which bias operators
determine the galaxy perturbations at the initial time. Our bias basis is designed so that we separate operators which are local in second derivatives of the linear potential (what we call local evolution operators, or LE), and those which are nonlocal (induced by nonlocal evolution, hence NLE operators) that involve the nonlinear Lagrangian potentials. Other bases in the literature mix these properties, see Appendix~\ref{sec:basis-relation}.  

The most drastic reduction in parameter space is achieved by putting initially all NLE operators to zero and all LE operators nonlocal in $\delta$ to zero, this is the local Lagrangian approximation: it reduces the 11 parameters in Table~\ref{tab:num_parameters} to just 3 free parameters ($b_1,b_2,b_3$) and the other eight parameters get determined from them by the time evolution arguments in Section~\ref{sec:bias-evo} (putting all of the eight Lagrangian values to zero). A hierarchy of more accurate approximations (but with more free parameters) can be made where one takes e.g. only the NLE operators to be zero in the far past (as in peak or excursion set models of bias), or just the 4th-order ones  that only enter through the loop correction of a single diagram in the bispectrum. The validity of such assumptions must be
tested numerically and compared to the full model, which we will present in~\cite{BiasLoops2}. For practical applications to galaxy surveys, given finite error bars, one should be able to reduce the parameter space sequentially going from most general to least general checking the final results are not impacted by the assumptions.

On the more technical side, the central concept explored in this work is the use of multipoint propagators, which correspond to the sum over all reducible diagrams,  as a way to writing the perturbative expansion in terms of explicit observables (that corresponds for Gaussian fluctuations to the commonly measured cross-correlation bias between galaxies and matter).   As a result of this reformulation we were able to circumvent the renormalization procedure --- a tedious
  redefinition of the standard bias parameters required to absorb diverging contributions in the computation of
  correlation functions at one-loop and beyond. To achieve this we showed that it is best to proceed in two steps: first, compute the multipoint propagators  at initial time where   nonlinearities in the matter field can be ignored, then evolve these initial
  propagators to the time of observation. Evolving the propagators conserving the number of tracers (the so-called ``coevolution" first given by~\cite{Fry9604}) cannot generate any diverging terms proportional to $\sigma^2 = \langle  \delta(\B{x})^2\rangle$ and therefore the late-time propagators are already renormalized (and any $k^2$ or higher-order renormalizations are simple to handle by adding the corresponding terms). Once the evolved multipoint propagators
  have been computed, they serve as simple building blocks for the general $N$-point correlation function.
  
We also discussed how our choice of basis, written in terms of Galileons and with a clear split between local and nonlocal functions of second derivatives of the linear potential, simplifies the calculation of the multipoint propagators: they are given by the tree-level
  expressions plus loop integrals over the NLE operators contributions only. Other bases that mix our operators (see Appendix~\ref{sec:basis-relation}) complicate the calculation, as one has to separate terms that get absorbed by the renormalization procedure.

  A significant part of the evolution of the propagators is absorbed into the evolution of the bias parameters
  in them. Our calculation of this time evolution was carried out using a new approach that takes full advantage
  of Galilean invariance (see Appendix~\ref{sec:GIbiasevol}) and we present results for the fourth-order
  parameters for the first time (see Section~\ref{sec:bias-evo}), generalizing the third-order results
  originally given in~\cite{Chan:2012}.  Up to quadratic bias we also show explicitly that renormalization
  by fourth-order operators does not change the time evolution of bias parameters obtained at tree level (see
  Appendix~\ref{sec:RENbiasevol}), as expected.

  Finally, to summarize the steps necessary for applying the results of this paper in practice, one would proceed as
  follows: first, one computes the Lagrangian multipoint propagators given in Eqs.~(\ref{eq:MP.Gamma1}),
  (\ref{eq:MP.Gamma2}) and (\ref{eq:MP.Gamma3}). These serve as initial conditions for the evolution equations
  yielding the Eulerian multipoint propagators, i.e. Eqs.~(\ref{eq:Grec.G2tree}-\ref{eq:Grec.G2loop}), which in
  turn determine the power spectrum and bispectrum at the time of observation via Eqs.~(\ref{eq:PB.Pg1loop}) and
  (\ref{eq:PB.Bg1loop}). The final expressions are obtained after adding the combination of stress-tensor and
  higher-derivative corrections from Eqs.~(\ref{eq:PB.Pvlasov}) and (\ref{eq:cha7_PB.Bsigmanabla}), as well as
  the stochastic contributions up to next-to-leading order (Eqs.~\ref{Pshot}-\ref{BshotNLO}).

  This will be demonstrated explicitly in our follow-up paper~\cite{BiasLoops2}. In particular, we carry out a
  detailed likelihood analysis of the power spectrum and bispectrum in numerical simulations of biased tracers
  to ascertain the performance of the results derived here, the importance of the different terms in the bias
  expansion, the accuracy of theoretical assumptions that reduce the number of free parameters, and the
  improvement of our results compared to current models in the literature.

\vspace*{-1em}
\acknowledgements

We thank M.~Crocce, A.~Sanchez, R.~Sheth for useful discussions. AE acknowledges support from the UK Science and
Technology Facilities Council (STFC) via Research Training Grant (grant number ST/M503836/1), as well as from
the European Research Council (grant number ERC-StG-716532-PUNCA). AE also thanks the Center for Cosmology and
Particle Physics of New York University for hospitality while part of this work was undertaken, and the STFC for
supporting this research stay. RES acknowledges support from the STFC (grant number ST/P000525/1).

\begin{widetext}

\appendix

\section{Further notes on Galilean basis for galaxy bias}
\label{sec:bias-further-notes}

\subsection{Basis operators in Fourier space}
\label{sec:basis-oper-fourier}

  Here we briefly summarize Fourier space expressions for our basis operators given in
  Table\ref{tab:basis}, which are being used in the computation of the multipoint propagators. In
  genereal, we write any $n$-th order operator ${\cal O}^{(n)}$ as an integral over $n$ linear matter
  perturbations:
  \begin{align}
    {\cal O}^{(n)}_B(\B{k}) = (2\pi)^3 \int\displaylimits_{\B{k}_1,\ldots,\B{k}_n}
    \left[\delta_D\right]_n\,{\cal K}^{(n)}_B(\B{k}_1,\ldots,\B{k}_n)\,\prod_{i=1}^n \delta_L(\B{k}_i)\,, 
  \end{align}
  where $B$ stands for any of the basis operators at that order and we use the notation $\left[\delta_D\right]_n
  \equiv \delta_D(\B{k}-\B{k}_{1\ldots n})$. For $n=2$ we have from
  Eq.~(\ref{eq:tracers.G2k}):
  \begin{align}
    {\cal K}^{(2)}_{\delta^2}(\B{k}_1,\B{k}_2) &= 1\,, \label{eq:appbasis.K2d2} \\
    {\cal K}^{(2)}_{{\cal G}_2}(\B{k}_1,\B{k}_2) &= K(\B{k}_1,\B{k}_2)\,, \label{eq:appbasis.K2G2}
  \end{align}
  and the kernel $K(\B{k}_1,\B{k}_2)$ was already defined in Eq.~(\ref{eq:tracers.K}). At third order the only
  nontrivial operator is ${\cal G}_2(\varphi_2,\varphi_1)$, which becomes upon Fourier transformation:
  \begin{align}
    \label{eq:appbasis.G2p2p1}
    {\cal G}_2(\varphi_2,\varphi_1\,|\,\B{k}) &= (2\pi)^3 \int\displaylimits_{\B{k}_1,\B{k}_2}
    \left[\delta_D\right]_2\,K(\B{k}_1,\B{k}_2)\,\delta_L(\B{k}_1)\,{\cal G}_2(\Phi_L\,|\,\B{k}_2) \nonumber \\ 
    &= (2\pi)^3 \int\displaylimits_{\B{k}_1,\B{k}_2,\B{k}_3}
   \left[\delta_D\right]_3\,K(\B{k}_1,\B{k}_{23})\,K(\B{k}_2,\B{k}_3)\,\delta_L(\B{k}_1)\,\delta_L(\B{k}_2)\,\delta_L(\B{k}_3)\,,
  \end{align}
  where we have made use of Eq.~(\ref{eq:tracers.G2k}) and made the redefinition
  $\B{k}_2\,\rightarrow\,\B{k}_{23}$ in the second step. After symmetrization we then obtain:
  \begin{align}
    {\cal K}^{(3)}_{\delta^3}(\B{k}_1,\B{k}_2,\B{k}_3) &= 1\,, \label{eq:appbasis.K3d3} \\
    {\cal K}^{(3)}_{\delta{\cal G}_2}(\B{k}_1,\B{k}_2,\B{k}_3) &= \frac{1}{3}\Big[K(\B{k}_1,\B{k}_2) +
    \text{cyc.} \Big]\,, \label{eq:appbasis.K3dG2} \\
    {\cal K}^{(3)}_{{\cal G}_3}(\B{k}_1,\B{k}_2,\B{k}_3) &=
    L(\B{k}_1,\B{k}_2,\B{k}_3)\,, \label{eq:appbasis.K3G3} \\
    {\cal K}^{(3)}_{{\cal G}_2(\varphi_2,\varphi_1)}(\B{k}_1,\B{k}_2,\B{k}_3) &=
    \frac{1}{3}\Big[K(\B{k}_1,\B{k}_{23})\,K(\B{k}_2,\B{k}_3) + \text{cyc.}\Big] \label{eq:appbasis.K3Gp2p1}
  \end{align}
  Next, let us consider the most complicated combination that appears at fourth order, ${\cal
    G}_2(\varphi_3,\varphi_1)$, all other operators will follow in a very similar manner. Starting from the
  definition in Eq.~(\ref{eq:tracers.G2p3p1_def}) and using the relations (\ref{eq:tracers.p3a_def}) to
  (\ref{eq:tracers.A3_def}) for the LPT potentials, we have
  \begin{align}
    {\cal G}_2(\varphi_3,\varphi_1\,|\,\B{k}) = (2\pi)^3 \int\displaylimits_{\B{k}_1,\B{k}_2}
    \left[\delta_D\right]_2 &\left[\frac{1}{18} K(\B{k}_1,\B{k}_2) \left({\cal G}_3(\varphi_1\,|\,\B{k}_2) +
        \frac{15}{7}{\cal G}_2(\varphi_2,\varphi_1\,|\,\B{k}_2)\right)\right. \nonumber \\ &\left.+\frac{1}{14}
      \frac{\left(\B{k}_1 \cdot \B{k}_2\right)\,k_{1,j}\,k_{2,l}}{k_1^2\,k_2^2}
      \left[\nabla_{lm}\varphi_1\,\nabla_{jm}\varphi_2 - \nabla_{lm}\varphi_2\,\nabla_{jm}\varphi_1\right](\B{k}_2)\right] \delta_L(\B{k}_1)\,.
  \end{align}
  Plugging in the Fourier expressions for the remaining potentials and Galileons (using
  Eq.~\ref{eq:appbasis.G2p2p1}), and replacing $\B{k}_2\,\rightarrow\,\B{k}_{234}$ we get
  \begin{align}
    \label{eq:appbasis.G2p3p1}
    {\cal G}_2(\varphi_3,\varphi_1\,|\,\B{k}) = (2\pi)^3 \int\displaylimits_{\B{k}_1,\ldots,\B{k}_4}
    &\left[\delta_D\right]_4 \left[\frac{1}{18} K(\B{k}_1,\B{k}_{234}) \left(\frac{15}{7}
        K(\B{k}_{23},\B{k}_4)\,K(\B{k}_2,\B{k}_3) - L(\B{k}_2,\B{k}_3,\B{k}_4)\right)\right. \nonumber \\
    &\left.+\frac{1}{14} \Big(M(\B{k}_1,\B{k}_{23},\B{k}_{4},\B{k}_{234}) - M(\B{k}_1,\B{k}_{234},\B{k}_{23},\B{k}_4)\Big)
      K(\B{k}_2,\B{k}_3)\right] \prod_{i=1}^4 \delta_L(\B{k}_i)\,.
  \end{align}
  where we have introduced the new kernel
  \begin{align}
    M(\B{k}_1,\B{k}_2,\B{k}_3,\B{k}_4) \equiv \frac{\left(\B{k}_1 \cdot \B{k}_2\right) \left(\B{k}_2 \cdot
        \B{k}_3\right) \left(\B{k}_3 \cdot \B{k}_4\right) \left(\B{k}_4 \cdot \B{k}_1\right)}{\left(k_1\,k_2\,k_3\,k_4\right)^2}\,,
  \end{align}
  which is symmetric under cyclic permutations of its four momenta. The fully symmetric kernels for the basis
  operators at fourth order are thus given by
  \begin{align}
    {\cal K}^{(4)}_{\delta^4}(\B{k}_1,\B{k}_2,\B{k}_3,\B{k}_4) &\equiv 1\,, \label{eq:appbasis.K4d4} \\
    {\cal K}^{(4)}_{\delta^2{\cal G}_2}(\B{k}_1,\B{k}_2,\B{k}_3,\B{k}_4) &\equiv
    \frac{1}{6}\Big[K(\B{k}_1,\B{k}_2) + \text{sym.}(6)\big]\,, \\
    {\cal K}^{(4)}_{\delta{\cal G}_3}(\B{k}_1,\B{k}_2,\B{k}_3,\B{k}_4) &\equiv
    \frac{1}{4}\Big[L(\B{k}_1,\B{k}_2,\B{k}_3) + \text{sym.}(4)\Big]\,, \\
    {\cal K}^{(4)}_{{\cal G}_2^2}(\B{k}_1,\B{k}_2,\B{k}_3,\B{k}_4) &\equiv
    \frac{1}{3}\Big[K(\B{k}_1,\B{k}_2)\,K(\B{k}_3,\B{k}_4) + \text{sym.}(3)\Big]\,, \\
    {\cal K}^{(4)}_{\delta{\cal G}_2(\varphi_2,\varphi_1)}(\B{k}_1,\B{k}_2,\B{k}_3,\B{k}_4) &\equiv
    \frac{1}{12}\Big[K(\B{k}_1,\B{k}_{23})\,K(\B{k}_2,\B{k}_3) + \text{sym.}(12)\Big]\,, \\
    {\cal K}^{(4)}_{{\cal G}_3(\varphi_2,\varphi_1,\varphi_1)}(\B{k}_1,\B{k}_2,\B{k}_3,\B{k}_4) &\equiv
    \frac{1}{6}\Big[L(\B{k}_1,\B{k}_2,\B{k}_{34})\,K(\B{k}_3,\B{k}_4) + \text{sym.}(6)\Big]\,, \\
    {\cal K}^{(4)}_{{\cal G}_2(\varphi_2,\varphi_2)}(\B{k}_1,\B{k}_2,\B{k}_3,\B{k}_4) &\equiv
    \frac{1}{3}\Big[K(\B{k}_{12},\B{k}_{34})\,K(\B{k}_1,\B{k}_2)\,K(\B{k}_3,\B{k}_4) + \text{sym.}(3)\Big]\,, \\
    {\cal K}^{(4)}_{{\cal G}_2(\varphi_3,\varphi_1)}(\B{k}_1,\B{k}_2,\B{k}_3,\B{k}_4) &\equiv
    \frac{1}{12}\left[\frac{1}{18}K(\B{k}_1,\B{k}_{234})\left(\frac{15}{7}\,K(\B{k}_{23},\B{k}_4)\,K(\B{k}_2,\B{k}_3) -
        L(\B{k}_2,\B{k}_3,\B{k}_4)\right)\right. \nonumber \\
    &\hspace{3em}\left.+\frac{1}{14}\Big(M(\B{k}_1,\B{k}_{23},\B{k}_{4},\B{k}_{234}) -
      M(\B{k}_1,\B{k}_{234},\B{k}_{23},\B{k}_4) \Big) K(\B{k}_2,\B{k}_3) +
      \text{sym.}(12)\right]\,, \label{eq:appbasis.K4Gp3p1}
  \end{align}
  where $\text{sym.}(n)$ denotes the total number of terms the expressions have to be symmetrized over with
  respect to the four wave vectors.

\subsection{Relation to other bias bases in the literature}
\label{sec:basis-relation}

In~\cite{Desjacques:2018}, the authors extend the bias basis from~\cite{Mirbabayi:2015}. Let us establish the connection between their basis of operators and ours up to fourth order. They use operators that correspond to tracing objects denoted as $\Pi^{[n]}$ defined from convective derivatives. The calculation of the connection between both bases is conceptually simple but in practice long and tedious beyond second order, so we only provide the final results. To linear order,  we simply have ${\rm Tr}[ \Pi^{[1]}]=\delta$. The only non-trivial (i.e. beyond $\delta^2$) at quadratic order is  ${\rm Tr}[ (\Pi^{[1]})^2 ]= \delta^2 + {\cal G}_2 $. At cubic order we have for the non-trivial ones,

\beq
{\rm Tr} \Big[(\Pi^{[1]})^3\Big] = \delta^3 + \frac{3}{2} \, {\cal G}_2 \, \delta + \frac{1}{2}\,{\cal G}_3  , \ \ \ \ \ 
\mathrm{Tr}\left[\Pi^{[1]} \Pi^{[2]}\right] = \delta^3 + \frac{11}{14} \, {\cal G}_2 \, \delta + \frac{1}{2}\, {\cal G}_3 -\frac{5}{7}\,  {\cal G}_2(\varphi_2,\varphi_1)   
\eeq
whereas at fourth order we have,

\beqa
{\rm Tr}\Big[\left(\Pi^{[1]}\right)^4\Big] &=& \delta^4+2 \,{\cal G}_2\, \delta^2+ {2 \over 3}\, {\cal G}_3 \, \delta + {1\over 2} \,{\cal G}_2^2\\
{\rm Tr}\Big[\Pi^{[1]} \Pi^{[1]} \Pi^{[2]}\Big] &=& \delta^4+ {9\over 7} \,{\cal G}_2\, \delta^2+ {2 \over 3}\, {\cal G}_3 \, \delta + {1\over 7} \,{\cal G}_2^2
-\frac{5}{7}\,  {\cal G}_2(\varphi_2,\varphi_1)\, \delta - {5\over 14}\, {\cal G}_3(\varphi_2,\varphi_1,\varphi_1) \\
{\rm Tr}\Big[\Pi^{[2]} \Pi^{[2]}\Big] &=& \delta^4+ {4\over 7}\, {\cal G}_2\, \delta^2+ {2 \over 3} \,{\cal G}_3 \, \delta  + {29\over 98} \,{\cal G}_2^2
-\frac{10}{7}\,  {\cal G}_2(\varphi_2,\varphi_1)\, \delta - {5\over 7} \,{\cal G}_3(\varphi_2,\varphi_1,\varphi_1) +\frac{25}{49}\,  {\cal G}_2(\varphi_2,\varphi_2) \\
{\rm Tr}\Big[ \Pi^{[1]} \Pi^{[3]} \Big] &=& 2\, \delta^4+ {15\over 7}\, {\cal G}_2\, \delta^2 -{7\over 9}\, {\cal G}_3\, \delta + {1\over 14}\, {\cal G}_2^2 
- {4\over 21}\, {\cal G}_2(\varphi_2,\varphi_1)\, \delta+ {13\over 14}\, {\cal G}_3(\varphi_2,\varphi_1,\varphi_1) + 14 \,{\cal G}_2(\varphi_3,\varphi_1) \ \ \ \ \ 
\eeqa
Note that these relations show that this basis mixes operators that are local and nonlocal in second derivatives of the linear potential.

Another set of papers use a yet different basis~\cite{Sen1406,AngFasSen1503,Fujita:2016,Nadler:2018}. The comparison to these cases is more complicated since they also include velocity bias operators mixed together with standard ones. The simplest to compare with is~\cite{Nadler:2018}, which when constrained to the no-velocity bias case shows one more operator (five in total) at third order than in our basis. In fact, this set is precisely the same set used in~\cite{McDonald:2009}; however, as pointed out in~\cite{Chan:2012} not all of these five operators are independent (only four of them are). In terms of our basis, apart from $\delta^3$ they have at cubic order the following operators

\beq
s^3 = {1\over 2} {\cal G}_3 + {1\over 2} \delta\, {\cal G}_2 + {2\over 9} \delta^3, \ \ \ \ \ \delta s^2 = \delta\, {\cal G}_2 + {2\over 3} \delta^3, \ \ \ \ \ 
s t = {2\over 7}\, {\cal G}_2(\varphi_2,\varphi_1) + {4\over 21} \delta\, {\cal G}_2, \ \ \ \ \ 
\psi = {5 \over 7} {\cal G}_3 + \delta\, {\cal G}_2 -{30\over 49}\, {\cal G}_2(\varphi_2,\varphi_1)
\label{MDvars}
\eeq
which means that  we can express $\delta s^2, s^3, st,\psi$ each in terms of the four other basis elements in their basis, e.g.
\beq
\psi = -{15\over 7} st + {10\over 7} s^3 +{34\over 49} \delta s^2 -{344\over 441} \delta^3.
\label{psiMD}
\eeq
Again, it is worth noting that Eq.~(\ref{MDvars}) shows that this basis mixes operators that are local and nonlocal in second derivatives of the linear potential.

The papers~\cite{AngFasSen1503,Fujita:2016,Nadler:2018} do not consider fourth-order operators, but they do include higher-derivative operators.  For example, in the bispectrum~\cite{Nadler:2018} include terms corresponding to our $\beta_1,\beta_{2,1},\beta_{2,2}$ (but they don't include $\beta_{2,3},\beta_{2,4}$) and for the noise terms they include $\eta_{2,1},\eta_{2,2}$ and seem to be missing $\epsilon_2$ (though they do include such terms in the power spectrum, which for reasons that are unclear also includes a $k^4$ noise term); in addition they have the usual $\epsilon_0,\eta_0$ shot noise terms. See our Table~\ref{tab:basis} and Eqs.~(\ref{Bshot}) and~(\ref{BshotNLO}) for reference of the operators associated with these coefficients.

Finally, the approach described in~\cite{Matsubara:2014,Matsubara:2016} uses a number of basis of operators
obtained from different phenomenological models of biasing (e.g. halos, peaks).  Note that the 'renormalized
bias functions' they define correspond to our initial multipoint propagators, whereas what they refer to as the
multipoint propagators correspond to our evolved multipoint propagators. The LPT formalism of
\cite{Matsubara:2011,Matsubara:2014} was later extended by \cite{CarReiWhi1302} and in the context of the
two-point correlation function the authors of \cite{VlaCasWhi1612} introduced second order bias operators
equivalent to those given in~\cite{McDonald:2009}, as well as a higher-derivative term.  However, they neglect
third order bias parameters and thus do not cover the complete bias model at one-loop order.

\subsection{Large-scale limit of 4th-order NLE operators}
\label{sec:lowk_NLE}

As shown in Sec.~\ref{sec:galaxy-propagators}, loop corrections of the multipoint propagators are exclusively
given in terms of the NLE operators. In the large-scale limit these loop integrals can be Taylor expanded in
powers of the participating wave vectors. For the 4th-order contributions to the two-point propagator we thus
obtain:
\begin{align}
  \int_{\B{q}} {\cal K}^{(4,{\rm F})}_{\delta{\cal G}_2(\varphi_2,\varphi_1)}(\B{k}_1,\B{k}_2,\B{q},-\B{q})\,P_L(q) &= \frac{1}{1890}\Big\{8 k_{12}^2
  -16\B{k}_1\cdot\B{k}_2 - \Big[ 9k_{12}^2 -
  18\B{k}_1\cdot\B{k}_2 \Big]K(\B{k}_1,\B{k}_2)\Big\}\,\int_{q} \frac{P_L(q)}{q^2} + {\cal O}(k_i^4)\,,\\
  \int_{\B{q}} {\cal K}^{(4)}_{{\cal G}_3(\varphi_2,\varphi_1,\varphi_1)}(\B{k}_1,\B{k}_2,\B{q},-\B{q})\,P_L(q) &= \frac{1}{135}\Big[k_{12}^2 -
  2\B{k}_1\cdot\B{k}_2 \Big]K(\B{k}_1,\B{k}_2)\,\int_{q} \frac{P_L(q)}{q^2} + {\cal O}(k_i^4)\,,\\
  \int_{\B{q}} {\cal K}^{(4)}_{{\cal G}_2(\varphi_2,\varphi_2)}(\B{k}_1,\B{k}_2,\B{q},-\B{q})\,P_L(q) &= -\frac{8}{945}\Big\{3 k_{12}^2
    + \Big[ k_{12}^2 -
  \B{k}_1\cdot\B{k}_2 \Big]K(\B{k}_1,\B{k}_2)\Big\}\,\int_{q} \frac{P_L(q)}{q^2} + {\cal
O}(k_i^4)\,, \\
  \int_{\B{q}} {\cal K}^{(4)}_{{\cal G}_2(\varphi_3,\varphi_1)}(\B{k}_1,\B{k}_2,\B{q},-\B{q})\,P_L(q) &= -\frac{1}{39690}\Big\{69 k_{12}^2
    - \Big[12 k_{12}^2 -
  26\B{k}_1\cdot\B{k}_2 \Big]K(\B{k}_1,\B{k}_2)\Big\}\,\int_{q} \frac{P_L(q)}{q^2} + {\cal O}(k_i^4)\,,
\end{align}
where ${\cal O}(k_i^4)$ denotes terms which are of order $k_1^4$, $k_2^4$ or $k_{12}^4$ and higher. These
expressions serve to illustrate that the nonlocal evolution terms in the large-scale limit are given by
combinations of the higher-derivative operators. The converse is equally true as can be easily verified by
inverting the relations above.

\section{Time evolution of multipoint propagators}
\label{sec:derivation-gamma-recursion}

  In this appendix we give a detailed derivation for the time evolution of the multipoint propagators. We proceed
  in two steps: first, we evaluate expectation values of products of two or three Wiener-Hermite functionals,
  and second, by using these results we directly show how to obtain the recursion relations reported in
  Eq.~(\ref{eq:MP.Gamma_recursion}).

  \subsection{Orthogonality relations for generalized Wiener-Hermite functionals}
  \label{sec:hermite_products}

  Let us consider the PDF of $\delta_L$, shifted by a generic source term $\alpha(\B{k})$, which we take to be
  an arbitrary function of wavenumber $\B{k}$. A Taylor expansion around $\alpha=0$ yields:
  \begin{align}
    {\cal P}[\alpha-\delta_L] = \sum_{n=0}^{\infty} \frac{1}{n!} \int\displaylimits_{\B{k}_1,\ldots,\B{k}_n} \left.\frac{\partial^n
        {\cal P}[\alpha-\delta_L]}{\partial\alpha_1 \cdots \partial\alpha_n}\right|_{\alpha=0}
    \alpha_1 \cdots \alpha_n\,,
  \end{align}
  where $\alpha_i \equiv \alpha(\B{k}_i)$. Swapping the derivatives from $\alpha$ to $\delta_L$ and using the
  definition of the Wiener-Hermite functionals from Eq.~(\ref{eq:MP.defhermite}) we get
  \begin{align}
    \left.\frac{\partial^n{\cal P}[\alpha-\delta_L]}{\partial\alpha_1 \cdots \partial\alpha_n}\right|_{\alpha=0} 
    = (-1)^n \left.\frac{\partial^n{\cal
          P}[\delta_L]}{\partial\delta_{L,1}\cdots \partial\delta_{L,n}}\right|_{\alpha=0} 
    = \frac{{\cal P}[\delta_L]\,{\cal H}_n(\B{k}_1,\ldots,\B{k}_n)}{P_L(k_1) \cdots P_L(k_n)}\,,
  \end{align}
  and thus:
  \begin{align}
    \label{eq:Grec.PDF_hermite_expansion}
    \frac{{\cal P}[\alpha-\delta_L]}{{\cal P}[\delta_L]} = \sum_{n=0}^{\infty} \frac{1}{n!}
    \int\displaylimits_{\B{k}_1,\ldots,\B{k}_n} \frac{{\cal H}_n(\B{k}_1,\ldots,\B{k}_n)}{P_L(k_1)
      \cdots P_L(k_n)}\,\alpha_1 \cdots \alpha_n\,.
  \end{align}
  To derive the orthogonality relation between two Wiener-Hermite functionals of orders $m$ and $n$, we first
  compute the following integral
  \begin{align}
    \int {\cal D}[\delta_L]\,{\cal P}[\delta_L]\,\frac{{\cal P}[\alpha-\delta_L]}{{\cal P}[\delta_L]}\,
    \frac{{\cal P}[\beta-\delta_L]}{{\cal P}[\delta_L]} = \exp{\Bigg[\int_{\B{q}} \frac{\alpha(\B{q})\,\beta(-\B{q})}{P_L(q)}\Bigg]}\,,
  \end{align}
  where we have plugged in Eq.~(\ref{eq:MP.PDF}). However, using Eq.~(\ref{eq:Grec.PDF_hermite_expansion}) to
  replace the PDF's, we must also have:
  \begin{align}
    \sum_{m,n}\frac{1}{m!\,n!}\,\int\displaylimits_{\B{k}_1,\ldots,\B{k}_m}\,\int\displaylimits_{\B{q}_1,\ldots,\B{q}_n}
    \frac{\langle {\cal H}_m(\B{k}_1,\ldots,\B{k}_m)\,{\cal
        H}_n(\B{q}_1,\ldots,\B{q}_n)\rangle}{P_L(k_1) \cdots P_L(k_m)\,P_L(q_1) \cdots P_L(q_n)}\,
    \alpha_1 \cdots \alpha_m\,\beta_1 \cdots \beta_n
    = \exp{\Bigg[\int_{\B{q}} \frac{\alpha(\B{q})\,\beta(-\B{q})}{P_L(q)}\Bigg]}\,,
  \end{align}
  and by Taylor expanding the righthand side of the expression above, we see that we need to match up all
  $\B{k}$ and $\B{q}$ modes, which is only possible if $m=n$. From that observation it immediately follows that
  \begin{align}
    \label{eq:Grec.2Hproduct}
    \langle {\cal H}_m(\,\B{k}_1,\ldots,\B{k}_m)\,{\cal H}_n(\B{q}_1,\ldots,\B{q}_n)\rangle =
    (2\pi)^{3m}\,\delta_{mn}^{\rm K}\,\delta_D\left(\B{k}_{\{1,m\}},\B{q}_{\{1,m\}}\right) \prod_{i=1}^m P_L(k_i)\,,
  \end{align}
  where we have used the short-hand notation,
  \begin{align}
    \delta_D\left(\B{k}_{\{1,m\}},\B{q}_{\{1,m\}}\right) \equiv \delta_D(\B{k}_1+\B{q}_1) \cdots \delta_D(\B{k}_m+\B{q}_m) + \text{sym.}\,,
  \end{align}
  for writing all possible ways ($m!$ in total) of matching up the two sets of modes.

  Let us now compute the expectation value of three Wiener-Hermite functionals, i.e. $\langle {\cal H}_m\,{\cal
    H}_n\,{\cal H}_l \rangle$. Similar to the above procedure, we first evaluate an integral over PDF's, now
  with the three different sources $\alpha$, $\beta$ and $\gamma$:
  \begin{align}
    \label{eq:Grec.three_sources}
    \int {\cal D}[\delta_L]\,{\cal P}[\delta_L]\,\frac{{\cal P}[\alpha-\delta_L]}{{\cal P}[\delta_L]}\,
    \frac{{\cal P}[\beta-\delta_L]}{{\cal P}[\delta_L]}\,\frac{{\cal P}[\gamma-\delta_L]}{{\cal P}[\delta_L]} =
    \exp{\Bigg[\int_{\B{q}} \frac{\alpha(\B{q})\,\beta(-\B{q}) + \alpha(\B{q})\,\gamma(-\B{q}) +
          \beta(\B{q})\,\gamma(-\B{q})}{P_L(q)}\Bigg]}\,.
  \end{align}
  Expanding both sides of Eq.~(\ref{eq:Grec.three_sources}), we get
  \begin{align}
    \label{eq:Grec.tree_sources_expleft}
    (\text{\ref{eq:Grec.three_sources}, LHS}) = \sum_{m,n,l}\frac{1}{m!\,n!\,l!}
    \int\displaylimits_{\B{k}_1,\ldots,\B{k}_m}\,\int\displaylimits_{\B{q}_1,\ldots,\B{q}_n}\,\int\displaylimits_{\B{p}_1,\ldots,\B{p}_l} 
    &\langle {\cal H}_m(\B{k}_1,\ldots,\B{k}_m)\, {\cal H}_n(\B{q}_1,\ldots,\B{q}_n)\,
    {\cal H}_l(\B{p}_1,\ldots,\B{p}_l) \rangle \nonumber \\
    &\times\, \left[\prod_{i=1}^{m} \frac{\alpha(\B{k}_{i})}{P_L(k_{i})}\right]\, \left[\prod_{i=1}^{n}
    \frac{\beta(\B{q}_{i})}{P_L(q_{i})}\right]\, \left[\prod_{i=1}^{l} \frac{\gamma(\B{p}_{i})}{P_L(p_{i})}\right]\,,
  \end{align}
  and
  \begin{align}
    \label{eq:Grec.tree_sources_expright}
    (\text{\ref{eq:Grec.three_sources}, RHS}) = \sum_{a,b,c} \frac{1}{a!\,b!\,c!}
    \int\displaylimits_{\B{k}_1,\ldots,\B{k}_a}\,\int\displaylimits_{\B{q}_1,\ldots,\B{q}_b}\,\int\displaylimits_{\B{p}_1,\ldots,\B{p}_c}
    \left[\prod_{i=1}^a \frac{\alpha(\B{k}_i)\,\beta(-\B{k}_i)}{P_L(k_i)}\right]\, \left[\prod_{i=1}^b
      \frac{\alpha(\B{q}_i)\,\gamma(-\B{q}_i)}{P_L(q_i)}\right]\, \left[\prod_{i=1}^c
      \frac{\beta(\B{p}_i)\,\gamma(-\B{p}_i)}{P_L(p_i)}\right]\,.
  \end{align}
  In order for Eqs.~(\ref{eq:Grec.tree_sources_expleft}) and (\ref{eq:Grec.tree_sources_expright}) to be equal,
  they need to contain the same number of source terms, which requires that $l+m+n = 2(a+b+c)$ and thus, $l+m+n
  \in 2\mathbb{N}$. Moreover, the indices must satisfy the conditions
  \begin{align}
    \label{eq:Grec.conditions}
    \left.\begin{array}{ccc}
      a+b & = & m \\[0.8em]
      a+c & = & n \\[0.8em]
      b+c & = & l
    \end{array}\right\} \,\Leftrightarrow\,
    \left\{\begin{array}{ccc}
        a & = & \displaystyle \frac{m+n-l}{2} \\[0.8em]
        b & = & \displaystyle \frac{l+m-n}{2} \\[0.8em]
        c & = & \displaystyle \frac{n+l-m}{2}
    \end{array}\right.\,,
  \end{align}
  from which follows that $m+n \geq l$, and cyclic permutations thereof. According to
  Eq.~(\ref{eq:Grec.conditions}), we can divide all $\B{k}$-, $\B{q}$-, and $\B{p}$-modes into two subsets each,
  either of size $a$, $b$ or $c$. For instance, the $\B{k}$-modes will be split into a group containing $a$
  modes, and another containing $b$, such that $a+b=m$. Each mode in the former group can then be assigned a
  mode from an equally sized group of $\B{q}$'s, i.e. $\B{q}_1 = -\B{k}_1,\,\ldots,\,\B{q}_a = -\B{k}_a\,,$
  while the latter are matched in a similar manner with a set of $\B{p}$-modes, which guarantees that we obtain
  Eq.~(\ref{eq:Grec.tree_sources_expright}). Altogether, this means we need to require that the expectation
  value of the three Wiener-Hermite functionals is given by
  \begin{align}
    \label{eq:Grec.3Hproduct}
    \langle {\cal H}_m(\B{k}_1,\ldots,\B{k}_m)\, {\cal H}_n(\B{q}_1,\ldots,\B{q}_n)\, {\cal
      H}_l(\B{p}_1,\ldots,\B{p}_l) \rangle = (2\pi)^{\frac{3}{2}(m+n+l)} \left[\prod_{i=1}^{b}
      P_L(k_{a+i})\right] \left[\prod_{i=1}^{a}P_L(q_{i})\right] \left[\prod_{i=1}^{c}P_L(p_{b+i})\right]
    \nonumber \\
    \times\,\Big[\delta_D\left(\B{k}_{\{1,a\}},\B{q}_{\{1,a\}}\right)\,
    \delta_D\left(\B{k}_{\{a+1,m\}},\B{p}_{\{1,b\}}\right)\, \delta_D\left(\B{q}_{\{a+1,n\}},\B{p}_{\{b+1,l\}}\right) +
    \text{sym.}\Big]\,,
  \end{align}
  and it must vanish if the conditions above are not satisfied. The number of terms that have to be added in
  order to symmetrize Eq.~(\ref{eq:Grec.3Hproduct}) is given by the number of possibilities of selecting subsets
  of $\B{k}$-, $\B{q}$-, and $\B{p}$-modes that are of size $b$, $a$ and $c$, respectively. The total number
  of terms in the square brackets is thus:
  \begin{align}
    \binom{m}{b} \times \binom{n}{a} \times \binom{l}{c} \times a!\,b!\,c! = \frac{m!\,n!\,l!}{a!\,b!\,c!}\,.
  \end{align}

\subsection{The $\Gamma$-recursion relation}
\label{sec:gamma-recursion-relation}
  
  We are interested in the time evolution of the $n$-th order multipoint propagator. Exploiting the orthogonality
  of Wiener-Hermite functionals, we can single out its contribution to the series expansion in
  Eq.~(\ref{eq:MP.Psi_hermite}) by multiplying both sides with ${\cal H}_n$ and taking the ensemble average:
  \begin{align}
    \langle {\cal H}_n(\delta_L\,|\,\B{k}_1,\ldots,\B{k}_n)\,\Psi_a(\B{k},\eta)\rangle =
    (2\pi)^3\,\delta_D(\B{k}-\B{k}_{1\cdots n})\,\Gamma_a^{(n)}(\B{k}_1,\ldots,\B{k}_n\,;\,\eta)\,\prod_{i=1}^n P_L(k_i)
  \end{align}
  Applying the same procedure to the integral solution for $\Psi_a$ from Eq.~(\ref{eq:MP.Psi_intsolution}) we
  get
  \begin{align}
    \Gamma_a^{(n)}(\B{k}_1,\ldots,\B{k}_n\,;\,\eta)
    =\;&g_{ab}(\eta)\,\Gamma_{b,{\cal L}}^{(n)}(\B{k}_1,\ldots,\B{k}_n) +
    \sum_{m,\,l=1}\frac{1}{m!\,l!} \int_0^{\eta} \D{\eta'}\,g_{ab}(\eta-\eta')
    \nonumber \\
    &\times\,\int\displaylimits_{\B{q}_1,\ldots,\B{q}_m}\,\int\displaylimits_{\B{p}_1,\ldots,\B{p}_l}\gamma_{bcd}(\B{q}_{1\cdots m},\B{p}_{1 \cdots l})\,
    \Gamma_c^{(m)}(\B{q}_1,\ldots,\B{q}_m\,;\,\eta')\,\Gamma_d^{(l)}(\B{p}_1,\ldots,\B{p}_l\,;\,\eta') \nonumber \\ &\times\,\langle {\cal
      H}_n(\B{k}_1,\ldots,\B{k}_n) {\cal H}_m^*(\B{q}_1,\ldots,\B{q}_m) {\cal 
      H}_l^*(\B{p}_1,\ldots,\B{p}_l)\rangle \prod_{i=1}^n \frac{1}{P_L(k_i)}\,,
  \end{align}
  where $\Gamma_{a,{\cal L}}$ denotes multipoint propagators at initial time $\eta=0$. Upon inserting
  Eq.~(\ref{eq:Grec.3Hproduct}) we perform the first $a=(n+m-l)/2$ integrations over $\B{q}$-modes as well as
  over all of the $\B{p}$-modes, such that the second term above reduces to
  \begin{align}
    \sum_{\substack{m,\,l=1 \\ n+m+l \in 2\mathbb{N} \\ n+m \geq l\,, \hspace{0.3em}\text{cyc.}}}
    &\frac{1}{m!\,l!}\,\kappa\,\int_0^{\eta} \D{\eta'}\,g_{ab}(\eta-\eta')
    \int\displaylimits_{\B{q}_{a+1},\ldots,\B{q}_m} \Big[\gamma_{bcd}(\B{k}_{1\ldots
      a}+\B{q}_{a+1\ldots m},\B{k}_{a+1\ldots n}-\B{q}_{a+1 \ldots m})\,\Big. \nonumber \\
    &\times\,\Big.\Gamma_c^{(m)}(\B{k}_1,\ldots,\B{k}_a,\B{q}_{a+1},\ldots,\B{q}_m\,;\,\eta')\,
    \Gamma_d^{(l)}(\B{k}_{a+1},\ldots,\B{k}_n,-\B{q}_{a+1},\ldots,-\B{q}_m\,;\,\eta') + \text{sym.}\Big]\,\prod_{i=a+1}^{m}P_L(q_{i})\,.
  \end{align}
  The symmetrization is carried out by summing over all $\binom{n}{a}$ subsets of $\B{k}$-modes, which implies
  that the combinatorial factor $\kappa$ is given by:
  \begin{align}
    \kappa = \frac{m!\,n!\,l!}{a!\,b!\,c!}\,\binom{n}{a}^{-1} = \frac{m!\,l!}{\left(m-a\right)!}\,.
  \end{align}
  Next, let us change the summation index from $l$ to $r = (l+m-n)/2 = m-a$ and relabel the remaining integrations
  over $\B{q}$'s into $\B{q}_1,\ldots,\B{q}_r$. The conditions $n+m \geq l$ (and cyclic permutations) then
  transform into $r_* \leq r \leq m$, where
  \begin{align}
    \label{eq:rstar}
    r_* = \left\{
    \begin{array}{cl}
      0\,, & n > m \\
      1\,, & n = m \\
      m-n\,, & n < m
    \end{array}\right.\,.
  \end{align}
  Defining the quantity
  \begin{align}
    \Gamma^{(n,m,r)}_a(\B{k}_1,\ldots,\B{k}_n\,;\,\eta) \equiv \int\displaylimits_{\B{q}_1,\ldots,\B{q}_r}
    &\Bigg[\gamma_{abc}(\B{k}_{1\ldots 
      m-r}+\B{q}_{1\ldots r},\B{k}_{m-r+1\ldots n}-\B{q}_{1\ldots r})\,
    \Gamma_b^{(m)}(\B{k}_1,\ldots,\B{k}_{m-r},\B{q}_1,\ldots,\B{q}_r\,;\,\eta) \Bigg.\nonumber \\
    &\Bigg.\times\,\Gamma_c^{(2r+n-m)}(\B{k}_{m-r+1},\ldots,\B{k}_{n},-\B{q}_1,\ldots,-\B{q}_r\,;\,\eta)\prod_{i=1}^r    
    P_L(\B{q}_i) + \text{sym}.(\B{k}_i) \Bigg]\,,
  \end{align}
  we finally obtain the desired recursion relation for multipoint propagators:
  \begin{align}
    \label{eq:Gammarec}
    \Gamma_a^{(n)}(\B{k}_1,\ldots,\B{k}_n\,;\,\eta) =
    g_{ab}(\eta)\,\Gamma_{b,{\cal L}}^{(n)}(\B{k}_1,\ldots,\B{k}_n) + \sum_{r=0}\,\sum_{m=\max\{r,1\}}^{r+n-\delta^K_{r,0}} \frac{1}{r!}\,\int_0^\eta
    \D{\eta'}\,g_{ab}(\eta-\eta')\,\Gamma^{(n,m,r)}_b(\B{k}_1,\ldots,\B{k}_n\,;\,\eta')\,,
  \end{align}
  where we have arranged the two summations in the second contribution such that the outer sum runs over the
  number of loop integrals, $r$, and used Eq.~(\ref{eq:rstar}) to determine the allowed range of the index $m$.

\subsection{Computation of time-evolved propagators up to one-loop order}
\label{sec:time-evolved-prop}

We now use the recursion relation from Sec.~\ref{sec:gamma-recursion-relation} to derive explicit expressions
for the building blocks of the one-loop bispectrum, which requires us to evaluate the first two multipoint
propagators at one-loop order and the three-point propagator at tree-level. In order to make the connection to
the initial conditions evident, where convenient, we write them completely in terms of the Lagrangian multipoint
propagators, which we computed in Sec.~\ref{sec:galaxy-propagators}.

To simplify the resulting expressions it is useful to define the following two quantities that arise through
contractions of the linear propagator $g_{ab}$ with the vertices $\gamma_{abc}$:
\begin{align}
  g^{(2)}_{abc}(\B{k}_1,\B{k}_2\,;\eta) &\equiv \int_0^{\eta} \D{\eta'}
  g_{ad}(\eta-\eta')\,\gamma_{def}(\B{k}_1,\B{k}_2)\,g_{eb}(\eta')\,g_{fc}(\eta')\,, \label{eq:Grec.g2} \\
  g^{(3)}_{abcd}(\B{k}_1,\B{k}_2\,\B{k}_3\,;\eta) &\equiv \int_0^{\eta} \D{\eta'}
  g_{ae}(\eta-\eta')\,\gamma_{efg}(\B{k}_1,\B{k}_{23})\,g_{fb}(\eta')\,g^{(2)}_{gcd}(\B{k}_2,\B{k}_3\,;\eta')\,, \label{eq:Grec.g3}
\end{align}
where we let the superscript indicate the number of external lines that are attached to these objects, i.e. two
for $g^{(2)}_{abc}$ etc.  The tree-level expression ($r=0$) for $\Gamma_a^{(1)}$ is trivial since the second
piece in Eq.~(\ref{eq:Gammarec}) makes no contribution. For that reason we start with the two-point propagator,
and by using the kernel from Eq.~(\ref{eq:Grec.g2}) we obtain
\begin{equation}
  \label{eq:Grec.G2tree}
  \Gamma_a^{(2)}(\B{k}_1,\B{k}_2\,;\eta)\Big|_{\text{tree}} = g_{ab}(\eta)\,\Gamma_{b,{\cal L}}^{(2)}(\B{k}_1,\B{k}_2) + 2
  g_{abc}^{(2)}(\B{k}_1,\B{k}_2\,;\eta)\,\Gamma_{b,{\cal L}}^{(1)}(\B{k}_1)\,\Gamma_{c,{\cal L}}^{(1)}(\B{k}_2)\,,
\end{equation}
with all initial multipoint propagators being evaluated at tree-level. The tree-level three-point propagator
receives two contributions from the second term in Eq.~(\ref{eq:Gammarec}), such that
\begin{equation}
  \label{eq:Grec.G3tree}
  \begin{split}
      \Gamma_a^{(3)}(\B{k}_1,\B{k}_2,\B{k}_3\,;\eta)\Big|_{\text{tree}} &= g_{ab}(\eta)\,\Gamma_{b,{\cal L}}^{(3)}(\B{k}_1,\B{k}_2,\B{k}_3) + \left[2
  g_{abc}^{(2)}(\B{k}_1,\B{k}_{23}\,;\eta)\,\Gamma_{b,{\cal L}}^{(1)}(\B{k}_1)\,\Gamma_{c,{\cal
      L}}^{(2)}(\B{k}_2,\B{k}_3) \right. \\ &\left. + 4g_{abcd}^{(3)}(\B{k}_1,\B{k}_2,\B{k}_3\,;\eta)\,\Gamma_{b,{\cal
      L}}^{(1)}(\B{k}_1)\,\Gamma_{c,{\cal L}}^{(1)}(\B{k}_2)\,\Gamma_{d,{\cal L}}^{(1)}(\B{k}_3) + \text{cyc.}\right]\,,
  \end{split}
\end{equation}
where we have plugged in the solution for $\Gamma^{(2)}_a$ from above. Similarly, Eq.~(\ref{eq:Grec.G2tree}) can
be used to determine the nonlinear correction to the one-point propagator at one-loop order (or
$r=1$), giving
\begin{equation}
  \label{eq:Grec.G1loop}
  \begin{split}
      \Gamma_a^{(1)}(\B{k}\,;\eta)\Big|_{\text{loop}} &= g_{ab}(\eta)\,\Gamma_{b,{\cal L}}^{(1)}(\B{k})\Big|_{\text{loop}} +
      2\int_{\B{q}} g_{abc}^{(2)}(\B{q},\B{k}-\B{q}\,;\eta)\,\Gamma_{b,{\cal L}}^{(1)}(\B{q})\,\Gamma_{c,{\cal
          L}}^{(2)}(\B{k},-\B{q})\,P_L(q) \\ &+ 4\int_{\B{q}}
      g_{abcd}^{(3)}(\B{q},\B{k},-\B{q}\,;\eta)\,\Gamma_{b,{\cal L}}^{(1)}(\B{q})\, \Gamma_{c,{\cal
          L}}^{(1)}(\B{k})\,\Gamma_{d,{\cal L}}^{(1)}(-\B{q})\,P_L(q)\,.
  \end{split}
\end{equation}
Equations~(\ref{eq:Grec.G2tree}-\ref{eq:Grec.G1loop}) are the full results including all transients and the
evolution of the Lagrangian bias parameters. For comparison to observational data only the fastest growing mode
is relevant and we can further absorb all terms due to bias evolution into the final Eulerian bias parameters
(see Sec.~\ref{sec:bias-evo}), allowing us to greatly simplify these expressions. Moreover, if we separate the
contributions that get newly generated by nonlinear evolution (i.e. they are not due to bias evolution of terms
already present in the initial multipoint propagators), and expressing them in terms of the SPT kernels, we
arrive at Eqs.~(\ref{eq:MP.Gamma1loop_evo}-\ref{eq:MP.Gamma3tree_evo}).

While Eqs.~(\ref{eq:MP.Gamma1loop_evo}-\ref{eq:MP.Gamma3tree_evo}) can be more easily obtained from the
nonlinear evolution of the matter and velocity fields directly, starting from $\Gamma^{(2)}$ at one-loop order
the evolution of the multipoint propagators becomes useful, since we do not have to deal with cumbersome
renormalizations as shown in Sec.~\ref{sec:bias-exp-with-evo}. Because the result is still complicated we have
chosen to present it in terms of the previously derived expressions for the evolved propagators,
\begin{alignat}{2}
  \label{eq:Grec.G2loop}
    \Gamma_a^{(2)}(\B{k}_1,\B{k}_2;\eta)\Big|_{\text{loop}} &=
    g_{ab}(\eta)\,\Gamma_{b,{\cal L}}^{(2)}(\B{k}_1,\B{k}_2)\Big|_{\text{loop}} + 2\int_0^{\eta} \D{\eta'}\,
    &&g_{ab}(\eta-\eta')\, \gamma_{bcd}(\B{k}_1,\B{k}_2) \left[\Gamma_c^{(1)}(\B{k}_1;\eta')\Big|_{\text{loop}}\,
    \Gamma_d^{(1)}(\B{k}_2;\eta')\Big|_{\text{tree}}\right. \nonumber \\ & &&\left. +\;\Gamma_c^{(1)}(\B{k}_1;\eta')\Big|_{\text{tree}}\,
    \Gamma_d^{(1)}(\B{k}_2;\eta')\Big|_{\text{loop}}\right] \nonumber \\ &+ 2\int_{\B{q}} P_L(q) \int_{0}^{\eta}
  \D{\eta'}\,g_{ab}(\eta-\eta') &&\hspace{-2.5em}\bigg[\gamma_{bcd}(\B{q},\B{k}_{12}-\B{q})\,
    \Gamma_c^{(1)}(\B{q};\eta')\Big|_{\text{tree}}\,
    \Gamma_d^{(3)}(\B{k}_1,\B{k}_2,-\B{q};\eta')\Big|_{\text{tree}} \nonumber \\
  & &&\hspace{-7.1em} + \gamma_{bcd}(\B{k}_1+\B{q},\B{k}_{2}-\B{q})\,
    \Gamma_c^{(2)}(\B{k}_1,\B{q};\eta')\Big|_{\text{tree}}\,
    \Gamma_d^{(2)}(\B{k}_2,-\B{q};\eta')\Big|_{\text{tree}}\bigg]\,,
\end{alignat}
where we have indicated at which order the propagators are evaluated.

\section{Using Galilean Invariance to compute time evolution of bias}
\label{sec:GIbiasevol}

\subsection{SPT in Galilean basis for galaxy bias}
\label{sec:SPTgal}

Let us consider the Galilean invariant contributions to SPT at each order in perturbation theory (PT). This will
allow us to calculate in a more straightforward way the time evolution of bias parameters up to any order in PT
(we will compute up to fourth-order here). Since our bias basis contains the LPT potentials, we start from the
expression for density fluctuations in terms of the Lagrangian displacement field:

\beq
1 + \delta(\B{x}) ={1\over \text{det}[\delta_{ij}^{\rm K}+\Psi_{i,j}(\B{q})]}= \frac{1}{1+G_1({ \Psi}) -{1\over 2} G_2({\Psi}) + {1\over 6} G_3(\Psi)}\bigg|_{\B{q}}
\label{deltaFromLPT}
\eeq where $\Psi_{i,j}\equiv \partial \Psi_i/\partial q_j$ ($\delta_{ij}^{\rm K}$ denotes the Kronecker delta)
and the $G_i$'s are the Galileons built out of the deformation tensor, working up to fourth-order in PT they
read: \beqa G_1(\Psi)&\equiv& \Psi_{i,i} = -\delta +{3\over 14} {\cal G}_2 + {1\over 18} {\cal G}_3 -{5\over 42}
{\cal G}_2(\varphi_2,\varphi_1) + {7\over 11} {\cal G}_2(\varphi_3,\varphi_1) +{51\over 4312} {\cal
  G}_2(\varphi_2,\varphi_2) - {13\over 308} {\cal G}_3(\varphi_2,\varphi_1,\varphi_1)
\label{G1disp} \\
G_2(\Psi)&\equiv&(\Psi_{i,j})^2-(\Psi_{i,i} )^2 =  {\cal G}_2  -{3\over 7} {\cal G}_2(\varphi_2,\varphi_1) + 2 {\cal G}_2(\varphi_3,\varphi_1) +{9\over 196} {\cal G}_2(\varphi_2,\varphi_2) 
\label{G2disp} \\
G_3(\Psi)&\equiv&(\Psi_{i,i} )^3+ 2 \Psi_{i,j}\Psi_{j,k}\Psi_{k,i}-3 (\Psi_{i,j})^2(\Psi_{i,i} ) =  - {\cal G}_3   +{9\over 14}  {\cal G}_3(\varphi_2,\varphi_1,\varphi_1) 
\label{G3disp}
\eeqa

By using these results in Eq.~(\ref{deltaFromLPT}), and using that $\B{x}=\B{q}+\B{\Psi}$, one can derive the standard Eulerian PT expansion. However, we are interested in finding out the bias relation in the absence of velocity bias, therefore any dipole terms coming from displacement from $\B{q}$ to $\B{x}$ will not appear in any of the final expressions, and our basis being Galilean invariant does not include any such operators. Therefore, we can simply work with the Galilean invariant SPT expression at each order, since  dipole terms will cancel at the end of the calculation. Thus we have,

\beq
1 + \delta_\text{GI}(\B{x}) = \frac{1}{1+G_1({ \Psi}) -{1\over 2} G_2({\Psi}) + {1\over 6} G_3(\Psi)}\bigg|_{\B{x}}
\label{deltaGI}
\eeq

Using Eqs.~(\ref{G1disp}-\ref{G3disp}) in Eq.~(\ref{deltaGI}) we obtain the Galilean invariant piece of the SPT expansion at each order in our basis of bias operators,

\beq
\delta_\text{GI}^{(1)}= \delta , \ \ \ \ \ \ \ \ \ \ \delta_\text{GI}^{(2)}= \delta^2 + {2\over 7} {\cal G}_2, \ \ \ \ \ \ \ \ \ \ 
\delta_\text{GI}^{(3)}= \delta^3 + {4\over 7} \delta\,{\cal G}_2 - {2\over 21} {\cal G}_2(\varphi_2,\varphi_1) + {1\over 9}{\cal G}_3,
\label{deltaGI23}
\eeq
\beq
\delta_\text{GI}^{(4)}= \delta^4 + {6\over 7} \delta^2\,{\cal G}_2 +{4\over 49} {\cal G}_2^2 - {4\over 21} \delta\,{\cal G}_2(\varphi_2,\varphi_1) + {2\over 9} \delta\,{\cal G}_3 +{6\over 539} {\cal G}_2(\varphi_2,\varphi_2) + {4\over 11}  {\cal G}_2(\varphi_3,\varphi_1) -{5\over 77} {\cal G}_3(\varphi_2,\varphi_1,\varphi_1).
\label{deltaGI4}
\eeq

A simple check on these expressions is provided by the spherical collapse dynamics, which is also Galilean invariant. To do this all we need is the spherical average of our basis operators, which is straightforward
\beqa
\overline{{\cal G}_2}&=& -{2\over 3} \delta^2, \ \ \ \ \  \overline{{\cal G}_2(\varphi_2,\varphi_1)}=2\, \overline{{\cal G}_3}={4\over 9} \delta^3, \ \ \ \ \  \overline{{\cal G}_2(\varphi_2,\varphi_2)}=2\, \overline{{\cal G}_3(\varphi_2,\varphi_1,\varphi_1) }=-{8\over 27} \delta^4, \ \ \ \ \  \overline{{\cal G}_2(\varphi_3,\varphi_1)}=-{46\over 1701}  \delta^4.
\label{SCaverage} \nonumber \\
\eeqa
Using these results in Eqs.~(\ref{deltaGI23}-\ref{deltaGI4}) we recover the well-known vertices $\nu_n$ from the spherical collapse dynamics~\cite{Ber92}, i.e. 
\beq
\delta_\text{SC}=\sum_n {\nu_n \over n!} \, \delta^n, \ \ \ \ \ \nu_1={1}, \ \ \ \ \ \nu_2={34\over 21}, \ \ \ \ \ \nu_3={682\over 189}, \ \ \ \ \ \nu_4={446440\over 43659}
\label{nuSC}
\eeq

Now we proceed to derive the analogous results for the velocity field, which is what is needed for the nonlinear evolution of $\Phi_v$ or the normalized divergence $\theta=\nabla^2\Phi_v$. By definition, the Eulerian velocity field agrees with the velocity of particles calculated from the time derivative of the displacement field at the same location. Taking the normalized divergence $- \nabla/{\cal H} f$ we then have that
\beq
\theta(\B{x}) = - \nabla_{\B{x}} \cdot \B{\Psi}'(\B{q})
\label{ThetaPsi}
\eeq
where $'\equiv \partial/\partial{ \ln D}$, with $D$ the linear growth factor. Again, we are only interested in the Galilean invariant piece of this exact relation. In order to obtain that, all we need is $\B{\Psi}'(\B{q})$ as a function of $\B{x}$, with {\em at most} one displacement free from derivatives acting on it, because other contributions cannot be rendered Galilean invariant by taking a single derivative when calculating $\theta_\text{GI}(\B{x})$. A straightforward Taylor expansion gives then,

\beq
\B{\Psi}'(\B{q}) = \B{\Psi}' - \Psi_i (\nabla_i \B{\Psi}')+  \Psi_j (\nabla_j\Psi_i) (\nabla_i \B{\Psi}')-   \Psi_k (\nabla_k\Psi_j)  (\nabla_j\Psi_i) (\nabla_i \B{\Psi}') + \ldots
\label{PsiToX}
\eeq
where all quantities on the RHS of this equation are now evaluated at $\B{x}$. The result for  $\theta_\text{GI}$ follows by acting with the divergence in Eq.~(\ref{ThetaPsi}) on the derivative-free displacement field in each term of Eq.~(\ref{PsiToX}) giving us the simple result,

\beq
\theta_\text{GI}(\B{x}) =\sum_{n=0}^\infty {(-1)^n \over n} \Big[ \text{Tr}  (\Psi_{i,j})^n \Big]'
\label{GItheta}
\eeq
The traces here can be easily rewritten in terms of the deformation tensor Galileons in Eq.~(\ref{G1disp}), and the derivative with respect to $\ln D$  just multiplies each contribution by its PT order. Therefore we obtain the Galilean invariant piece of the normalized divergence at each order,

\beq
\theta_\text{GI}^{(1)}= \delta , \ \ \ \ \ \ \ \ \ \ \theta_\text{GI}^{(2)}= \delta^2 + {4\over 7} {\cal G}_2, \ \ \ \ \ \ \ \ \ \ 
\theta_\text{GI}^{(3)}= \delta^3 + {6\over 7} \delta\,{\cal G}_2 - {2\over 7} {\cal G}_2(\varphi_2,\varphi_1) + {1\over 3}{\cal G}_3,
\label{thetaGI23}
\eeq
\beq
\theta_\text{GI}^{(4)}= \delta^4 + {8\over 7} \delta^2\,{\cal G}_2 +{8\over 49} {\cal G}_2^2 - {8\over 21} \delta\,{\cal G}_2(\varphi_2,\varphi_1) + {4\over 9} \delta\,{\cal G}_3 +{24\over 539} {\cal G}_2(\varphi_2,\varphi_2) + {16\over 11}  {\cal G}_2(\varphi_3,\varphi_1) -{20\over 77} {\cal G}_3(\varphi_2,\varphi_1,\varphi_1).
\label{thetaGI4}
\eeq

Again, we can double check these results comparing them to the spherical collapse dynamics, by performing the spherical average using Eq.~(\ref{SCaverage}) in Eqs.~(\ref{thetaGI23}-\ref{thetaGI4}) to compute the vertices $\mu_n$ for the normalized velocity divergence,
\beq
\theta_\text{SC}=\sum_n {\mu_n \over n!} \, \delta^n, \ \ \ \ \ \mu_1={1}, \ \ \ \ \ \mu_2={26\over 21}, \ \ \ \ \ \mu_3={142\over 63}, \ \ \ \ \ \mu_4={236872\over 43659}
\label{nuSC}
\eeq
which agree with the standard spherical collapse dynamics values~\cite{Ber92}.

\subsection{Time Evolution of Bias}
\label{sec:BiasEvol}

As already explained in Sec.~\ref{sec:bias-evo}, in order to determine the time evolution of bias, we can use a
simple trick that exploits the Galilean invariance of the galaxy overdensity in the absence of velocity
bias. That means all dipole terms in the matter density must cancel, such that
\beq
\delta_g(\B{x})= \delta_g^\text{IC}(\B{x}) +\delta^\text{GI}(\B{x}) +  \delta_g^\text{IC}(\B{x})\, \delta^\text{GI}(\B{x})\,,
\label{dgEvol2}
\eeq where $\delta^\text{GI}$ is the Galilean invariant part of the (nonlinear) matter density as determined in
the last section, and $\delta_g^\text{IC}$ is the galaxy density at initial time (where all fluctuations are
linear and Gaussian).  At linear order it is easy to check that Eq.~(\ref{dgEvol2}) simply gives
$b_1=1+b_{1,{\cal L}}$, which agrees with the full evolution~\cite{Fry9604} when neglecting transients. To find
the evolution of higher-order bias parameters, we must first subtract all contributions due to lower order
operators (see~\cite{Chan:2012}). For instance, by computing (suppressing $\B{x}$ arguments) \beq \delta_g-b_1\,
\delta^\text{GI} = \delta_g^\text{IC} - b_{1,{\cal L}}\, \delta^\text{GI} + \delta_g^\text{IC}\,
\delta^\text{GI}\,, \eeq and writing the RHS of this equation in our bias basis using Eqs.~(\ref{deltaGinitial})
and~(\ref{deltaGI23}) at second order, we are able to identify the coefficients of the $\delta^2$ and
${\cal G}_2$ operators, which leads to \beq b_2=b_{2,{\cal L}},\ \ \ \ \ \gamma_2=-{2\over 7}\, b_{1,{\cal L}} +
\gamma_{2,{\cal L}}\,.
\label{B2res}
\eeq
At cubic order we proceed in the same way and compute,
\beq
\delta_g^{(3)}- b_1\,  \delta_\text{GI}^{(3)}- {b_2\over 2} \, [\delta^2_\text{GI}]^{(3)}-\gamma_2\, [{\cal G}_2]^{(3)}
\label{subtract3}
\eeq
and we use Eqs.~(\ref{deltaGinitial}) and~(\ref{deltaGI23}) plus that 
\beq
[{\cal G}_2]^{(3)}={\cal G}_3+\delta\,{\cal G}_2 - {6\over 7} {\cal G}_2(\varphi_2,\varphi_1)
\label{G2cubicOrder}
\eeq
which follows from Eq.~(93) in~\cite{Chan:2012} when constrained to Galilean invariant operators (for which we need Eqs.~\ref{deltaGI23} and \ref{thetaGI23}). We thus identify the coefficients (i.e. bias parameters) in front of the operators in our bias basis at third order,
\beq
b_3=b_{3,{\cal L}}-3 b_2,\ \ \ \ \ \gamma_3 = -{1\over 9} b_{1,{\cal L}} - \gamma_2 + \gamma_{3,{\cal L}},\ \ \ \ \ \gamma_2^\times=-{2\over 7} b_2+\gamma_{2,{\cal L}}^\times,\ \ \ \ \ \gamma_{21}={2\over 21}b_{1,{\cal L}} +{6\over 7}\gamma_2 + \gamma_{21,{\cal L}}
\label{B3res}
\eeq
Equations~(\ref{B2res})-(\ref{B3res}) agree with those derived in~\cite{Chan:2012} up to decaying modes (which we neglect here) and a typo in their Eq.~(116) which should have a minus sign for its second term.  

Finally, at quartic order we need to subtract all fourth-order contributions from up to cubic bias, 
\beq
\delta_g^{(4)}- b_1\,  \delta_\text{GI}^{(4)}- {b_2\over 2} \, [\delta^2_\text{GI}]^{(4)}-\gamma_2\, [{\cal G}_2]^{(4)}-{b_3\over 6}\, [\delta^3_\text{GI}]^{(4)} - \gamma_3\, [{\cal G}_3]^{(4)} 
- \gamma_2^\times \, [\delta_\text{GI}\, {\cal G}_2]^{(4)} - \gamma_{21} \, [{\cal G}_2(\varphi_2,\varphi_1)]^{(4)}
\eeq
and then, using that (again, constrained to Galilean invariant operators)
\beq
[{\cal G}_2]^{(4)}= \delta^2\, {\cal G}_2 + {2\over 7} {\cal G}_2^2 - {6\over 7} \delta\, {\cal G}_2(\varphi_2,\varphi_1) + {9\over 49} {\cal G}_2(\varphi_2,\varphi_2) + 6 {\cal G}_2(\varphi_3,\varphi_1) + \delta\, {\cal G}_3 -{15\over 14}  {\cal G}_3(\varphi_2,\varphi_1,\varphi_1),
\eeq
\beq
[{\cal G}_3]^{(4)}= {9\over 7} {\cal G}_3(\varphi_2,\varphi_1,\varphi_1) + \delta\, {\cal G}_3,\ \ \ \ \ 
[{\cal G}_2(\varphi_2,\varphi_1)]^{(4)}= -{\cal G}_3(\varphi_2,\varphi_1,\varphi_1)+ \delta\, {\cal G}_2(\varphi_2,\varphi_1)
\eeq
we can identify the evolved bias parameters from the coefficients of each of the  fourth-order operators in our bias basis. We thus obtain,

\beqa
b_4&=&b_{4,{\cal L}} -12\, b_2 -8\, b_3,\ \ \ \ \ 
\gamma_2^{\times\times}=-{3\over 7}b_2-{1\over 7}b_3 -\gamma_2^\times + \gamma_{2,{\cal L}}^{\times\times},\ \ \ \ \ 
\gamma_3^\times=-{b_2\over 9}-\gamma_2^\times+\gamma_{3,{\cal L}}^\times ,\ \ \ \ \ 
\label{B4resA} \\
\gamma_2^\text{sq}&=&-{2\over 49}b_2 -{2\over 7} \gamma_2^\times + \gamma_{2,{\cal L}}^\text{sq},\ \ \ \ \ 
\gamma_{21}^\times={2\over 21}b_2+{6\over 7}\gamma_2^\times + \gamma_{21,{\cal L}}^\times,\ \ \ \ \ 
\gamma_{31}=-{4\over 11} b_{1,{\cal L}} - 6 \gamma_2 +\gamma_{31,{\cal L}},\ \ \ \ \  
\label{B4resB}\\
\gamma_{22}&=&-{6\over 539}b_{1,{\cal L}} - {9\over 49} \gamma_2 + \gamma_{22,{\cal L}} ,\ \ \ \ \  
\gamma_{211}={5\over 77}b_{1,{\cal L}}+{15\over 14}\gamma_2+\gamma_{21}-{9\over 7}\gamma_3 + \gamma_{211,{\cal L}}.
\label{B4resC}
\eeqa

\subsection{Spherical Average Limit}
\label{SphAvgSubtle}

Let us now briefly consider the spherical-average limit of these results, which is somewhat subtle due to the fact that operators beyond local bias do not vanish under spherical average. As a result of this, the spherically averaged bias parameters $b_n^\text{sph}$ are linear combinations of all the bias parameters (local and nonlocal) in our basis. 

Consider quadratic bias, where the impact of spherical averaging first appears (linear bias is invariant under spherical average). From Eq.~(\ref{SCaverage}) one can take the spherical average of the quadratic bias relation $(b_2/2)\, \delta^2 + \gamma_2\, {\cal G}_2$ and write the spherically averaged quadratic bias parameter in terms of our bias parameters as,

\beq
b_2^\text{sph} = b_2 - {4\over 3} \gamma_2= b_{2,{\cal L}} - {4\over 3} \gamma_{2,{\cal L}} + {8\over 21} b_{1,{\cal L}} \equiv b_{2,{\cal L}}^\text{sph} + {8\over 21} b_{1,{\cal L}}^\text{sph}
\label{b2sph}
\eeq
where in the second equality we have used our time-evolution result from Eq.~(\ref{B2res}) and then rewritten it in terms of the Lagrangian spherical averaged quadratic bias parameter, giving the well-known spherical collapse model bias evolution at quadratic order~\cite{MoJinWhi97}.

At cubic order, we see for the first time something subtler. The reason is that the spherical average of Eq.~(\ref{subtract3}) does not correspond to
\beq
\delta_g^{(3)}- b_1\,  \delta_\text{SC}^{(3)}- {b_2^\text{sph}\over 2} \, [\delta^2_\text{SC}]^{(3)}
\label{subtract3s}
\eeq
due to the fact that at third-order the spherically averaged Galileon operator obeys from Eqs.~(\ref{SCaverage}) and~(\ref{G2cubicOrder})
\beq
\overline{ [{\cal G}_2]^{(3)}}=-{52\over 63}\, \delta^3 \neq -{2\over 3} \overline{[\delta^2_\text{GI}]^{(3)}} = -{2\over 3} \nu_2\, \delta^3 = -{68 \over 63}\, \delta^3
\label{G2cubicSph}
\eeq
Therefore, at cubic order in our bias basis part of what would be considered $b_3^\text{sph}$ is encoded by the deviation of the spherical average of $[{\cal G}_2]^{(3)}$ from that of $(-2/3) [\delta^2]^{(3)}$, the precise relation is then 
\beq
b_3^\text{sph} = b_3 - 4\gamma_2^\times + {4\over 3} \gamma_3 + {8\over 3} \gamma_{21} + {32\over 21} \gamma_2
\label{b3sph}
\eeq
Note this apparent dependence of $b_3^\text{sph}$ on a quadratic bias parameter such as $\gamma_2$ only happens at the final conditions, since at the initial conditions matter fluctuations are linear and thus  $[{\cal G}_2]^{(3)}=[\delta^2]^{(3)}=0$. Therefore $\gamma_{2,{\cal L}}$ cannot appear in  $b_{3,{\cal L}}^\text{sph}$, and in fact all this means that the only role of the $\gamma_2$ term in Eq.~(\ref{b3sph}) is to cancel the appearance of such terms in the evolved values of $\gamma_3$ and $\gamma_{21}$ in Eq.~(\ref{B3res}) and in  $b_3$ and $\gamma_2^\times$ through $b_2=b_2^\text{sph}+4\gamma_2/3$. Indeed, evaluating Eq.~(\ref{b3sph}) yields,
\beq
b_3^\text{sph} = b_{3,{\cal L}}^\text{sph} - {13\over 7} b_{2,{\cal L}}^\text{sph} - {796\over 1323} b_{1,{\cal L}}^\text{sph}
\label{b3sph2}
\eeq
where $b_{3,{\cal L}}^\text{sph}=b_{3,{\cal L}} - 4\gamma_{2,{\cal L}}^\times + (4/3) \gamma_{3,{\cal L}} + (8/ 3) \gamma_{21,{\cal L}}$. At quartic order, a similar subtlety arises with all nonlocal bias parameters of lower order than fourth. However, for the same reasons as in the cubic bias calculation, all such terms must cancel at the end. Therefore, a fast way of doing the calculation is to simply ignore all nonlocal quadratic and cubic bias parameters contributions to Eqs.~(\ref{B4resA}-\ref{B4resC}) and in the relation between spherical bias parameters and local bias parameters. A quick calculation then gives,
\beq
b_{4}^\text{sph}=b_{4,{\cal L}}^\text{sph}-{40\over 7} b_{3,{\cal L}}^\text{sph}+{7220\over 1323}b_{2,{\cal L}}^\text{sph}+{476320\over 305613}b_{1,{\cal L}}^\text{sph},
\label{b4sph}
\eeq
where $b_{4,{\cal L}}^\text{sph}=b_{4,{\cal L}}-16\gamma_{2,{\cal L}}^{\times\times} +(16/3)\gamma_{3,{\cal L}}^\times+(32/3)\gamma_{2,{\cal L}}^\text{sq}+(32/3)\gamma_{21,{\cal L}}^\times-(32/9)\gamma_{211,{\cal L}}-(64/9)\gamma_{22,{\cal L}}-(368/567)\gamma_{31,{\cal L}}$.
The results in Eqs.~(\ref{b3sph2}) and~(\ref{b4sph}) agree with those in~\cite{Matsubara:2011} who corrected the cubic and quartic spherical collapse bias results in~\cite{MoJinWhi97}.

\section{Evolution of Renormalized Bias}
\label{sec:RENbiasevol}

In Appendix~\ref{sec:GIbiasevol} we discussed time evolution of bias parameters in terms of the standard expansion, as opposed to the bias parameters that appear in the multipoint-propagator expansion, therefore we ignored renormalization. One may ask whether those results are somehow changed due to renormalization. In fact, \citep{Assassi:2014} contrasted the relation between renormalized and bare (final time) bias parameters to the time evolution of bare bias parameters found in~\cite{Chan:2012}, and appeared to suggest  that the results of \cite{Chan:2012} may be affected by renormalization. 

What matters of course is the time evolution of the {\em observable} bias parameters, i.e. how the renormalized late-time bias parameters are related to the renormalized initial-time bias parameters (this is what is often measured in numerical simulations). We therefore start from the initial conditions written using the multipoint propagator expansion up to fourth order,
\beqa
\delta_g^\text{IC} &=& b_{1,{\cal L}}\, \delta + {b_{2,{\cal L}}\over 2}\, \left[\delta^2-\sigma^2\right]+ \gamma_{2,{\cal L}}\, {\cal G}_2 +{b_{3,{\cal L}}\over 6}\, \left[\delta^3-3\delta\sigma^2\right] + \gamma_{3,{\cal L}}\, {\cal G}_3 
+ \gamma_{2,{\cal L}}^\times \, \left[\delta\, {\cal G}_2+{4\over 3} \delta\sigma^2\right] + \gamma_{{21},{\cal L}} \, {\cal G}_2(\varphi_2,\varphi_1) \nonumber \\ & & 
+ {b_{4,{\cal L}}\over 24}\, \left[\delta^4-6\delta^2\sigma^2\right] + \gamma_{2,{\cal L}}^{\times\times}\, \left[\delta^2\, {\cal G}_2 + {8\over 3} \delta^2\sigma^2-{\cal G}_2\sigma^2\right]+
 \gamma_{3,{\cal L}}^{\times}\, \left[\delta\, {\cal G}_3 + {\cal G}_2\sigma^2\right] +  \gamma_{2,{\cal L}}^\text{sq}\, \left[{\cal G}_2^2 - {32\over 15} \delta^2\sigma^2- {8\over 15}{\cal G}_2\sigma^2\right] \nonumber \\ & &
 +  \gamma_{21,{\cal L}}^{\times}\, \left[\delta\, {\cal G}_2(\varphi_2,\varphi_1) - {16\over 5} \delta^2\sigma^2 + {2\over 5}{\cal G}_2\sigma^2\right]
 + \gamma_{211,{\cal L}} \,{\cal G}_3(\varphi_2,\varphi_1,\varphi_1) + \gamma_{22,{\cal L}}\, {\cal G}_2(\varphi_2,\varphi_2) + \gamma_{31,{\cal L}}\, {\cal G}_2 (\varphi_3,\varphi_1)\nonumber \\ & &
+ \ldots 
\label{deltaGinitialren}
\eeqa
where, unlike Eq.~(\ref{deltaGinitial}) the bias parameters here are renormalized, and we have neglected  ${\cal O}(\sigma^4)$ contributions. Also, we have used that the loops in the multipoint propagators coming from nonlocal evolution operators cancel against the corresponding terms in the Gamma expansion to make the expression more compact.

The next step is to use Eq.~(\ref{dgEvol2}) to evolve this to find the late-time evolved $\delta_g$, and find the evolved renormalized bias parameters. To compare with and address the issue suggested by~\citep{Assassi:2014} we proceed as they did and write the non-linear renormalization formulae for linear and quadratic bias parameters to ${\cal O}(\sigma^2)$, following the calculations we discussed in Section~\ref{sec:bias-exp-with-evo}, that is
\beq
b_1 =  \bar{b}_1 + \left({1\over 2} \bar{b}_3 \right)\sigma^2 +\left({34\over 21} \bar{b}_2 \right)\sigma^2+ {\cal O}(\sigma^4)
\label{b1renNL}
\eeq
for linear bias~\cite{McDonald:2009} and 
\beqa
b_2 &=&\bar{b}_2 + \left({1\over 2}\bar{b}_4-{16\over 3} \bar{\gamma}^{\times\times}_2 + {32\over 15} \bar{\gamma}_{21}^\times + {64\over 15} \bar{\gamma}_2^\text{sq}\right)
 \sigma^2+  \left( {8126 \over 2205} \bar{b}_2 - {208\over 35} \bar{\gamma}_2^\times +{68\over 21} \bar{b}_3 \right) \sigma^2
 + {\cal O}(\sigma^4)
\label{b2renNL} \\ & & \nonumber  \\
\gamma_2 &=&\bar{\gamma}_2 +  \left( \bar{\gamma}^{\times\times}_2 - {2\over 5} \bar{\gamma}_{21}^\times- \bar{\gamma}_{3}^\times + {8\over 15} \bar{\gamma}_2^\text{sq}\right)
 \sigma^2 + \left( {127 \over 2205} \bar{b}_2 + {92\over 105} \bar{\gamma}_2^\times  \right) \sigma^2
 + {\cal O}(\sigma^4)
 \label{g2renNL}
\eeqa
for the quadratic bias parameters. Note that all bias parameters in Eqs.~(\ref{b1renNL}-\ref{g2renNL}) are final-time quantities. The first parenthesis contributions in each of these expressions correspond to the renormalizations by operators that depend on the linear Gaussian  fluctuations (that is, those that are straightforward to compute) and the second parenthesis contributions describe renormalizations induced by non-linear evolution (discussed in Section~\ref{sec:bias-exp-with-evo}) that were bypassed in the method we advocate elsewhere in the main text.  Apart from some sign typos and the renormalizations induced by $\bar{\gamma}_2^\times$ (what they call $b_{{\cal G}_2\delta}^{(0)}$), Eqs.~(\ref{b2renNL}-\ref{g2renNL}) agree with Eqs.~(3.17-18) in~\citep{Assassi:2014}. 

Using Eq.~(\ref{deltaGinitialren}) in Eq.~(\ref{dgEvol2}), we can identify the evolved  bare bias parameters as the coefficients of the operators in our bias basis up to quadratic order, and then renormalize according to Eqs.~(\ref{b1renNL})-(\ref{g2renNL}), finding the evolved renormalized bias parameters in terms of the initial renormalized bias parameters. In doing so we have taken into account all renormalizations induced by operators up to fourth order (both in the initial bias parameters and in the final ones). As expected, all $\sigma^2$ dependencies go away (a nontrivial check on the nonlinear renormalization calculations in Eqs.~\ref{b2renNL}-\ref{g2renNL}), and we recover
\beq
b_1 = 1+ b_{1,{\cal L}},\ \ \ \ \ 
b_2 = b_{2,{\cal L}},\ \ \ \ \
\gamma_2 = -{2\over 7} b_{1,{\cal L}} + \gamma_{2,{\cal L}}
\label{evolB1B2G2renorm}
\eeq
i.e. the same results obtained from the evolution of the bare bias parameters, showing explicitly that the bare evolution results in~\cite{Chan:2012} are inmune to renormalization. Therefore final-time renormalization formulae such as Eqs.~(\ref{b1renNL}-\ref{g2renNL}) are not in conflict with time evolution of bare bias parameters, but rather guarantee that the time evolution results also hold for the renormalized bias parameters. In our approach advocated in the main text, the initial-time  renormalization (equivalent to the first parenthesis in Eqs.~\ref{b1renNL}-\ref{g2renNL}) is automatically carried out by the multipoint propagator expansion, while the `second parenthesis' renormalizations are never needed because time evolution of conserved tracers cannot generate such divergences (i.e. the evolved propagators already avoid  spurious contributions proportional to $\sigma^2$).

\end{widetext}


\bibliography{refs_new}

\begin{thebibliography}{100}

\bibitem{Kaiser:1984}
N.~{Kaiser}.
\newblock {On the spatial correlations of Abell clusters}.
\newblock {\em \apjl}, 284:L9--L12, September 1984.

\bibitem{PolWis8410}
H.~D. {Politzer} and M.~B. {Wise}.
\newblock {Relations between spatial correlations of rich clusters of
  galaxies}.
\newblock {\em \apj}, 285:L1--L3, Oct 1984.

\bibitem{BarBonKai8605}
J.~M. {Bardeen}, J.~R. {Bond}, N.~{Kaiser}, and A.~S. {Szalay}.
\newblock {The Statistics of Peaks of Gaussian Random Fields}.
\newblock {\em \apj}, 304:15, May 1986.

\bibitem{MoWhi9609}
H.~J. {Mo} and S.~D.~M. {White}.
\newblock {An analytic model for the spatial clustering of dark matter haloes}.
\newblock {\em \mnras}, 282:347--361, September 1996.

\bibitem{SheTor9909}
R.~K. {Sheth} and G.~{Tormen}.
\newblock {Large-scale bias and the peak background split}.
\newblock {\em \mnras}, 308:119--126, September 1999.

\bibitem{Desjacques:2018}
Vincent {Desjacques}, Donghui {Jeong}, and Fabian {Schmidt}.
\newblock {Large-scale galaxy bias}.
\newblock {\em \physrep}, 733:1--193, February 2018.

\bibitem{McDonald:2006}
P.~{McDonald}.
\newblock {Clustering of dark matter tracers: Renormalizing the bias
  parameters}.
\newblock {\em \prd}, 74(10):103512, November 2006.

\bibitem{McDonald:2009}
P.~{McDonald} and A.~{Roy}.
\newblock {Clustering of dark matter tracers: generalizing bias for the coming
  era of precision LSS}.
\newblock {\em \jcap}, 8:020, August 2009.

\bibitem{Chan:2012}
K.~C. {Chan}, R.~{Scoccimarro}, and R.~K. {Sheth}.
\newblock {Gravity and large-scale nonlocal bias}.
\newblock {\em \prd}, 85(8):083509, April 2012.

\bibitem{Baldauf:2012}
Tobias {Baldauf}, Uro{\v{s}} {Seljak}, Vincent {Desjacques}, and Patrick
  {McDonald}.
\newblock {Evidence for quadratic tidal tensor bias from the halo bispectrum}.
\newblock {\em \prd}, 86:083540, October 2012.

\bibitem{Schmidt:2013}
Fabian {Schmidt}, Donghui {Jeong}, and Vincent {Desjacques}.
\newblock {Peak-background split, renormalization, and galaxy clustering}.
\newblock {\em \prd}, 88:023515, July 2013.

\bibitem{Assassi:2014}
V.~{Assassi}, D.~{Baumann}, D.~{Green}, and M.~{Zaldarriaga}.
\newblock {Renormalized halo bias}.
\newblock {\em \jcap}, 8:056, August 2014.

\bibitem{Mirbabayi:2015}
M.~{Mirbabayi}, F.~{Schmidt}, and M.~{Zaldarriaga}.
\newblock {Biased tracers and time evolution}.
\newblock {\em \jcap}, 7:030, July 2015.

\bibitem{FonRegSee1805}
L.~{Fonseca de la Bella}, D.~{Regan}, D.~{Seery}, and D.~{Parkinson}.
\newblock {Impact of bias and redshift-space modelling for the halo power
  spectrum: Testing the effective field theory of large-scale structure}.
\newblock {\em ArXiv e-prints}, May 2018.

\bibitem{FriGaz94}
J.{}A. {Frieman} and E.~{Gaztanaga}.
\newblock {The three-point function as a probe of models for large-scale
  structure}.
\newblock {\em \apj}, 425:392--402, 1994.

\bibitem{GazFri94}
E.~{Gaztanaga} and J.{}A. {Frieman}.
\newblock {Bias and high-order galaxy correlation functions in the APM galaxy
  survey}.
\newblock {\em \apjl}, 437:L13--L16, 1994.

\bibitem{Fry:1994}
J.~N. {Fry}.
\newblock {Gravity, bias, and the galaxy three-point correlation function}.
\newblock {\em \prl}, 73:215--219, July 1994.

\bibitem{Matarrese:1997}
S.~{Matarrese}, L.~{Verde}, and A.~F. {Heavens}.
\newblock {Large-scale bias in the Universe: bispectrum method}.
\newblock {\em \mnras}, 290:651--662, October 1997.

\bibitem{ScoColFry9803}
R.~{Scoccimarro}, S.~{Colombi}, J.~N. {Fry}, J.~A. {Frieman}, E.~{Hivon}, and
  A.~{Melott}.
\newblock {Nonlinear Evolution of the Bispectrum of Cosmological
  Perturbations}.
\newblock {\em \apj}, 496:586, March 1998.

\bibitem{Sco00}
R.~{Scoccimarro}.
\newblock {The Bispectrum: From Theory to Observations}.
\newblock {\em \apj}, 544:597--615, 2000.

\bibitem{SefCroPue0607}
E.~{Sefusatti}, M.~{Crocce}, S.~{Pueblas}, and R.~{Scoccimarro}.
\newblock {Cosmology and the bispectrum}.
\newblock {\em \prd}, 74(2):023522--+, July 2006.

\bibitem{Pollack:2012}
J.~E. {Pollack}, R.~E. {Smith}, and C.~{Porciani}.
\newblock {Modelling large-scale halo bias using the bispectrum}.
\newblock {\em \mnras}, 420:3469--3489, March 2012.

\bibitem{Bel:2015}
J.~{Bel}, K.~{Hoffmann}, and E.~{Gazta{\~n}aga}.
\newblock {Non-local bias contribution to third-order galaxy correlations}.
\newblock {\em \mnras}, 453:259--276, October 2015.

\bibitem{SaiBalVla1405}
Shun {Saito}, Tobias {Baldauf}, Zvonimir {Vlah}, Uro{\v{s}} {Seljak}, Teppei
  {Okumura}, and Patrick {McDonald}.
\newblock {Understanding higher-order nonlocal halo bias at large scales by
  combining the power spectrum with the bispectrum}.
\newblock {\em \prd}, 90:123522, December 2014.

\bibitem{GilNorVer1407}
H{\'e}ctor {Gil-Mar{\'\i}n}, Jorge {Nore{\~n}a}, Licia {Verde}, Will~J.
  {Percival}, Christian {Wagner}, Marc {Manera}, and Donald~P. {Schneider}.
\newblock {The power spectrum and bispectrum of SDSS DR11 BOSS galaxies - I.
  Bias and gravity}.
\newblock {\em \mnras}, 451:539--580, July 2015.

\bibitem{GilPerVer1606}
H{\'e}ctor {Gil-Mar{\'\i}n}, Will~J. {Percival}, Licia {Verde}, Joel~R.
  {Brownstein}, Chia-Hsun {Chuang}, Francisco-Shu {Kitaura}, Sergio~A.
  {Rodr{\'\i}guez-Torres}, and Matthew~D. {Olmstead}.
\newblock {The clustering of galaxies in the SDSS-III Baryon Oscillation
  Spectroscopic Survey: RSD measurement from the power spectrum and bispectrum
  of the DR12 BOSS galaxies}.
\newblock {\em \mnras}, 465:1757--1788, February 2017.

\bibitem{Coles:1993}
P.~{Coles}.
\newblock {Galaxy formation with a local bias}.
\newblock {\em \mnras}, 262:1065--1075, June 1993.

\bibitem{FryGaz9308}
J.~N. {Fry} and E.~{Gaztanaga}.
\newblock {Biasing and hierarchical statistics in large-scale structure}.
\newblock {\em \apj}, 413:447--452, August 1993.

\bibitem{Catelan:2000}
P.~{Catelan}, C.~{Porciani}, and M.~{Kamionkowski}.
\newblock {Two ways of biasing galaxy formation}.
\newblock {\em \mnras}, 318:L39--L44, November 2000.

\bibitem{Fujita:2016}
Tomohiro {Fujita}, Valentin {Mauerhofer}, Leonardo {Senatore}, Zvonimir {Vlah},
  and Raul {Angulo}.
\newblock {Very Massive Tracers and Higher Derivative Biases}.
\newblock {\em ArXiv e-prints}, page arXiv:1609.00717, September 2016.

\bibitem{Nadler:2018}
Ethan~O. {Nadler}, Ashley {Perko}, and Leonardo {Senatore}.
\newblock {On the bispectra of very massive tracers in the Effective Field
  Theory of Large-Scale Structure}.
\newblock {\em Journal of Cosmology and Astro-Particle Physics}, 2018:058,
  February 2018.

\bibitem{PueSco0908}
S.~{Pueblas} and R.~{Scoccimarro}.
\newblock {Generation of vorticity and velocity dispersion by orbit crossing}.
\newblock {\em \prd}, 80(4):043504--+, August 2009.

\bibitem{PieManSav1108}
M.~{Pietroni}, G.~{Mangano}, N.~{Saviano}, and M.~{Viel}.
\newblock {Coarse-Grained Cosmological Perturbation Theory}.
\newblock {\em ArXiv e-prints}, August 2011.

\bibitem{Baumann:2012}
Daniel {Baumann}, Alberto {Nicolis}, Leonardo {Senatore}, and Matias
  {Zaldarriaga}.
\newblock {Cosmological non-linearities as an effective fluid}.
\newblock {\em Journal of Cosmology and Astro-Particle Physics}, 2012:051, July
  2012.

\bibitem{Carrasco:2012}
John Joseph~M. {Carrasco}, Mark~P. {Hertzberg}, and Leonardo {Senatore}.
\newblock {The effective field theory of cosmological large scale structures}.
\newblock {\em Journal of High Energy Physics}, 2012:82, September 2012.

\bibitem{BalMerMir1406}
T.~{Baldauf}, L.~{Mercolli}, M.~{Mirbabayi}, and E.~{Pajer}.
\newblock {The Bispectrum in the Effective Field Theory of Large Scale
  Structure}.
\newblock {\em ArXiv e-prints}, June 2014.

\bibitem{AngForSch1406}
Raul~E. {Angulo}, Simon {Foreman}, Marcel {Schmittfull}, and Leonardo
  {Senatore}.
\newblock {The one-loop matter bispectrum in the Effective Field Theory of
  Large Scale Structures}.
\newblock {\em Journal of Cosmology and Astro-Particle Physics}, 2015:039,
  October 2015.

\bibitem{Scherrer:1998}
Robert~J. {Scherrer} and David~H. {Weinberg}.
\newblock {Constraints on the Effects of Locally Biased Galaxy Formation}.
\newblock {\em \apj}, 504:607--611, September 1998.

\bibitem{Dekel:1999}
Avishai {Dekel} and Ofer {Lahav}.
\newblock {Stochastic Nonlinear Galaxy Biasing}.
\newblock {\em \apj}, 520:24--34, July 1999.

\bibitem{CroSco0603a}
M.~{Crocce} and R.~{Scoccimarro}.
\newblock {Renormalized cosmological perturbation theory}.
\newblock {\em \prd}, 73(6):063519--+, March 2006.

\bibitem{Bernardeau:2008}
F.~{Bernardeau}, M.~{Crocce}, and R.~{Scoccimarro}.
\newblock {Multipoint propagators in cosmological gravitational instability}.
\newblock {\em \prd}, 78(10):103521, November 2008.

\bibitem{Bernardeau:2010}
F.~{Bernardeau}, M.~{Crocce}, and E.~{Sefusatti}.
\newblock {Multipoint propagators for non-Gaussian initial conditions}.
\newblock {\em \prd}, 82(8):083507, October 2010.

\bibitem{Bernardeau:2012}
F.~{Bernardeau}, M.~{Crocce}, and R.~{Scoccimarro}.
\newblock {Constructing regularized cosmic propagators}.
\newblock {\em \prd}, 85(12):123519, June 2012.

\bibitem{Crocce:2012}
M.~{Crocce}, R.~{Scoccimarro}, and F.~{Bernardeau}.
\newblock {MPTBREEZE: a fast renormalized perturbative scheme}.
\newblock {\em \mnras}, 427:2537--2551, December 2012.

\bibitem{TarNisBer1304}
A.~{Taruya}, T.~{Nishimichi}, and F.~{Bernardeau}.
\newblock {Precision modeling of redshift-space distortions from a multipoint
  propagator expansion}.
\newblock {\em \prd}, 87(8):083509, April 2013.

\bibitem{Matsubara:2011}
Takahiko {Matsubara}.
\newblock {Nonlinear perturbation theory integrated with nonlocal bias,
  redshift- space distortions, and primordial non-Gaussianity}.
\newblock {\em \prd}, 83:083518, April 2011.

\bibitem{Matsubara:2014}
Takahiko {Matsubara}.
\newblock {Integrated perturbation theory and one-loop power spectra of biased
  tracers}.
\newblock {\em \prd}, 90:043537, August 2014.

\bibitem{Matsubara:2016}
Takahiko {Matsubara} and Vincent {Desjacques}.
\newblock {Impacts of biasing schemes in the one-loop integrated perturbation
  theory}.
\newblock {\em \prd}, 93:123522, June 2016.

\bibitem{Smith:2007}
Robert~E. {Smith}, Rom{\'a}n {Scoccimarro}, and Ravi~K. {Sheth}.
\newblock {Scale dependence of halo and galaxy bias: Effects in real space}.
\newblock {\em \prd}, 75:063512, March 2007.

\bibitem{CroSco0603b}
M.~{Crocce} and R.~{Scoccimarro}.
\newblock {Memory of initial conditions in gravitational clustering}.
\newblock {\em \prd}, 73(6):063520--+, March 2006.

\bibitem{Szalay:1988}
A.~S. {Szalay}.
\newblock {Constraints on the biasing of density fluctuations}.
\newblock {\em \apj}, 333:21--23, October 1988.

\bibitem{Jin9904}
Y.~P. {Jing}.
\newblock {Accurate Determination of the Lagrangian Bias for the Dark Matter
  Halos}.
\newblock {\em \apjl}, 515:L45--L48, April 1999.

\bibitem{SheLem99}
R.{}K. {Sheth} and G.~{Lemson}.
\newblock {Biasing and the distribution of dark matter haloes}.
\newblock {\em \mnras}, 304:767--792, 1999.

\bibitem{PorCatLac9903}
C.~{Porciani}, P.~{Catelan}, and C.~{Lacey}.
\newblock {How Dark Matter Halos Cluster in Lagrangian Space}.
\newblock {\em \apjl}, 513:L99--L102, March 1999.

\bibitem{EliLudPor1204}
A.~{Elia}, A.~D. {Ludlow}, and C.~{Porciani}.
\newblock {The spatial and velocity bias of linear density peaks and
  protohaloes in the {$\Lambda$} cold dark matter cosmology}.
\newblock {\em \mnras}, 421:3472--3480, April 2012.

\bibitem{SheChaSco1304}
R.~K. {Sheth}, K.~C. {Chan}, and R.~{Scoccimarro}.
\newblock {Nonlocal Lagrangian bias}.
\newblock {\em \prd}, 87(8):083002, April 2013.

\bibitem{BiaChuDes1310}
Matteo {Biagetti}, Kwan~Chuen {Chan}, Vincent {Desjacques}, and Aseem
  {Paranjape}.
\newblock {Measuring non-local Lagrangian peak bias}.
\newblock {\em \mnras}, 441:1457--1467, June 2014.

\bibitem{ChaSheSco1711}
K.~C. {Chan}, R.~K. {Sheth}, and R.~{Scoccimarro}.
\newblock {Effective window function for Lagrangian halos}.
\newblock {\em \prd}, 96(10):103543, November 2017.

\bibitem{ModCasSel1612}
Chirag {Modi}, Emanuele {Castorina}, and Uro{\v{s}} {Seljak}.
\newblock {Halo bias in Lagrangian space: estimators and theoretical
  predictions}.
\newblock {\em \mnras}, 472:3959--3970, December 2017.

\bibitem{LazSch1712}
Titouan {Lazeyras} and Fabian {Schmidt}.
\newblock {Beyond LIMD bias: a measurement of the complete set of third-order
  halo bias parameters}.
\newblock {\em Journal of Cosmology and Astro-Particle Physics}, 2018:008,
  September 2018.

\bibitem{AbiBal1807}
M.~M. {Abidi} and T.~{Baldauf}.
\newblock {Cubic halo bias in Eulerian and Lagrangian space}.
\newblock {\em \jcap}, 7:029, July 2018.

\bibitem{SchElsJas1901}
Fabian {Schmidt}, Franz {Elsner}, Jens {Jasche}, Nhat~Minh {Nguyen}, and
  Guilhem {Lavaux}.
\newblock {A rigorous EFT-based forward model for large-scale structure}.
\newblock {\em Journal of Cosmology and Astro-Particle Physics}, 2019:042, Jan
  2019.

\bibitem{SchSimAss1811}
Marcel {Schmittfull}, Marko {Simonovi{\'c}}, Valentin {Assassi}, and Matias
  {Zaldarriaga}.
\newblock {Modeling Biased Tracers at the Field Level}.
\newblock {\em arXiv e-prints}, page arXiv:1811.10640, Nov 2018.

\bibitem{Bernardeau:2002}
F.~{Bernardeau}, S.~{Colombi}, E.~{Gazta{\~n}aga}, and R.~{Scoccimarro}.
\newblock {Large-scale structure of the Universe and cosmological perturbation
  theory}.
\newblock {\em \physrep}, 367:1--248, September 2002.

\bibitem{Bouchet:1992}
F.~R. {Bouchet}, R.~{Juszkiewicz}, S.~{Colombi}, and R.~{Pellat}.
\newblock {Weakly nonlinear gravitational instability for arbitrary Omega}.
\newblock {\em \apjl}, 394:L5--L8, July 1992.

\bibitem{Bouchet:1995}
F.~R. {Bouchet}, S.~{Colombi}, E.~{Hivon}, and R.~{Juszkiewicz}.
\newblock {Perturbative Lagrangian approach to gravitational instability.}
\newblock {\em \aap}, 296:575, April 1995.

\bibitem{Fry84}
J.{}N. {Fry}.
\newblock {The Galaxy correlation hierarchy in perturbation theory}.
\newblock {\em \apj}, 279:499--510, 1984.

\bibitem{GorGriRey86}
M.{}H. {Goroff}, B.~{Grinstein}, S.-J. {Rey}, and M.{}B. {Wise}.
\newblock {Coupling of modes of cosmological mass density fluctuations}.
\newblock {\em \apj}, 311:6--14, 1986.

\bibitem{Fry9604}
J.~N. {Fry}.
\newblock {The Evolution of Bias}.
\newblock {\em \apjl}, 461:L65+, April 1996.

\bibitem{Sen1406}
Leonardo {Senatore}.
\newblock {Bias in the effective field theory of large scale structures}.
\newblock {\em Journal of Cosmology and Astro-Particle Physics}, 2015:007,
  November 2015.

\bibitem{Scoccimarro:1996}
R.~{Scoccimarro} and J.~{Frieman}.
\newblock {Loop Corrections in Nonlinear Cosmological Perturbation Theory}.
\newblock {\em \apjs}, 105:37, July 1996.

\bibitem{HofKun71}
K.~M. {Hoffman} and R.~{Kunze}.
\newblock {Linear Algebra}.
\newblock {\em Pearson}, 1971.

\bibitem{MoJinWhi97}
H.~J. {Mo}, Y.~P. {Jing}, and S.~D.~M. {White}.
\newblock {High-order correlations of peaks and haloes: a step towards
  understanding galaxy biasing}.
\newblock {\em \mnras}, 284:189--201, 1997.

\bibitem{ParShe1211}
Aseem {Paranjape} and Ravi~K. {Sheth}.
\newblock {Peaks theory and the excursion set approach}.
\newblock {\em \mnras}, 426:2789--2796, November 2012.

\bibitem{KofPog9503}
L.~{Kofman} and D.~{Pogosyan}.
\newblock {Dynamics of gravitational instability is nonlocal}.
\newblock {\em \apj}, 442:30--38, March 1995.

\bibitem{BiasLoops2}
A.~{Eggemeier}, R.~{Scoccimarro}, M.~{Crocce}, A.~G. {Sanchez}, and R.~E.
  {Smith}.
\newblock In preparation (2018).

\bibitem{Buc9301}
T.~{Buchert}.
\newblock {Lagrangian perturbation theory - A key-model for large-scale
  structure}.
\newblock {\em \aap}, 267:L51--L54, January 1993.

\bibitem{Buc9404}
T.~{Buchert}.
\newblock {Lagrangian Theory of Gravitational Instability of Friedman-Lemaitre
  Cosmologies - a Generic Third-Order Model for Nonlinear Clustering}.
\newblock {\em \mnras}, 267:811, April 1994.

\bibitem{Matsubara:1995}
T.~{Matsubara}.
\newblock {Diagrammatic Methods in Statistics and Biasing in the Large-Scale
  Structure of the Universe}.
\newblock {\em \apjs}, 101:1, November 1995.

\bibitem{ParSefChu1305}
Aseem {Paranjape}, Emiliano {Sefusatti}, Kwan~Chuen {Chan}, Vincent
  {Desjacques}, Pierluigi {Monaco}, and Ravi~K. {Sheth}.
\newblock {Bias deconstructed: unravelling the scale dependence of halo bias
  using real-space measurements}.
\newblock {\em \mnras}, 436:449--459, November 2013.

\bibitem{SchBalSel1411}
Marcel {Schmittfull}, Tobias {Baldauf}, and Uro{\v{s}} {Seljak}.
\newblock {Near optimal bispectrum estimators for large-scale structure}.
\newblock {\em \prd}, 91:043530, February 2015.

\bibitem{HofBelGaz1702}
K.~{Hoffmann}, J.~{Bel}, and E.~{Gazta{\~n}aga}.
\newblock {Linear and non-linear bias: predictions versus measurements}.
\newblock {\em \mnras}, 465:2225--2235, February 2017.

\bibitem{BiaDesKeh1408}
Matteo {Biagetti}, Vincent {Desjacques}, Alex {Kehagias}, and Antonio {Riotto}.
\newblock {Halo velocity bias}.
\newblock {\em \prd}, 90:103529, November 2014.

\bibitem{CatLucMat9807}
P.~{Catelan}, F.~{Lucchin}, S.~{Matarrese}, and C.~{Porciani}.
\newblock {The bias field of dark matter haloes}.
\newblock {\em \mnras}, 297:692--712, July 1998.

\bibitem{Scoccimarro:1998b}
R.~{Scoccimarro}.
\newblock {Transients from initial conditions: a perturbative analysis}.
\newblock {\em \mnras}, 299:1097--1118, October 1998.

\bibitem{SchJeoDes1307}
Fabian {Schmidt}, Donghui {Jeong}, and Vincent {Desjacques}.
\newblock {Peak-background split, renormalization, and galaxy clustering}.
\newblock {\em \prd}, 88:023515, Jul 2013.

\bibitem{Avi1810}
Alejandro {Aviles}.
\newblock {Renormalization of Lagrangian bias via spectral parameters}.
\newblock {\em \prd}, 98:083541, Oct 2018.

\bibitem{PajZal1308}
E.~{Pajer} and M.~{Zaldarriaga}.
\newblock {On the renormalization of the effective field theory of large scale
  structures}.
\newblock {\em \jcap}, 8:37, August 2013.

\bibitem{BlaGarKon1309}
D.~{Blas}, M.~{Garny}, and T.~{Konstandin}.
\newblock {Cosmological perturbation theory at three-loop order}.
\newblock {\em ArXiv e-prints}, September 2013.

\bibitem{BerTarNis1401}
F.~{Bernardeau}, A.~{Taruya}, and T.~{Nishimichi}.
\newblock {Cosmic propagators at two-loop order}.
\newblock {\em \prd}, 89(2):023502, January 2014.

\bibitem{BalMerZal1507}
T.~{Baldauf}, L.~{Mercolli}, and M.~{Zaldarriaga}.
\newblock {The Effective Field Theory of Large Scale Structure at Two Loops:
  the apparent scale dependence of the speed of sound}.
\newblock {\em ArXiv e-prints}, July 2015.

\bibitem{Sco97}
R.~{Scoccimarro}.
\newblock {Cosmological Perturbations: Entering the Nonlinear Regime}.
\newblock {\em \apj}, 487:1--+, 1997.

\bibitem{LazWagBal1602}
Titouan {Lazeyras}, Christian {Wagner}, Tobias {Baldauf}, and Fabian {Schmidt}.
\newblock {Precision measurement of the local bias of dark matter halos}.
\newblock {\em Journal of Cosmology and Astro-Particle Physics}, 2016:018, Feb
  2016.

\bibitem{StressTensor}
G.~{D'Amico}, M.~{Crocce}, and R.~{Scoccimarro}.
\newblock In preparation (2018).

\bibitem{Pue0810}
S.~{Pueblas}.
\newblock {\em {Dark matter clustering in precision cosmology}}.
\newblock PhD thesis, New York University, 2008.

\bibitem{DesCroSco1011}
V.~{Desjacques}, M.~{Crocce}, R.~{Scoccimarro}, and R.~K. {Sheth}.
\newblock {Modeling scale-dependent bias on the baryonic acoustic scale with
  the statistics of peaks of Gaussian random fields}.
\newblock {\em \prd}, 82(10):103529--+, November 2010.

\bibitem{Heavens:1998}
A.~F. {Heavens}, S.~{Matarrese}, and L.~{Verde}.
\newblock {The non-linear redshift-space power spectrum of galaxies}.
\newblock {\em \mnras}, 301:797--808, December 1998.

\bibitem{Schmidt:2016}
Fabian {Schmidt}.
\newblock {Towards a self-consistent halo model for the nonlinear large-scale
  structure}.
\newblock {\em \prd}, 93:063512, March 2016.

\bibitem{Ginzburg:2017}
Dimitry {Ginzburg}, Vincent {Desjacques}, and Kwan~Chuen {Chan}.
\newblock {Shot noise and biased tracers: A new look at the halo model}.
\newblock {\em \prd}, 96:083528, October 2017.

\bibitem{Pee80}
P.{}J.{}E. {Peebles}.
\newblock {\em {The large-scale structure of the universe}}.
\newblock Princeton University Press, 1980.

\bibitem{BiaDesKeh1405}
Matteo {Biagetti}, Vincent {Desjacques}, Alex {Kehagias}, and Antonio {Riotto}.
\newblock {Nonlocal halo bias with and without massive neutrinos}.
\newblock {\em \prd}, 90:045022, August 2014.

\bibitem{SanScoCro1701}
A.~G. {S{\'a}nchez}, R.~{Scoccimarro}, M.~{Crocce}, J.~N. {Grieb},
  S.~{Salazar-Albornoz}, C.~{Dalla Vecchia}, M.~{Lippich}, F.~{Beutler}, J.~R.
  {Brownstein}, C.-H. {Chuang}, D.~J. {Eisenstein}, F.-S. {Kitaura}, M.~D.
  {Olmstead}, W.~J. {Percival}, F.~{Prada}, S.~{Rodr{\'{\i}}guez-Torres}, A.~J.
  {Ross}, L.~{Samushia}, H.-J. {Seo}, J.~{Tinker}, R.~{Tojeiro},
  M.~{Vargas-Maga{\~n}a}, Y.~{Wang}, and G.-B. {Zhao}.
\newblock {The clustering of galaxies in the completed SDSS-III Baryon
  Oscillation Spectroscopic Survey: Cosmological implications of the
  configuration-space clustering wedges}.
\newblock {\em \mnras}, 464:1640--1658, January 2017.

\bibitem{AngFasSen1503}
Raul {Angulo}, Matteo {Fasiello}, Leonardo {Senatore}, and Zvonimir {Vlah}.
\newblock {On the statistics of biased tracers in the Effective Field Theory of
  Large Scale Structures}.
\newblock {\em Journal of Cosmology and Astro-Particle Physics}, 2015:029,
  September 2015.

\bibitem{CarReiWhi1302}
Jordan {Carlson}, Beth {Reid}, and Martin {White}.
\newblock {Convolution Lagrangian perturbation theory for biased tracers}.
\newblock {\em \mnras}, 429:1674--1685, Feb 2013.

\bibitem{VlaCasWhi1612}
Zvonimir {Vlah}, Emanuele {Castorina}, and Martin {White}.
\newblock {The Gaussian streaming model and convolution Lagrangian effective
  field theory}.
\newblock {\em Journal of Cosmology and Astro-Particle Physics}, 2016:007, Dec
  2016.

\bibitem{Ber92}
F.~{Bernardeau}.
\newblock {The gravity-induced quasi-Gaussian correlation hierarchy}.
\newblock {\em \apj}, 392:1--14, 1992.

\end{thebibliography}

\end{document}